\documentclass[a4paper,fleqn,usenatbib]{mnras}

\usepackage{newtxtext,newtxmath}

\usepackage[T1]{fontenc}
\usepackage{ae,aecompl}
\usepackage{gensymb}

\usepackage{graphicx}	
\usepackage{amsmath}	
\usepackage{amssymb}	
\usepackage{pdflscape}	
\usepackage{float}
\usepackage[lofdepth,lotdepth]{subfig}
\usepackage{ulem}       

\usepackage{CJK}

\newcommand{\hii}{H~{\scshape ii}~}
\newcommand{\um}{$\mu$m}

\title[Milky Way Project DR2]{The Milky Way Project Second Data Release: \\ Bubbles and Bow Shocks }

\author[T. Jayasinghe et al.]{Tharindu Jayasinghe$^{1,2,3}$\thanks{E-mail: jayasinghearachchilage.1@osu.edu},
Don Dixon$^{1,4,5}$,
Matthew S. Povich$^{1,6}$, 
Breanna Binder$^{1}$,
\newauthor Jose Velasco$^{1}$, 
Denise M. Lepore$^{1}$,
Duo Xu$^{7}$,
Stella Offner$^{7}$, Henry A. Kobulnicky$^{8}$,
\newauthor  Loren D. Anderson$^{9}$, Sarah Kendrew$^{10}$, and Robert J. Simpson$^{11}$\\
$^{1}$Department of Physics and Astronomy, California State Polytechnic University Pomona, 3801 West Temple Avenue, Pomona CA 91768, USA\\
$^{2}$Department of Astronomy, The Ohio State University, 140 West 18th Avenue, Columbus, OH 43210, USA \\ 
$^{3}$Centre for Cosmology and Astroparticle Physics, The Ohio State University, 191 W. Woodruff Avenue, Columbus, OH 43210, USA\\
$^{4}$Vanderbilt University, Department of Physics and Astronomy, 6301 Stevenson centre Lane, Nashville TN, 37235, USA \\ 
$^{5}$Fisk University, Department of Physics, 1000 17th Avenue N., Nashville, TN 37208, USA\\
$^{6}$Visitor in Astronomy, California Institute of Technology, Pasadena, CA 91125, USA \\
$^{7}$Department of Astronomy, The University of Texas at Austin, Austin, TX 78712, USA \\
$^{8}$Department of Physics \& Astronomy, University of Wyoming, Dept 3905, Laramie, WY 82070-1000, USA \\
$^{9}$Department of Physics and Astronomy, West Virginia University, Morgantown, WV 26506, USA\\
$^{10}$European Space Agency, Space Telescope Science Institute, 3700 San Martin Drive, Baltimore MD 21218, USA\\
$^{11}$Oxford Astrophysics, Denys Wilkinson Building, Keble Road, Oxford OX1 3RH, UK
}

\date{Accepted XXX. Received YYY; in original form ZZZ}

\pubyear{2019}

\begin{document}
\label{firstpage}
\pagerange{\pageref{firstpage}--\pageref{lastpage}}
\maketitle

\begin{abstract}
Citizen science has helped astronomers comb through large data sets to identify patterns and objects that are not easily found through automated processes. The Milky Way Project (MWP), a citizen science initiative on the Zooniverse platform, presents internet users with infrared (IR) images from \textit{Spitzer Space Telescope} Galactic plane surveys. MWP volunteers make classification drawings on the images to identify targeted classes of astronomical objects. We present the MWP second data release (DR2) and an updated data reduction pipeline written in Python. We aggregate ${\sim}3$ million classifications made by MWP volunteers during the years 2012--2017 to produce the DR2 catalogue, which contains 2600 IR bubbles and 599 candidate bow-shock driving stars. 
The reliability of bubble identifications, as assessed by comparison to visual identifications by trained experts and scoring by a machine-learning algorithm, is found to be a significant improvement over DR1. We assess the reliability of IR bow shocks via comparison to expert identifications and the colours of candidate bow-shock driving stars in the 2MASS point-source catalogue. We hence identify highly-reliable subsets of 1394 DR2 bubbles and 453 bow-shock driving stars.
Uncertainties on object coordinates and bubble size/shape parameters are included in the DR2 catalog. Compared with DR1, the DR2 bubbles catalogue provides more accurate shapes and sizes. The DR2 catalogue identifies 311 new bow shock driving star candidates, including three associated with the giant \hii regions NGC 3603 and RCW 49.
\end{abstract}



\section{Introduction}

Massive, O and early B-type (OB) stars comprise no more than a few percent of the stellar population in star-forming galaxies. In spite of their rarity, feedback effects from the powerful radiation fields, stellar winds, and eventual supernova explosions of OB stars dominate the observed morphology of star-forming galaxies across the electromagnetic spectrum, sculpt the interstellar medium (ISM), and drive galaxy evolution \citep{2003ARA&A..41...15M}. Because OB stars are short-lived, \hii regions ionised by their UV radiation trace sites of recent and ongoing star formation. The total size and spatial distribution of the OB population and hence the star-formation rate in the Milky Way have long been inferred from the observed distribution of radio \hii regions \citep{SBM78,MR10,CP11}. However, measurements of the Galactic ionising photon budget still must correct for absorption by dust that significantly reduces the radio brightness of Galactic \hii regions \citep{MW97,BP18}.
Additionally, a significant fraction of OB stars may escape their natal \hii regions as runaways \citep{1967BOTT....4...86P}, traveling many kpc during their lifetimes and depositing their feedback in far-flung locations throughout the Galaxy. The spatial distribution of Galactic OB stars and luminosity function of young massive star clusters remain poorly known.

Dust bubbles blown by the stellar winds and/or radiation pressure from individual OB stars or massive star clusters \citep{W77,Draine11} provide readily-identifiable mid-infrared (IR) morphologies for Galactic \hii regions. 
The molecular photo-dissociation regions (PDRs) surrounding \hii regions are traced by bright 8~$\mu$m emission from polycyclic aromatic hydrocarbons (PAHs). PAH molecules are excited by UV photons that leak out of \hii regions. The IRAC 8~$\mu$m band encompasses two PAH emission lines at 7.7~$\mu$m and 8.3~$\mu$m, with the 7.7~$\mu$m emission line being the stronger of the two \citep{1987ppic.proc..305A}. Inside these PAH rims, dust mixed with ionised gas and heated by the hard radiation field produces bright 24~$\mu$m nebulosity that closely matches the radio continuum emission \citep{cp06,2008ApJ...681.1341W,2010ApJ...716.1478W}. 
This morphology of a bright 8 $\mu$m ring surrounding a central arc/torus of 24 $\mu$m emission is also characteristic of giant Galactic \hii regions \citep{2007ApJ...660..346P}.

In the first systematic search for dust bubbles in the Milky Way, \citet[][hereafter CP06 and CWP07]{cp06,cwp07} catalogued nearly 600 IR bubbles in the inner 130\degree\ of the Galactic plane by visually reviewing 3.6 $\mu$m -- 8.0 $\mu$m images from the Galactic Legacy Infrared Mid-Plane Survey Extraordinaire survey (GLIMPSE; \citealt{2003PASP..115..953B,2009PASP..121..213C}). The resultant catalogs, although reliable, were described by the authors as very incomplete. The Milky Way Project (MWP; \citealt{Simpson2012}, hereafter SPK12), launched in 2010 on the Zooniverse platform for online citizen science, scoured the same search area in 3-colour 4.5~$\mu$m and 8~$\mu$m GLIMPSE plus 24~$\mu$m MIPSGAL \citep{2009PASP..121...76C} images over a period of two years. The MWP first data release (DR1) produced a catalogue of over 5000 bubbles, including 86\% of the bubbles catalogued by CP06 and CWP07 (SPK12). \citet[][hereafter A14]{at11,A14} further expanded the search for IR bubbles beyond the boundaries of the GLIMPSE survey by visually inspecting images from the {\textit Wide-Field Infrared Survey Explorer} (\textit{WISE}; \citealt{2010AJ....140.1868W}), ultimately cataloging over 8000 IR bubbles and other IR-bright \hii regions.

CP06 noted that the spatial distributions in longitude and latitude of IR bubbles and OB stars were similar, found strong correlations between the locations of individual bubbles and known Galactic \hii regions \citep{2003A&A...397..213P}, and hence concluded that most IR bubbles are produced by OB stars. Half of the \textit{WISE} regions identified by A14 are associated with radio continuum emission, making them confirmed or candidate \hii regions. 

As the relative velocity between an individual OB star and the ambient ISM increases, a wind-blown bubble becomes increasingly deformed from a circular shape, presenting a more elliptical or even arc-like morphology. In cases of supersonic relative velocity a bow shock can form with a parabolic, arc-like morphology \citep{1988ApJ...329L..93V}.

The strong winds or radiation pressure from OB stars sweep up interstellar dust that becomes heated by the strong stellar radiation field, producing characteristic arc-shaped nebulae in IR images \citep{1995AJ....110.2914V,K+10,SPS15,HA19a}. In this work we will refer to all nebular IR arcs as candidate bow shocks, which presumes the wind-driven mechanism that appears to dominate the majority of such objects scrutinized to date \citep{HA19b}.
Broadly separated into two classes, the driving stars of bow shocks are either (1) stars with high peculiar velocities sweeping up ambient gas and dust as they move \citep{GB08,G+11}, or (2) {\it in situ}, where gas and dust from an expanding \hii region flows around a stationary star \citep{Povich+08}. 
The largest catalogue of Galactic IR bow shocks includes 709 candidates, the great majority of which are in isolated locations far from known star-forming regions, suggesting high peculiar velocities (\citealt{2016ApJS..227...18K}; hereafter K16). These objects have expanded our view of the spatial distribution of Galactic massive stars and provide a new, independent method for measuring the poorly-constrained mass-loss rates of OB stars \citep{KCP18,HA19b}.

In this paper, we present the second data release (DR2) for the MWP, which includes an updated catalogue of 2600 IR bubbles identified by the collective work of ${>}31 000$ citizen scientists visually inspecting survey images from the GLIMPSE, MIPSGAL, SMOG \citep{2008sptz.prop50398C} and Cygnus-X \citep{2009AAS...21335601H} surveys of the Galactic plane. 
Six major changes in the bubble identification and data analysis processes makes this catalogue the replacement for the now-deprecated DR1 catalogue:

\begin{enumerate}

\item Using an updated bubble drawing tool, MWP users fit ellipses at the interiors of bubbles instead of fitting elliptical annuli to bubble morphologies seen in MWP image cutouts. This resulted in improved bubble shape and size measurements.
\item MWP users had access to a much larger set of image cutouts with a maximum zoom level that was twice that employed in DR1. This resulted in the identification of small bubbles to a greater degree of precision.
\item DR2 presents a single, unified catalogue of bubbles across all angular sizes, which eliminates the problem of duplication between the DR1 large and small bubble catalogs.
\item Each bubble candidate is assigned a reliability flag based on their ``hit rate" (SPK12). Employing the distributions of hit rates across the two versions of the MWP, we flag highly-reliable bubbles as distinct from bubble candidates in the more complete sample.
\item The DR2 catalogue was cross-matched to the A14 \textit{WISE} catalogue of \hii regions to minimize the number of spurious bubbles. We only retained the DR2 bubbles that had a match in the A14 catalogue and the unmatched bubbles that passed a visual review.
\item The DR2 catalogue includes measurement uncertainties on bubble location, size, shape, and orientation parameters.

\end{enumerate}

MWP DR2 also includes the first citizen-science catalogue of 599 IR bow shock driving star candidates (BDSCs), including 311 newly-discovered objects. The combined K16 and MWP DR2 catalogs now comprise the most comprehensive list of Galactic IR bow shock candidates currently available, including 1,019 unique BDSCs, the great majority of which are expected to be OB stars. 

The locations of BDSCs were automatically cross-matched with those of DR2 bubbles, avoiding the subjectivity of the K16 catalogue for identifying isolated bow shock candidates and also revealing a number of morphologies that appear to be transitional objects with properties of both bubbles and bow shocks. As with bubbles, the DR2 BDSC catalogue flags a subset of highly-reliable candidates. Unlike the case of bubbles, but analogous to the procedure of K16, the sizes and orientations of the MWP bow shocks themselves were measured by hand by one of the co-authors (DD).

The remaining sections of this paper are organized as follows: in Sections 2 and 3 we present updates to the MWP online Zooniverse platform and our back-end data analysis procedures for constructing catalogues of astronomical objects from the data provided by citizen scientists. In sections 4 and 5 we present the MWP DR2 bubbles and BDSC catalogues, respectively. In Section 6 we discuss our results, emphasizing the performance of citizen scientists versus visual searches by `expert' astronomers and new insights gained from a unified analysis of bubbles and bow shocks. We summarize our conclusions in Section 7.

\section{Revisiting the Milky Way Project}
While successful in cataloging the spatial locations and general shapes of Galactic IR bubbles, both the MWP DR1 and A14 catalogs lacked precision when measuring the shapes and sizes of these objects and did not include uncertainties on these parameters. The MWP revisited this task following the release of DR1. In the second version of the MWP (V2), immediately following the release of DR1 in 2012, we uploaded 3-colour image assets produced exclusively with IRAC data, thereby allowing users to search for bubbles in the same RGB colour scheme used by CP06 and CWP07. Since the availability of 24~$\mu$m MIPSGAL data was not a requirement for V2, we were able to expand our search to the GLIMPSE3D and Vela-Carina surveys. We also searched for objects of interest in the Cygnus-X  and SMOG surveys. Key statistics and descriptive information for all three versions of the MWP are summarized in Table~\ref{mwpstats}. 

\begin{figure*}
	\includegraphics[height=0.9\textheight]{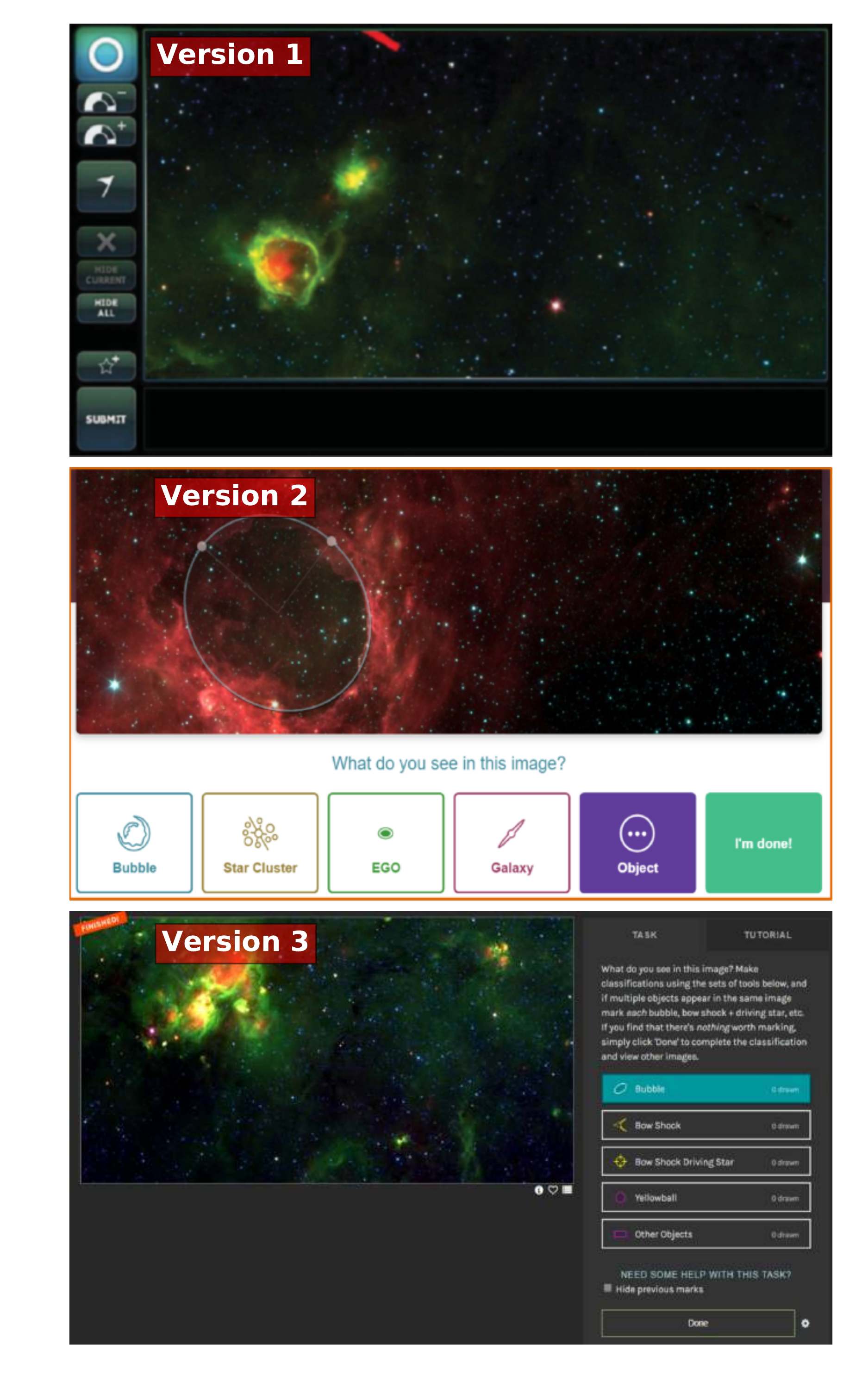}
	\caption{Screenshots of the user classification interface from all three versions of the MWP, from \textit{top} to \textit{bottom}: V1, V2 and V3.}
	\label{mwp2interface}
\end{figure*}

The MWP user interface was redesigned completely for each subsequent version of the project (Figure \ref{mwp2interface}). For V3, we used Zooniverse's project builder tool\footnote{\url{https://github.com/zooniverse/Panoptes}} and returned to the V1 RGB colour scheme with JPEG image assets including 24~$\mu$m data.  Two workflows were available to volunteers: a training workflow (`Learning the Ropes') that guided them through the identification of MWP objects of interest  guided by the results of expert classifications by one of us (MSP), and the primary workflow (`Mapping the Milky Way') for producing classifications used in subsequent science analysis. A short tutorial showed users how to use the different drawing tools, alongside examples of bubbles, bow shocks and yellowballs. In addition to the tutorial, users were able to access a dedicated help tool that provided further examples of these objects. Once the user had finished with a particular image, they were prompted to take part in the MWP Talk forum to discuss their thoughts on that image. As of March 2019, the MWP V3 talk forum hosted ${>}16~000$ discussion threads with ${>}25~000$ comments from ${\gtrsim}1000$ participants\footnote{\url{https://www.zooniverse.org/projects/povich/milky-way-project/talk/}}.

\begin{table*}
	\caption{Overview of the Milky Way Project across the years}
	\label{mwpstats}
	\begin{tabular}{|l|c|c|c|}
		\hline
		& \textbf{Version 1} & \textbf{Version 2} & \textbf{Version 3} \\
		\hline
		Years active & 2010--2012 &2012--2015 & 2016--2018\\
		Registered users & ${\sim}16$ 000 & 23 858 & 7 293 \\
		Bubble classifications & 520 120 & 504 933 & 243 478\\
   		Bow shock classifications & --- & --- & 25 233\\
 		Total classifications & ${>}1\times 10^6$ & ${>}2\times 10^6$ & ${>}1\times 10^6$	\\
        \hline
		Surveys included &  GLIMPSE+MIPSGAL & \begin{tabular}{@{}c@{}}GLIMPSE, Vela-Carina, GLIMPSE3D, \\ SMOG, Cygnus-X\end{tabular} & \begin{tabular}{@{}c@{}}GLIMPSE+MIPSGAL, SMOG, \\ Cygnus-X\end{tabular}	\\
        \hline
		Longitude coverage &$0\degree<l<65 \degree $, $295\degree<l<360 \degree $ &\begin{tabular}{@{}c@{}}$0\degree<l<65 \degree $, $295\degree<l<360 \degree $ ,\\ $255\degree<l<295 \degree$,\\ $10\degree<l<65 \degree$, \\ $102\degree<l<109 \degree$, \\ $76\degree<l<82 \degree $\end{tabular} & \begin{tabular}{@{}c@{}}$0\degree<l<65 \degree $, $295\degree<l<360 \degree $ , \\ $102\degree<l<109 \degree$, \\ $76\degree<l<82 \degree $\end{tabular}\\
        \hline
		Latitude coverage & $-1\degree<b<1 \degree $ & \begin{tabular}{@{}c@{}}$-1\degree<b<1 \degree $ , \\$-1.5\degree<b<1.5 \degree $ ,\\$|b|>1\degree$,\\ $0\degree<b<3 \degree$, \\ $-2.3\degree<b<4.1 \degree $\end{tabular} & \begin{tabular}{@{}c@{}}$-1\degree<b<1 \degree $ , \\ $0\degree<b<3 \degree$, \\ $-2.3\degree<b<4.1 \degree $\end{tabular} \\
        \hline
	JPEG RGB bandpasses & [24], [8.0], [4.5] & [8.0], [4.5], [3.6] & [24], [8.0], [4.5] \\
    Max JPEG image zoom & $0.3\degree \times 0.15\degree$ & $0.15\degree \times 0.075\degree$ & $0.15\degree \times 0.075\degree$ \\
	Total JPEG image assets & 12 000 & 121 310 & 77 017 \\	
		\hline
	\end{tabular}
	
\end{table*}

\subsection{MWP Image Assets}
As in MWP V1, our image assets in V2 and V3 were multiband, three-colour $800\times 400$~pixel JPEG image cutouts from the original large FITS mosaics produced by the GLIMPSE team. Image asset cutouts were produced in overlapping grids to allow all parts of the survey mosaics to be visible across all zoom levels. An object of interest in the Galactic plane will appear more than once in any given zoom level, guaranteeing that users will be shown at least one image with the object close to the centre of the image. Grids of image assets were made at 3 different `zoom' levels: $1.0\degree \times 0.5\degree$, $0.5\degree \times 0.25\degree$ and $0.15\degree \times 0.075\degree$. The highest zoom level in the MWP V1 image assets was $0.3\degree \times 0.15\degree$, a factor of four lower on-screen pixel resolution than the highest zoom level available in V2 and V3. The primary motivation to change the image zoom levels was to capture smaller objects, including small bubbles, bow shocks and yellowballs, to a higher degree of completeness and precision. This increased zoom and area coverage in V2 and V3 increased the number of images by more than a factor of 4 (Table~\ref{mwpstats}), increasing the time necessary to collect classifications and requiring the retirement of MWP image assets after 30 views.

A square-root stretch function (with the faintest 5\% of the pixels set to black and the brightest 2\% set to white) was applied independently to each of the three colour channels to provide an optimal dynamic range for the identification of both bubbles and bow shocks. For images in V2, we assigned GLIMPSE band images to a colour channel as follows: red = [8.0], green = [4.5], blue = [3.6]. For V3, we assigned GLIMPSE and MIPSGAL images as follows: red = [24], green = [8.0], blue = [4.5]. Saturated pixels were set to white to retain visual appeal in the JPEG images. This issue was prevalent in massive star-forming regions with bright 8.0~$\mu$m and 24~$\mu$m nebulosity.

\subsection{Bubble classifications}

MWP V2 and V3 provided a simple ellipse drawing tool to identify bubbles, replacing the far more complicated elliptical annulus tool used in MWP V1. Using the updated bubble drawing tool, MWP users were directed to capture the sharp inner 8~$\mu$m bubble rings only. This greatly simplified the task of producing a bubble classification, as users no longer needed to define an outer boundary for a bubble, which is in any case poorly-defined and dependent on the image stretch chosen, as the 8~$\mu$m nebulosity decays quasi-exponentially with increasing distance from the centre of a star-forming region \citep{2008ApJ...681.1341W, BP18}.

\begin{figure}
\includegraphics[width=0.45\textwidth]{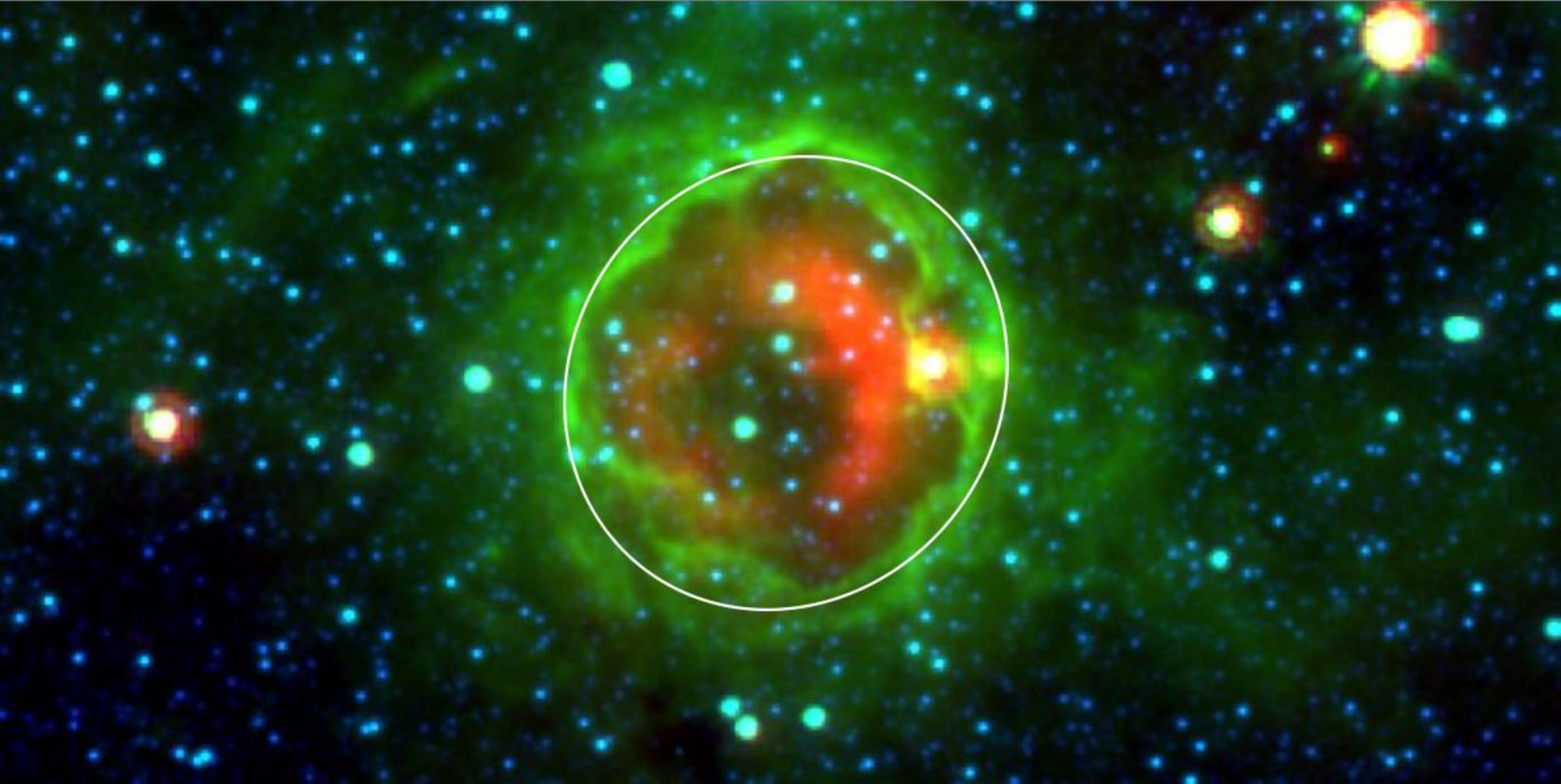}
\includegraphics[width=0.45\textwidth]{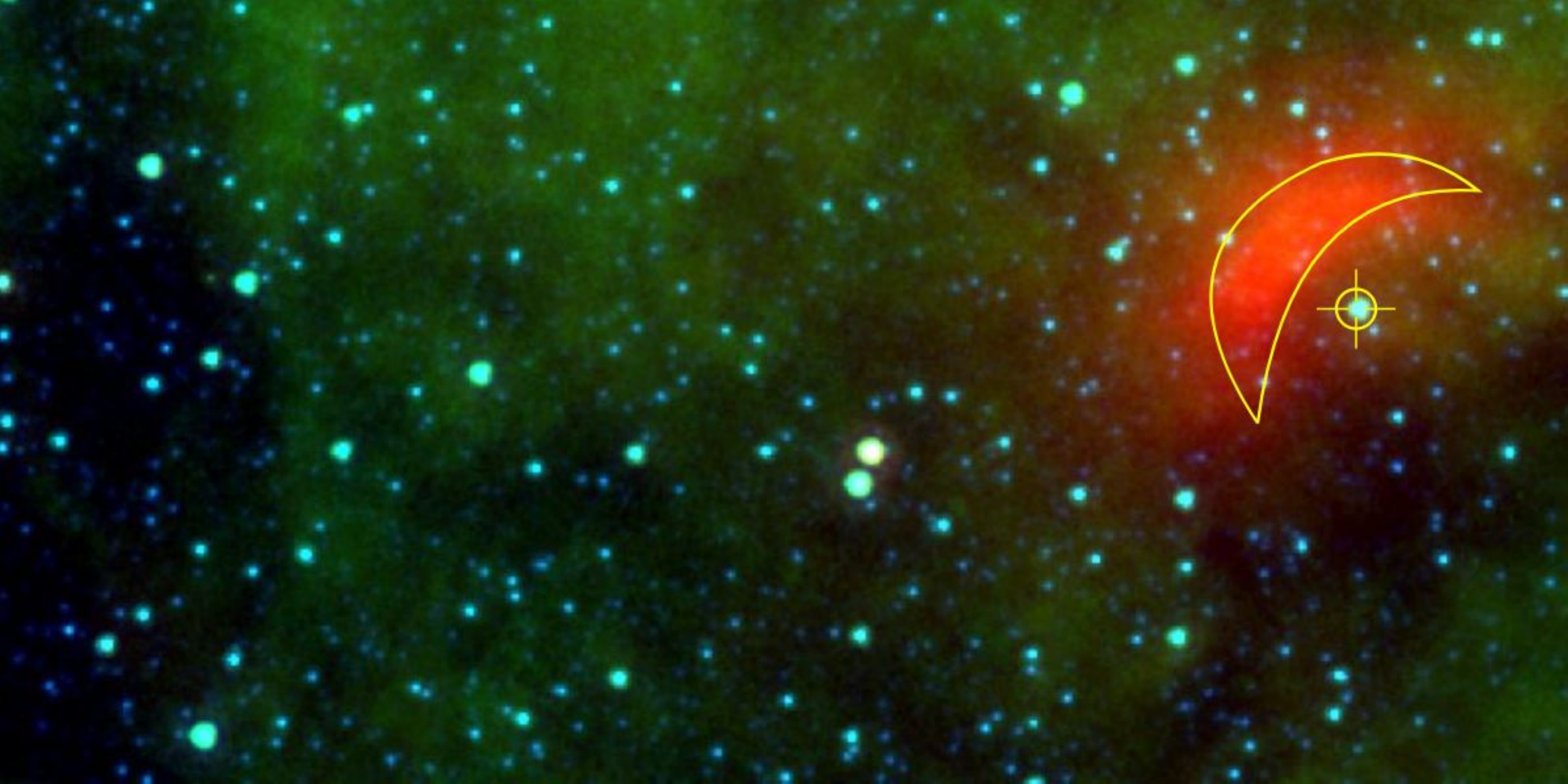}
	\caption{\textit{Top}: Example of a bubble classification on a MWP V3 image asset (blue = 4.5~\um, green = 8.0~\um, red = 24~\um). \textit{Bottom}: Example of a complete bow shock plus driving star classification. Colour figure available online.}
	\label{classifex}
\end{figure}

To mark a bubble, a user drew an ellipse that could be shifted, scaled in size and rotated to fit the location and shape of the bubble (Figure \ref{classifex}). Once drawn, the user could edit the parameters of the ellipse up until all classifications for that image asset are submitted. Users were also allowed to delete their drawings at any point before submitting classifications. By default, the semi-major and semi-minor axes of an ellipse had a ratio of 2:1. Bubble classifications with this ratio in their semi-major and semi-minor axes were deemed imprecise and assigned a lower weight. We discuss user weighting in Section 3.3 below. Of all the bubbles drawn (in V2 and V3 combined), 33.6$\%$ were imprecise.

\subsection{Bow shock and driving star classifications}
Bow shock classifications were made in V3 using the new \lq\lq B\'ezier\rq\rq\ drawing tool provided by the Zooniverse project builder. This tool gave volunteers the ability to draw a polygon and then bend its sides to achieve a \lq\lq best fit\rq\rq\ around the red (24 $\mu$m) arc of a bow shock. After closing a polygon users completed their classification by marking the star they judged to be driving the bow shock with a recticle. Only complete classifications, including both a polygon arc drawing and a reticle position, were nominated for inclusion in our bow shock catalog, as illustrated by the example in Figure~\ref{classifex}. We additionally required that complete bow shock classifications be generated from image assets at the highest zoom level only. For subsequent data analysis only the coordinates of the BDSC reticle were recorded and used, because of the imprecision and complexity of the B\'ezier polygon drawings. 

\subsection{Yellowballs and other objects}

\citet{Kerton+15} presented a set of compact, IR sources, bright in both the 8 $\mu$m and 24 $\mu$m bands, that had been frequently discussed on the Talk platform of MWP V1. In V3 we provided a dedicated circle tool for users to classify these `yellowballs.' The circle tool and higher maximum zoom level in V3 allowed users to provide more accurate information about the locations and sizes of yellowballs compared to V1. Classifications of V3 yellowballs will be presented in a future paper.

MWP users were always able to flag interesting objects that do not belong to the above mentioned categories using the `Other Objects' tool. This tool allowed users to draw a box around the region of interest in the image asset and then choose from a menu of object that included (for V3) `Star Cluster', `Galaxy', `Pillar', `Artifact', or `Other'. In V2, `Bow Shock' was included among a menu options for other objects. Frequently the other objects tool was used to stimulate discussion on MWP Talk, and in Section 6 we discuss two examples of serendipitous discoveries made by volunteers using the other objects tool.

\section{Catalogue construction}
The aggregation of user-made classifications is a critical component of any citizen science project. From our database of $\gtrsim3\times10^6$ total classifications collected over a 4-year period, we used 748 411 and 25 233 classifications (Table~\ref{mwpstats}) to create the DR2 bubble and bow shock catalogs, respectively. 
Classifications from both the second and third versions of the MWP were used in creating the bubble catalog. The bow shock classification tool was introduced in V3. 

\subsection{MWP user statistics}

The MWP has been one of the most popular citizen science projects on the Zooniverse platform over the years. Due to its continuing popularity, both V1 and V2 were translated into numerous languages including Spanish, German, French, Indonesian, Polish and Danish. As the project varied between its different iterations, so did its user base. An electronic table listing all the registered MWP volunteers who contributed to the DR2 catalogue is published along with this paper. The total number of registered users for each version of the MWP is listed in Table \ref{mwpstats}. Because users were not required to register to make classifications, these numbers are lower limits to the actual number of distinct users who accessed the website during each version. We also did not track the identities of individual registered users who contributed to more than one MWP version, although from MWP Talk discussions we know that numerous cross-version users exist. We hence cannot report a precise total number of unique individuals who have contributed to MWP over the lifetime of the project.

\subsection{Aggregating the work of MWP volunteers}

We created a data reduction pipeline in Python to handle the aggregation of MWP classifications. Classification data were obtained from the Zooniverse and parsed into an appropriate format before analysis. In the case of MWP V2, data were obtained in the  {\scshape MongoDB} format and then parsed in to code-readable text files. Data from V3 were obtained as a CSV file, which was then parsed into text files having the same data structure as V2. These parsed text files were used in the data reduction pipeline for our analysis. Raw data from both V2 and V3 report the central coordinates and the axes of bubble ellipses in the pixel coordinates (\textit{x,y}) of each image asset. For data analysis, these parameters were converted to Galactic coordinates (\textit{l,b}) during the parsing process, using the central coordinates and pixel dimensions of each image asset.

The driving feature of the MWP data reduction pipeline was the use of a density-based clustering algorithm. We used the Hierarchical Density-Based Spatial Clustering of Applications with Noise ({\scshape hdbscan}; \citealt{camp13}) algorithm to create the bubble catalog.  
Given sets of multi-dimensional sequences of values (`tuples'),  {\scshape hdbscan} groups together tuples that are close together based on a distance metric, labels this collection of tuples as a cluster and flags any other tuples that are not part of a cluster as outliers. Each tuple in the input is assigned an outlier score that ranges from 0 (least likely to be an outlier) to 1 (most likely to be an outlier).   {\scshape hdbscan} requires a single user-defined parameter, $\kappa$, which defines the minimum number of points in a cluster. We chose $\kappa=5$ to find bubble candidates that were seen and identified by at least five MWP users. To identify clusters of bubble classifications, the central coordinates and the effective radii of the bubble classifications were used as input tuples $(l,b,R_{\rm eff})$ to the {\scshape hdbscan} algorithm. This tuple was selected to identify bubble classifications with similar sizes clustered in the same spatial region.

{\scshape hdbscan} improves upon the more commonly used density-based clustering algorithm ({\scshape dbscan}; \citealt{ester1996}) in several ways that makes it preferable for the identification of bubbles. Real astrophysical bubbles display a large range in  angular sizes.  Bubbles with larger angular sizes display greater absolute dispersions in the central coordinates of their classification ellipses compared to smaller ones. This produced valid clusters of bubble classifications with varying absolute densities in the three-dimensional position+radius space defined by our tuples.  {\scshape hdbscan} finds clusters of varying densities with minimal parameter tuning, whereas {\scshape dbscan} would be biased towards clusters that are found using a single set of user-defined density thresholds.  In our new MWP data reduction pipeline, we employed the \verb"scikit-learn" \citep{2012arXiv1201.0490P} enabled implementation of  {\scshape hdbscan} by \citet{mcinnes}. 

 {\scshape hdbscan} starts its search for clusters by selecting an arbitrary tuple and searching for nearest neighbors inside a spherical region of radius $\epsilon$ around this tuple. If this region contains ${\ge}\kappa$ tuples, it is marked as a cluster. Otherwise, this starting tuple does not belong to a cluster and is labeled as an outlier with an outlier score of 1. All the tuples that fall within this $\epsilon$ neighborhood are assigned to this cluster and given an outlier score within the range [0,1]. Outlier scores are calculated using the GLOSH outlier detection algorithm \citep{camp15}. An unvisited tuple is visited once the density-connected cluster is completely found. This process also repeats, varying $\epsilon$ to find clusters that are stable over $\epsilon$. 

  {\scshape hdbscan} was run twice in order to identify clusters with minimal outliers. Following the first iteration, we calculated the 90th percentile in the distribution of outlier scores for clustered classifications and removed the classifications
that had outlier scores greater than this value (0.59 in V2, 0.64 in V3). This process increased the quality of the clusters by eliminating loosely clustered classifications. We ran  {\scshape hdbscan} once more on this reduced set of classifications and performed a similar clipping of classifications at the 90th percentile in outlier scores (0.45 in V2, 0.48 in V3) prior to the next step in our data reduction pipeline. Through this process of outlier removal, ${\sim}50~000$ and ${\sim}24~000$ bubble classifications were removed from V2 and V3, respectively.

Our pipeline can find the clusters within a database of one million classifications in ${\lesssim}5$ minutes on a standard desktop computer and is stable under multiple iterations, making it the best choice for creating a bubble catalog. 

\begin{figure}
\includegraphics[width=\columnwidth]{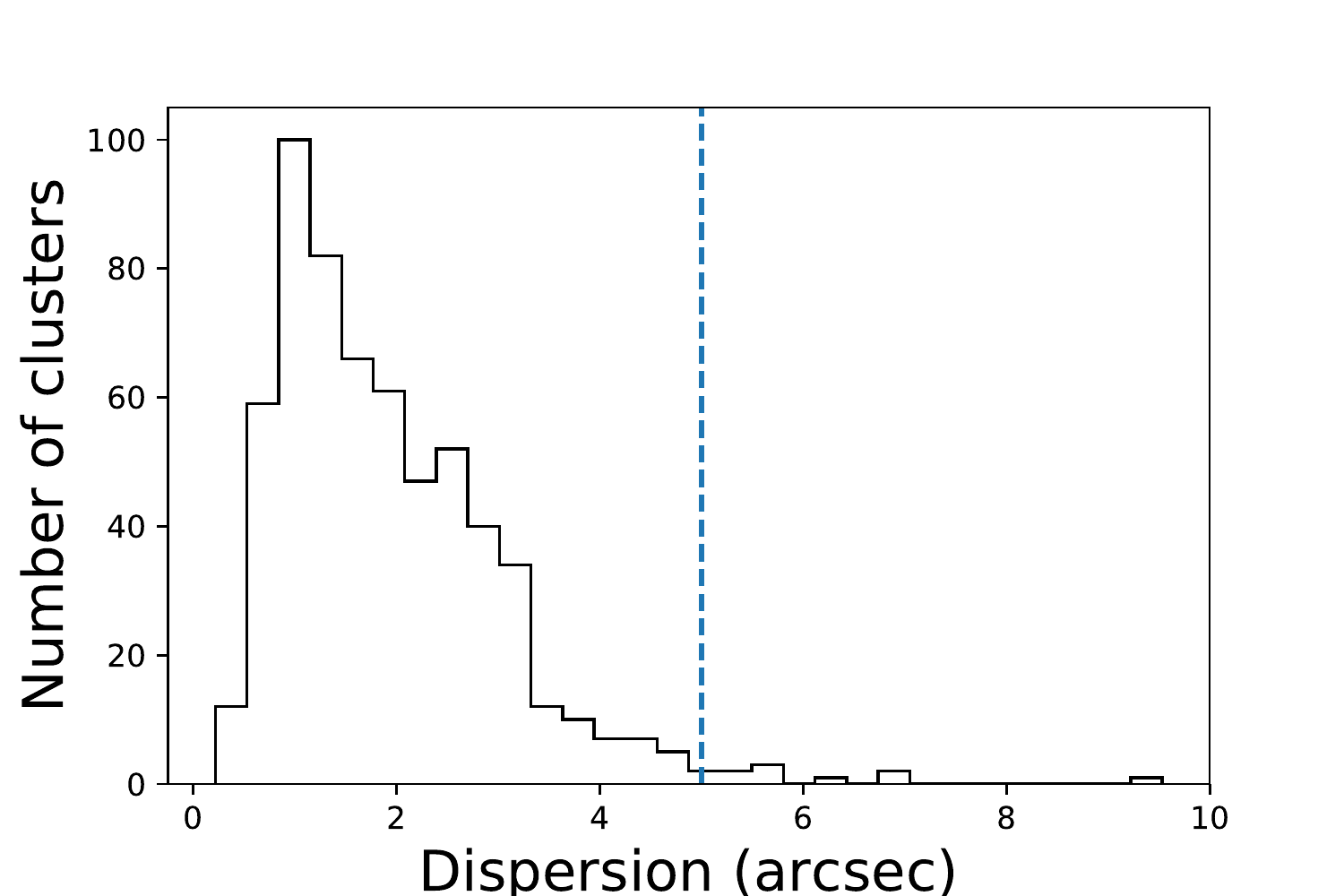} 
\centering
\caption{Dispersion distribution of all MWP bow shock clusters. These clusters are made with a minimum of 3 bow shock classifications and over 98.8\% of the clusters have a dispersion of < 5 arcseconds. This distribution supports using 5 arcseconds as our working clustering radius for bow shock classifications. \label{fig:DSCclusters}}
\end{figure}

Unlike the case of bubble clusters, clusters of BDSC classifications tended to have very little  dispersion in their central coordinates (Figure~\ref{fig:DSCclusters}). We therefore used {\scshape dbscan} not {\scshape hdbscan}, to identify clusters of  BDSC classifications, which allowed us to define a fixed (two-dimensional) clustering radius $\epsilon$. The minimum number of classifications needed to form a cluster was originally set to be the same as bubbles ($\kappa=5$), but experimentation and visual review (see Section 3.7.2 below) revealed that reducing this clustering threshold to $\kappa = 3$ increased the recovery rate of K16 bow shocks by 14\% without introducing significant false-positives.  We found $\epsilon = 5\arcsec$ to be the appropriate clustering radius when at least three volunteers agreed upon the same BDSC, which reflects the  ${\sim}2\arcsec$ resolution limit of IRAC and limitations of the accuracy of users placing BDSC reticles on the intended star in a MWP image.

 \subsection{User Weighting}
 
\begin{figure}
\includegraphics[width=\columnwidth]{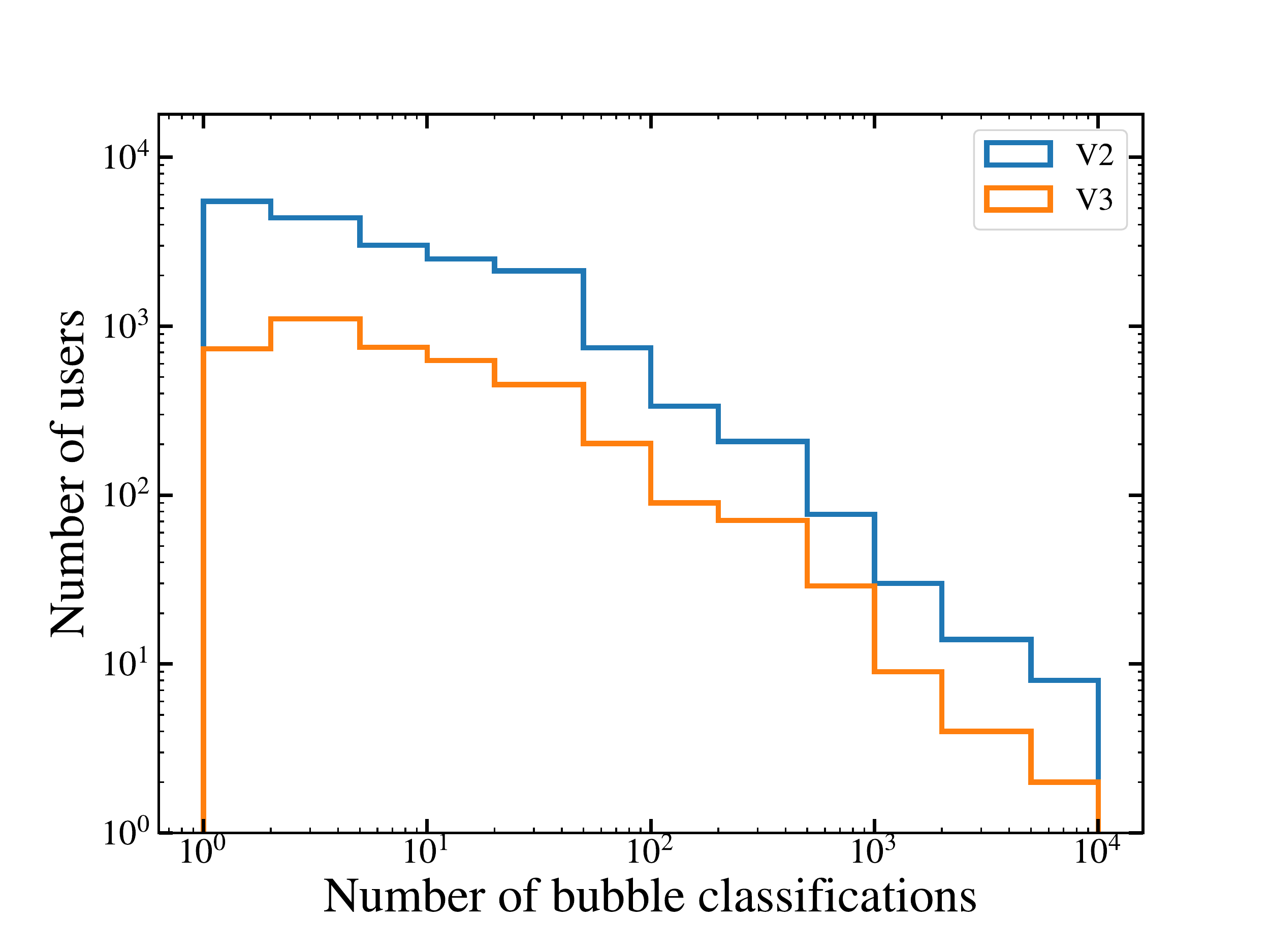} 
\centering
\caption{Distribution of bubble classifications by user.} \label{userstats}
\end{figure}

 Figure \ref{userstats} shows the distribution of bubble classification counts among MWP users. We see that a large majority of the users (V2: 68\%,V3: 63\%) trying the MWP, perhaps for the first time, make between 1 and 10 bubble classifications. Only 10\% of the user-base made more than 37 (50) bubble classifications in V2 (V3). There were 53 (21) individual users in V2 (V3) who each completed ${>}1000$ bubble classifications.

A precise bubble classification is more time-consuming to draw than an imprecise one, hence more careful users may make fewer classifications than users who are less careful. Following the approach of SPK12, we implemented user weighting designed to ensure that classification drawings made by more careful MWP users have higher weight in the final catalog.
For a bubble drawing made using the ellipse tool, the semi-major axis was twice the semi-minor axis by default. Any bubble drawing observed to have a 2:1 axis ratio was considered imprecise, because the semi-major and semi-minor axes were not adjusted independently by the user making the drawing. Precision bubble classifications were hence defined as those drawings with semi-major and semi-minor axes adjusted from the default ratio. The precision bubble fraction for a given user is the ratio between the total number of precision bubbles drawn to the total number of bubble classifications made by that user.

\begin{figure}
\includegraphics[width=\columnwidth]{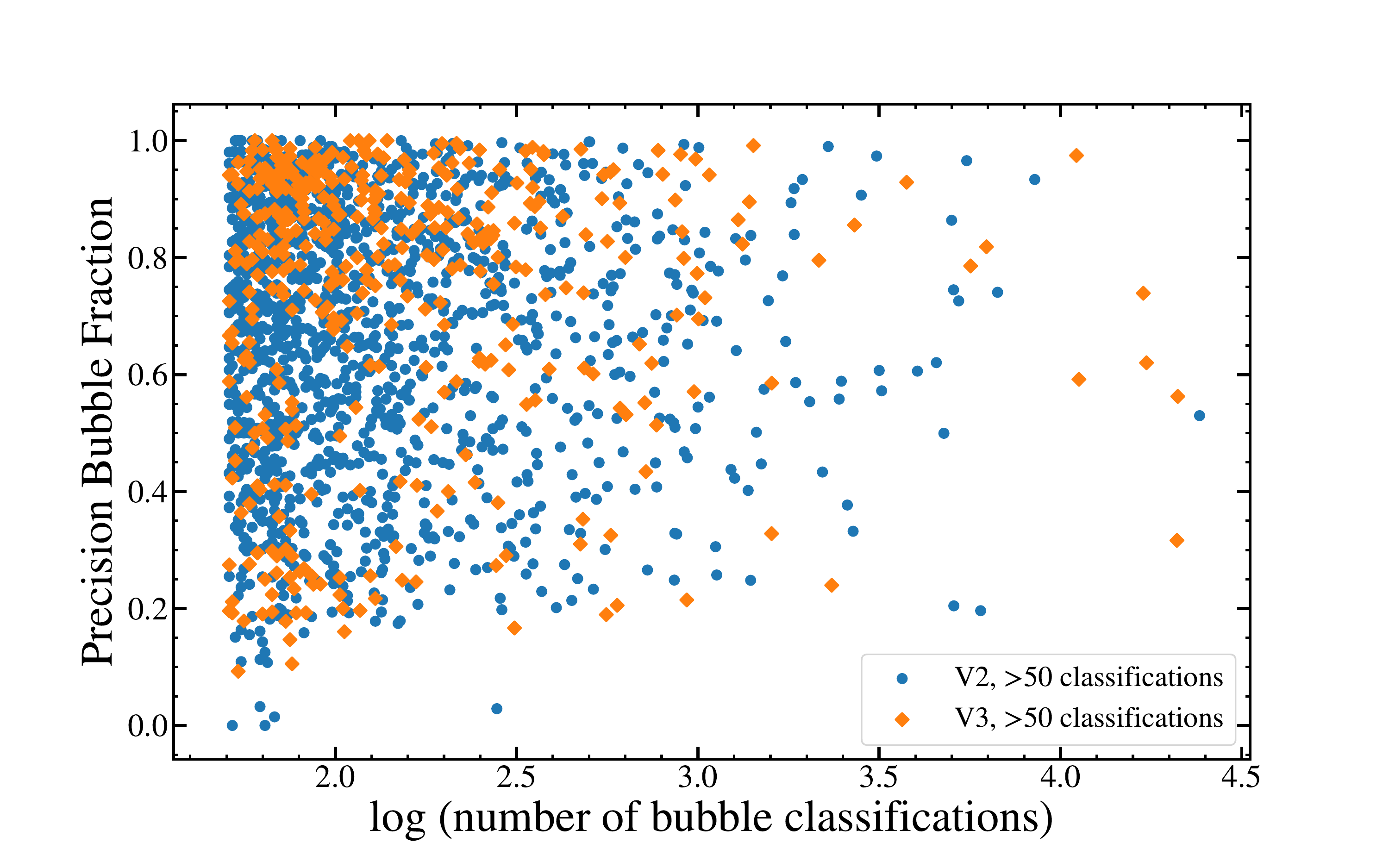} 
\centering
\caption{Precision bubble fraction versus number of bubble classifications for individual MWP users who made more than 50 classifications. } \label{userstats2}
\end{figure}
Plots of the precision bubble fraction versus number of bubble classifications for users who each made more than 50 classifications is shown in Figure \ref{userstats2}. It is clear that the raw number of bubble classifications by an individual citizen scientist is not a proxy for the accuracy of the classifications. The users with ${\gtrsim}10^3$ classifications have precision bubble fractions that reflect almost the full range observed. In contrast, the main locus of V3 users with 50--300 classifications have precision bubble fractions $>0.8$, which we consider excellent.

 \subsection{Averaging classifications within each `cluster'}

Once clusters of bubbles and bow shocks were obtained, we combined all the precision drawings in the cluster using a user-weighted mean. Each bubble drawing was weighted on the precision bubble fraction of the individual user who made that drawing. While imprecise classifications were included when establishing the existence of clusters, they were not used to derive the sizes and shapes of the bubbles. 

For each cluster of $N$ precision bubble classifications, the user-weighted average parameter is
\begin{equation}
\label{weight}
p=\frac{\Sigma_{i=0}^{N} p_iu_i}{\Sigma_{i=0}^{N}u_i},
\end{equation}
where $p_i$ is the individual parameter of each classification ($l$, $b$, $r_{\rm maj}$, or $r_{\rm min}$; Table \ref{bubcat}) and $u_i$ is the precision bubble fraction of the user responsible for making the $i$th classification.

We obtain the eccentricity of each bubble, using the semi-major axis $r_{\rm maj}$ and the semi-minor axis $r_{\rm min}$ (derived using Equation \ref{weight}), as

\begin{equation}
\label{eccent}
e=\sqrt{1-\frac{r_{\rm min}^2}{r_{\rm maj}^2 }},
\end{equation}
and the ellipsoidal quadratic mean radius ($R_{\rm eff}$) for each bubble is 
\begin{equation}
\label{reffc}
R_{\rm eff}=\sqrt{\frac{r_{\rm maj}^2 + r_{\rm min}^2}{2}}.
\end{equation}

Calculating the arithmetic mean of the orientation angle using Equation~\ref{weight} leads to incorrect results. Instead, we first convert the orientation angle ($\theta_i$) of a bubble classification into the corresponding points on a unit circle ($\cos\theta$, $\sin\theta$). The mean orientation angle ($\overline{\theta}$) is hence
\begin{equation}
\label{circmean}
\overline{\theta}={\rm atan2}\Big(\frac{1}{N}{\Sigma_{i=0}^{N} (\sin\theta_i)},{\frac{1}{N}\Sigma_{i=0}^{N}(\cos\theta_i})\Big),
\end{equation} 
where atan2 is the two-argument arctangent function.

In the case of bow shock classifications, we are solely interested in the location of the BDSC. The catalogued position of each driving star in a cluster is simply the mean of the Galactic latitude and longitude of each classification reticle for that cluster. User weights were not used to calculate the averaged location for a BDSC, since they rely on the precision bubble fraction, which is not directly relevant to bow shock classifications. We assume that users who had taken sufficient care to provide a complete bow shock classification (which includes both a beziel and reticle) have accurately placed the BSDC reticle on the intended driving star.
 
The uncertainties in each bubble parameter (size, position, angle) and BDSC position were calculated based on the standard deviations of these parameters among the classifications in the cluster. The dispersion of the central coordinates for both bubbles and BDSCs is 
\begin{equation}
\label{sigmalb}\
\sigma_{lb}=\sqrt{{\sigma_{l}}^2 + {\sigma_{b}}^2},
\end{equation}
where ${\sigma_{l}}$  and ${\sigma_{b}}$ are the standard deviations of the central position in Galactic longitude and latitude.

Propagating the errors ($dr_{\rm maj}$,$dr_{\rm min}$) in the semi-major ($r_{\rm maj}$) and semi-minor ($r_{\rm min}$) axes gives the uncertainty in bubble $R_{\rm eff}$ ($\sigma_{r}$),
\begin{equation}
\label{sigmar}\
\sigma_{r}=\sqrt{2 \left[ \left( \frac{r_{\rm maj}(dr_{\rm maj})}{\sqrt{r_{\rm maj}^2 + r_{\rm min}^2}} \right)^2 + \left( \frac{r_{\rm min}(dr_{\rm min})}{\sqrt{r_{\rm maj}^2 + r_{\rm min}^2}} \right)^2 \right]}.
\end{equation}
The uncertainty in the orientation angle ($\sigma_{\theta}$) is 
\begin{equation}
\label{thetae}\
\sigma_{\theta}=\sqrt{\bigg(-2\times \log \Big(\rm hypot(\overline{\sin\theta_i},\overline{\cos\theta_i})\Big)\bigg)},
\end{equation}
where hypot$(x,y)=\sqrt{(x^2+y^2)}$, and $\overline{\sin\theta_i},\overline{\cos\theta_i}$ are mean values computed from the individual classifications within the cluster.

\subsection{Cross-matches to the {\it WISE} catalogue of Galactic H II regions}
 We attempted to match each {\it WISE} \hii region candidate in the A14 catalogue with a single DR2 bubble. In general, the radius reported for an A14 \hii region candidate is several times larger than the $R_{\rm eff}$ of its matching MWP bubble, because A14 drew circles to enclose all apparent MIR emission associated with each region, while MWP aims to fit the best elliptical model to inner bubble rims.  An \hii region candidate from the A14 catalogue was identified as matching a DR2 bubble when the central coordinate of the A14 region lay within the radius of the DR2 bubble, and visual review of a subset of these cross-matches showed that this simple procedure generally worked very well. For atypical cases in which an A14 region was much smaller than its nominally-matched DR2 bubble ($R_{\rm eff,AT14}/R_{\rm eff,DR2}<0.25$), the match was only accepted if $dR_{\rm eff}/R_{\rm eff,min}<5$, where $R_{\rm eff,min}$ is the smallest of the two effective radii and $dR_{\rm eff}$ is the absolute difference in sizes. These matches to the A14 catalogue were used to reduce the number of spurious bubbles in the DR2 catalogue (see $\S3.6.1$).

\subsection{Visual review and reliability flags}

During the process of creating the bubble and BDSC catalogs, we visually verified the results of our pipeline at numerous stages in order to fine-tune and improve the different routines in the pipeline.
To quantitatively assess the reliability for each cluster we used a hit rate parameter, as defined by SPK12. The hit rate for a given cluster is simply the ratio of the number of classifications in that cluster to the number of times that images containing the classified object were viewed by MWP users. The hit rate distribution in V3 (HR3) is skewed to higher hit rates when compared to V2 (HR2). This is a result of the different image assets used in V3 and V2. Bubbles were easier to find in the V3 images with 24~$\mu$m data than in the V2 images that lacked 24~$\mu$m data.

\subsubsection{Bubbles}

Visual review of the clustering process was a critical step for creating the bubbles catalog. We examined the clustering results from both the  {\scshape hdbscan} and {\scshape dbscan} algorithms before ultimately deciding to use the  {\scshape hdbscan} algorithm. The cut made on the outlier scores of individual classifications in a cluster ($\S3.2$), our choice of the clustering parameter $\kappa$ in the {\scshape  {\scshape hdbscan}} algorithm and the cut made on the fractional uncertainty in $R_{\rm eff}$ were all decided through this visual review process. We also verified the performance of the user weighting ($\S3.3$) and averaging ($\S3.3$) processes. 

Independent lists of bubble clusters found in V2 and V3 were created using our data reduction pipeline. We performed two initial quality checks separately on each list. We first eliminated clusters with coordinate dispersion greater than the effective radius ($\sigma_{lb}>R_{\rm eff}$). This process cleaned the lists of spurious clusters generated by the  {\scshape  {\scshape hdbscan}} algorithm. We then identified duplicate clusters within each list (V2 and V3) as those with (1) central coordinates separated by a distance smaller than the higher of the two spatial dispersions ($\sigma_{lb}$), and (2) ratios of the semi-major and semi-minor axes smaller than ${\sim}50\%$. 

The above process was repeated to combine the V2 and V3 cleaned bubble candidate lists.
For matched bubbles, we recalculated the bubble parameters using a weighted average based on the V2 and V3 hit rates. Although we averaged the V2 and V3 bubble candidates to obtain final size and shape measurements when they were matched, we report HR2 and HR3 separately in the catalogue, as this information was used to define reliability cutoffs. 

\begin{figure}
\includegraphics[width=\columnwidth]{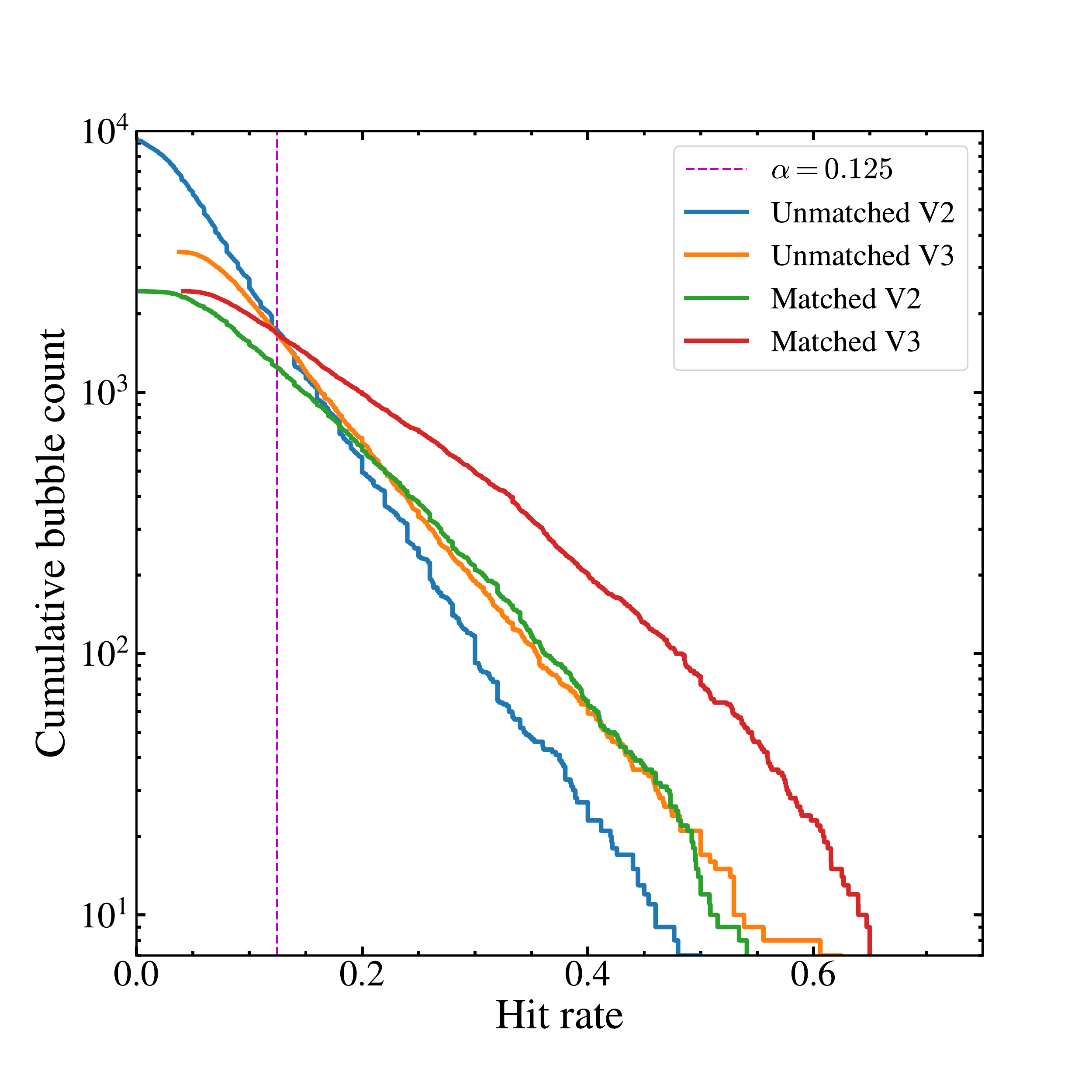}
\caption{Cumulative distribution functions for the MWP hit rate (V3=HR3 and V2=HR2). The parameter $\alpha=0.125$ is the hit rate at which the unmatched sample of V3 bubbles dominates over the matched sample of bubbles in both V2 and V3. }
\label{cdfhist}
\end{figure}

SPK12 used a somewhat arbitrary minimum hit rate cutoff of 0.1 for including bubbles in the DR1 catalog. Here we leverage our larger classification database from the combination of MWP V2 and V3 to implement a more nuanced approach. 
In Figure \ref{cdfhist}, we plot the CDFs for the hit rate distributions for both the matched and unmatched bubbles in our overall catalog. 
It is clear from this plot that bubbles identified and matched between V2 and V3 have systematically higher hit rates when measured using HR3 (red curve). For matched bubbles the CDF of HR2 (green curve) is systematically lower than HR3, and coincidentally looks most similar to the HR3 CDF for unmatched V3 bubbles (orange curve). Unmatched V2 bubbles (blue curve) are by far the most numerous in our database and are heavily skewed toward the lowest HR2 values, meaning they are dominated by spurious classifications. These CDFs validate our {\it a priori} assumptions that (1) bubbles were much easier for MWP users to to identify in V3 than in V2, and (2) matched bubbles identified independently in both MWP versions are the most reliable.
We therefore define the critical hit rate $\alpha=0.125$ at the intersection between the matched and unmatched V3 hit rate CDFs. 
This $\alpha$ parameter hence gives the hit rate value below which the unmatched sample of bubbles in V3 dominates over the matched sample of bubbles identified independently in both V2 and V3. 

We define three classes of reliability flags used to create the DR2 bubbles catalog:
\begin{enumerate}
\item \textit{More reliable subset}. Bubbles are assigned an `R' flag only if they have been independently discovered in both V2 and V3, and the highest hit rate (V2 or V3) is greater than $\alpha$. Only such matched bubbles receive an `R' flag, but not all matched bubbles are included in this subset.
\item \textit{More complete sample}: Matched bubbles are assigned a `C' reliability flag if the higher of the two hit rates HR2 or HR3 is less than $\alpha$. Unmatched V3 bubbles are also assigned to the more complete sample if ${\rm HR3} \ge\alpha$.
\item \textit{Reject}: All bubbles identified in V2 only were rejected from the DR2 catalog. Bubbles identified in V3 only were also rejected if they had ${\rm HR3}<\alpha$.
\end{enumerate}
Visual review by multiple co-authors confirmed that these choices of hit rate cutoff were appropriate for defining bubble reliability. 
 
Two of us (MSP and LDA) inspected the 1727 bubbles that lacked A14 matches. Of this group, we judged that 317 were real bubble candidates worth retaining in the DR2 catalog. The bubbles flagged for inclusion via this final review fell into three broad categories:
\begin{itemize}
    \item Bipolar bubbles or multi-bubble complexes where the one or more constituent DR2 did not enclose the centre of an associated A14 region.
    \item Highly symmetric/broken bubbles for which the DR2 ellipses were displaced from the centre of an otherwise matching A14 region.
    \item DR2 bubbles that identified star-forming regions missing from the A14 catalog.
\end{itemize}
The last group, newly-identified bubbles, contained approximately 20 objects.

\begin{figure*} 
	\centering
	\subfloat{
		\includegraphics[width=0.75\textwidth]{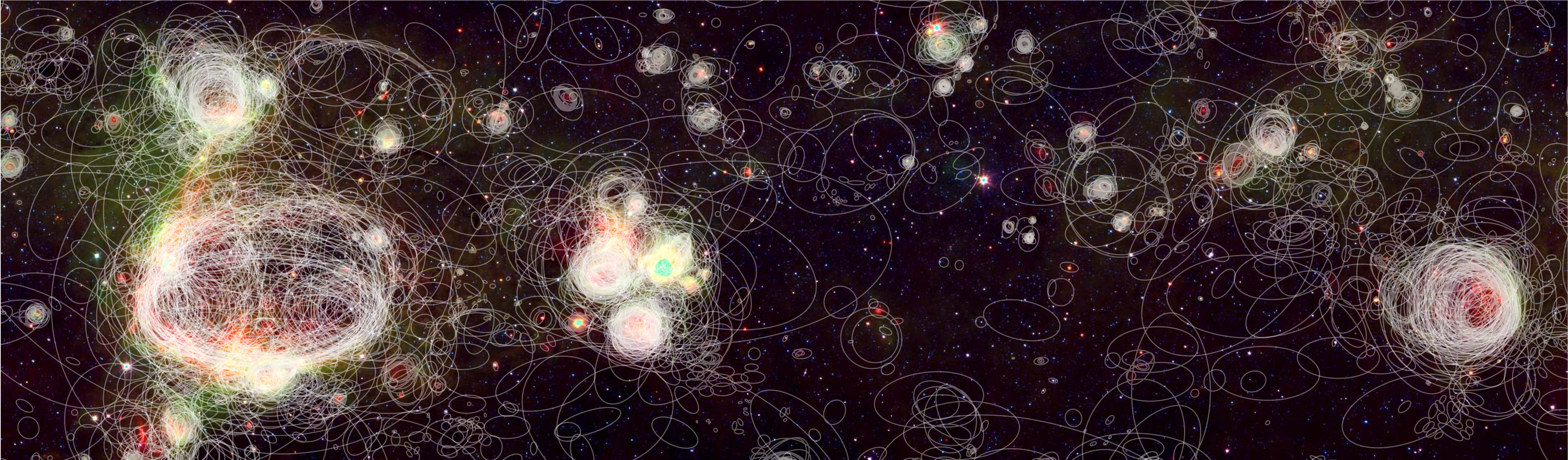}} \\

	\subfloat{
		\includegraphics[width=0.75\textwidth]{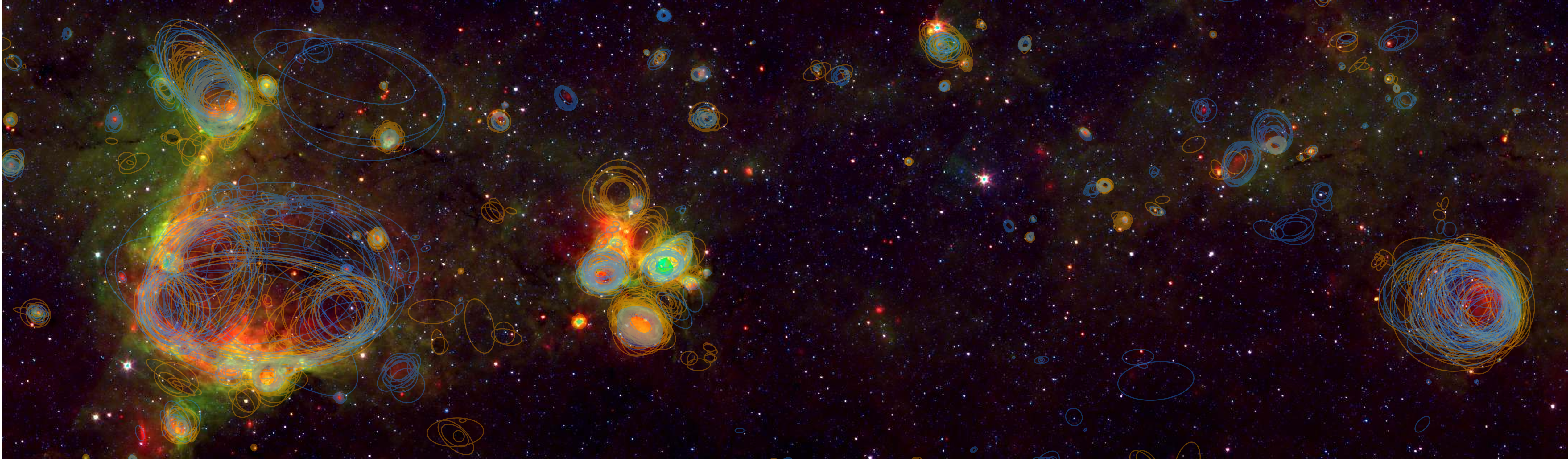}} \\

	\subfloat{
		\includegraphics[width=0.75\textwidth]{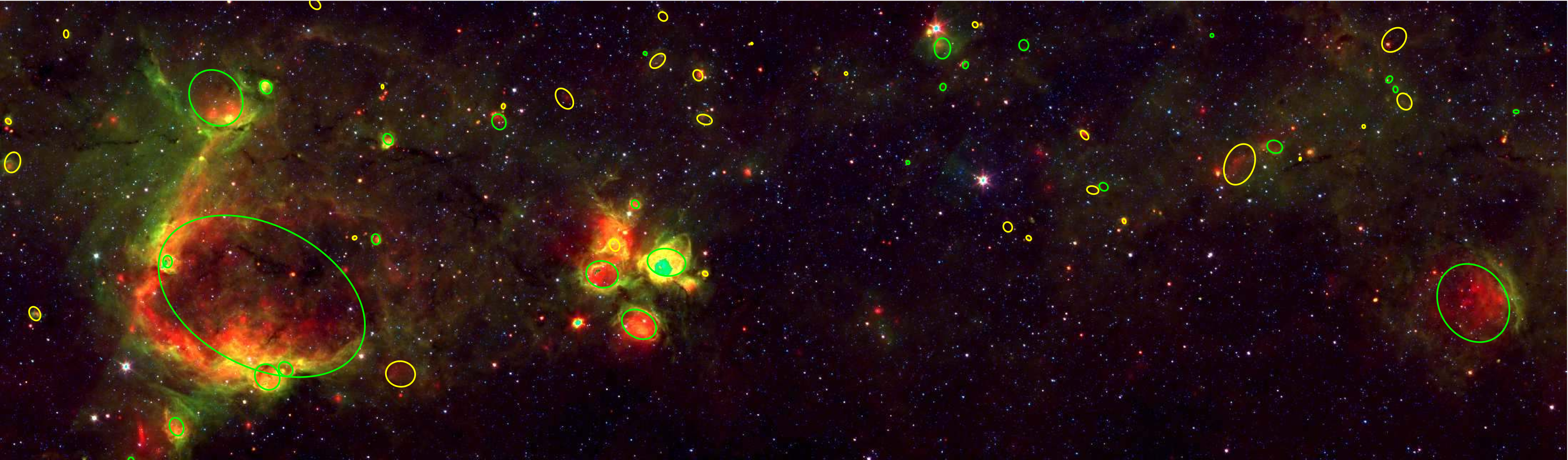}}	
	\caption{Illustration of the DR2 bubble catalogue construction process on a wide-field $(\sim 3\degree \times 1\degree)$ {\it Spitzer} mosaic image (blue = 4.5~\um, green = 8.0~\um, red = 24~\um) centered on $(l,b)=(18\degree,0\degree)$. Top: Image with all user drawings overlaid, Middle: Image with bubble clusters (V2: orange, V3: blue) overlaid, Bottom: Image with catalogued bubbles (More-reliable sample: green, more-complete sample: yellow) overlaid. 
	}
\label{catalogcreate}
\end{figure*}

Table~\ref{reliabilityclass} lists the numbers of bubble clusters in the final DR2 catalogue that fell into each reliability class. We present a visual summary of how the DR2 bubble catalogue was created in Figure \ref{catalogcreate}. For comparison purposes, we have shown the same field illustrated in Fig. 5 of SPK12, centred on $(l,b)=(18\degree,0\degree)$. The top panel illustrates all the user drawings. The {\scshape hdbscan} clustering effectively picks out `true' bubble clusters from the noise (middle panel) and we further refine these clusters by merging clusters across V2 and V3, implementing cuts on the hit rate and by visual review to arrive at a clean catalogue of bubbles (bottom panel).

\begin{table*}
	\caption{Reliability classifications for bubble and bow shock driving star clusters nominated for inclusion in the MWP DR2 catalog. 
	}
	\label{reliabilityclass}
	\begin{tabular}{ccc}
		\hline
		Reliability & Number of bubble clusters &  Number of bow shock driving star clusters \\
		\hline
		More reliable (`R') & 1394 & 453\\
		More complete (`C') & 1206 & 146 \\
		Reject (unpublished) & 12 571 & 542\\		
		\hline
	\end{tabular}
\end{table*}

\subsubsection{Bow shocks}
We cross-matched the MWP BDSCs to both the K16 and the 2MASS point-source \citep{2006AJ....131.1163S} catalogues. Generally the separations between K16 BDSCs are much larger than 5\arcsec, hence matching MWP BDSCs to the K16 catalogue using just the clustering radius was robust. The 2MASS data for the matched BDSCs were directly sourced from the K16 catalog. For the remaining BDSCs that do not have matches in K16, 2MASS matches were obtained through a cross-match. 

In some cases, multiple 2MASS sources were obtained within the matching radius, complicating the identification of the potential driving star. Occasionally the classifications that make up a BSDC are distributed between multiple stars, resulting in a significant displacement between the averaged BSDC coordinate and an actual star. To account for this, we applied different criteria for a star match based on the angular distance to the closest 2MASS source from a BSDC. We considered any BDSC with a matching 2MASS source within the resolution limit of IRAC (2\arcsec) as an accurate match. In some cases where the closest 2MASS source was between 2\arcsec and our clustering radius (5\arcsec), we often found the averaged BDSC coordinate between two nearby stars rather than on top of one, suggesting that MWP volunteers were undecided and voted for two possible driving stars. In such cases, we selected the brightest source in the $J$-band as the BDSC, because at similar distances it would more likely be the high luminosity OB star capable of driving bow shocks. The distance between the apsis of the bow shock arc and suspected driving star (i.e. the standoff distance) for our particularly large bow shocks is significantly larger than the clustering radius. At this size the distance from a BDSC to the nearest 2MASS source can exceed 5\arcsec if the composing volunteer classifications marked multiple different driving stars. Because a stronger preference of star can be resolved in BDSC's coordinates at larger sizes we match to the closest star for the 8 BDSCs that matched to 2MASS stars outside the clustering radius.

The existence of multiple stellar sources near the apsis of bow shock arcs can make the identification of the correct driving star more dependant upon user interpretation. In 31 such cases the driving star for a given bow shock chosen by MWP volunteers differed from the K16 source. We define these situations as arc matches to the K16 catalog, named for the singular infrared arc feature of the suspected bow shock candidates. To help identify these cases and avoid confusion with stellar matches, we automatically flagged a BDSC as a potential arc match to K16 when it had an angular separation between 5\arcsec\ and 60\arcsec\ to a K16 star. We later confirm or reject the arc matching based on the results of visual review.

MWP offers the first systematic check for spatial overlap between candidate bow shocks and bubbles. We define a BDSC-bubble match when a BDSC is located interior to a bubble rim defined by the MWP DR2 bubble catalogue. Most BDSCs that are matched to MWP bubbles reside in environments of high nebulousity, making them more likely to be in-situ bow shocks compared to runaways. We report the MWP IDs of bubbles along with their matched BDSC in the finalized bow shock catalog.

One of us (DD) visually inspected each bow shock and associated driving star candidate using  {\scshape SAOImage ds9}.
Following  K16, each bow shock was assigned an environment code based on the local surroundings of the BDSC. Definitions for environment codes are: FB = facing bright-rimmed cloud, FH = facing \hii~region, H = inside \hii~region and I = isolated, none of the other environment codes apply (K16). The FB environment code describes candidate {\it in situ} bow shocks facing photo-evaporating molecular cloud surfaces, which typically occurs within \hii regions. During the visual review process, if both flags were applicable to a given BDSC, the FB flag was favored as it is more physically descriptive.

To measure and record standoff distances and position angles of the bow shocks consistently with the practice of K16, we created vector {\scshape ds9} regionfiles for all the newly discovered MWP bow shocks. We simply adopted these parameters from K16 for all the matched bow shocks. We define the position angle for bow shocks in degrees east of celestial north, which differs from the orientation angle of bubbles but facilitates comparisons to proper motion data for BDSCs \citep[e.g.][]{KCP19}.

The results of our automated arc matching was visually reviewed and manually corrected as necessary. 
 
Since we chose a somewhat arbitrary radius for the arc matching (60\arcsec), there were a few occurrences where the size of a respective bow shock arc resulted in an incorrect match. 
In the false positive case, two bow shocks with different arcs would be within 60\arcsec , and in the false negative case the respective bow shock arc was large enough for two different stars to be separated by >60\arcsec . We performed the final step of visual review for a bow shock by creating a vector {\scshape ds9} region file with one endpoint on the 2MASS matched star and the other on the apsis of the arc. This file contains information on both the standoff distance and position angle of the bow shock.

Rarely, MWP users incorrectly identified an extended 24~\um~ nebular object as a BDSC. This was because the extended source was clipped by the edge of an image cutout and/or the image had a colour stretch such that the extended source resembled a bow shock. During our visual review, each bow shock candidate was checked to see if clipping and/or scaling had caused a spurious classification, and three BDSCs were manually removed from the catalog.

\begin{figure}
\includegraphics[width=\columnwidth]{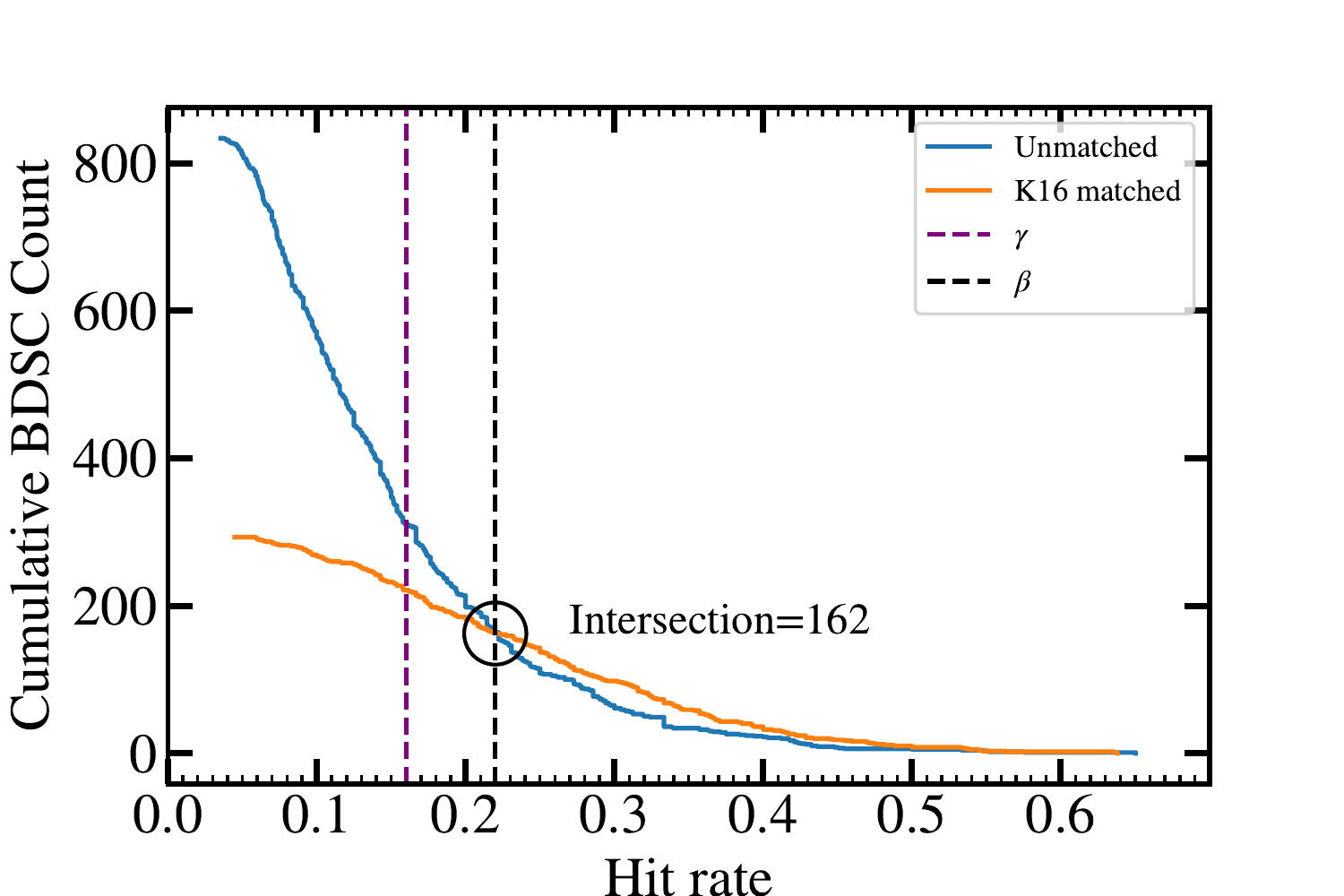}
\centering
\caption{Cumulative distribution functions of matched and unmatched bow shocks. Vertical dashed lines mark both the low ($\gamma$) and high ($\beta$) hit rate cutoffs for bow shock clusters. 
\label{fig:HRC_CDF}
}
\end{figure}

We select BDSCs for the DR2 catalogue and assign them reliability flags using a hit-rate based procedure analogous to that described above for the bubbles. In the case of BDSCs only HR3 is available, so we compared the HR3 CDFs for BDSCs that matched to K16 to those that were unmatched (Figure~\ref{fig:HRC_CDF}).
These CDFs reveal the critical HR3 value $\beta=0.22$ above which the matched sample dominated over the unmatched sample. The matched sample had been independently visually reviewed and confirmed by the K16 team, so this choice of $\beta$ imposes a similar quality standard for selecting more-reliable unmatched MWP BDSCs. 
 
\begin{figure*}
\includegraphics[width=\textwidth]{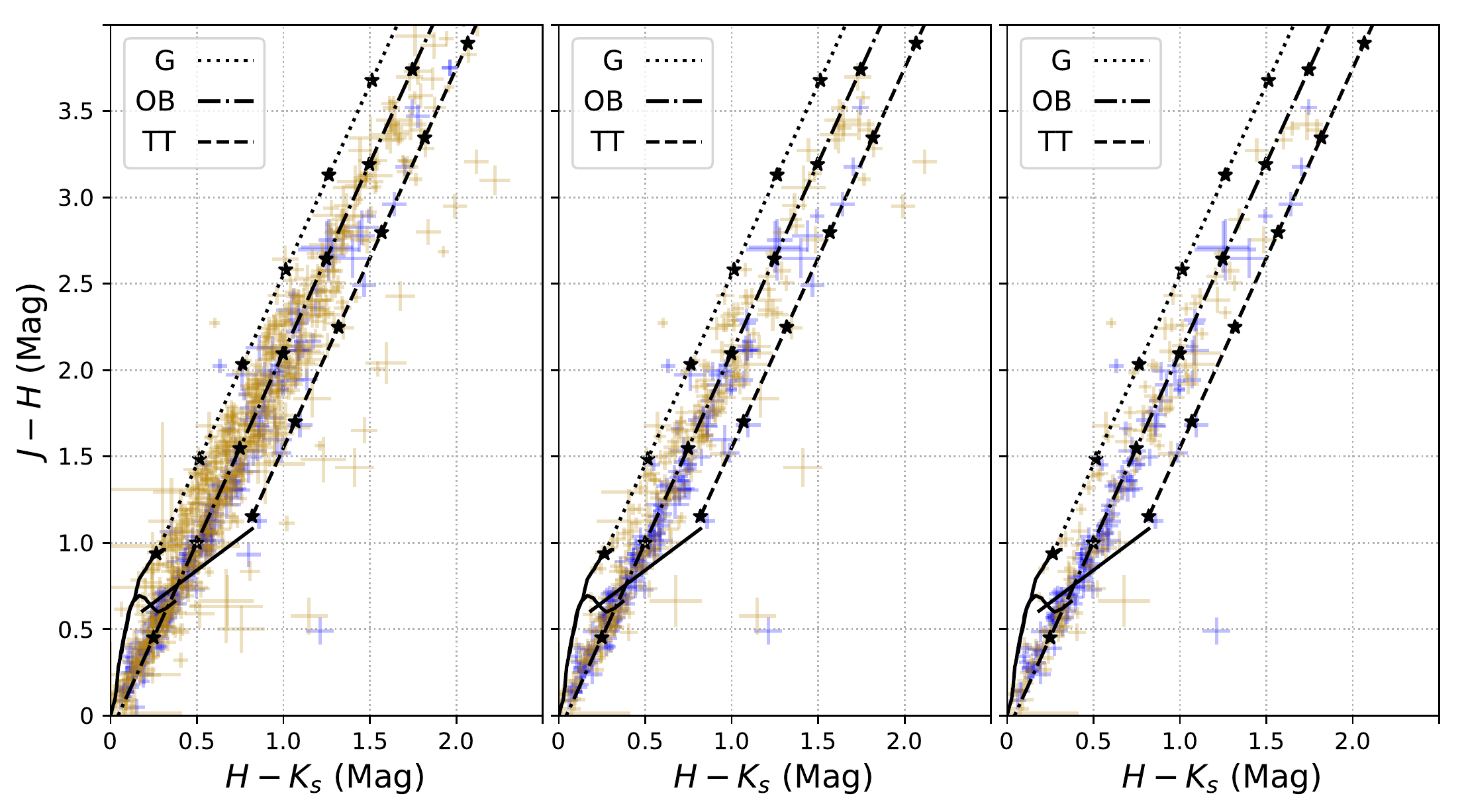}
\caption{2MASS colour-colour diagrams of MWP BDSCs with various minimum hit rate cutoffs imposed. {\it Left to right}: no cutoff, ${\rm HR3}\ge \gamma=0.16$, and ${\rm HR3} \ge\alpha = 0.22$. Blue crosses indicate matches between MWP and K16 BDSCs while golden-brown show unmatched MWP BDSCs. The size of each cross is the photometric error bar. In the plot legends G, OB, and TT refer to the reddening vectors originating from the ends of the giant sequence, OB main sequence and cool T Tauri locus, respectively. All reddening vectors are marked with $\star$ symbols at  intervals corresponding to $A_V=5$~mag.    2MASS sources with photometric upper limits reported or otherwise unconstrained errors in colour are not plotted.
\label{fig:CCDs}}
\end{figure*}

\begin{figure}
\includegraphics[width=\columnwidth]{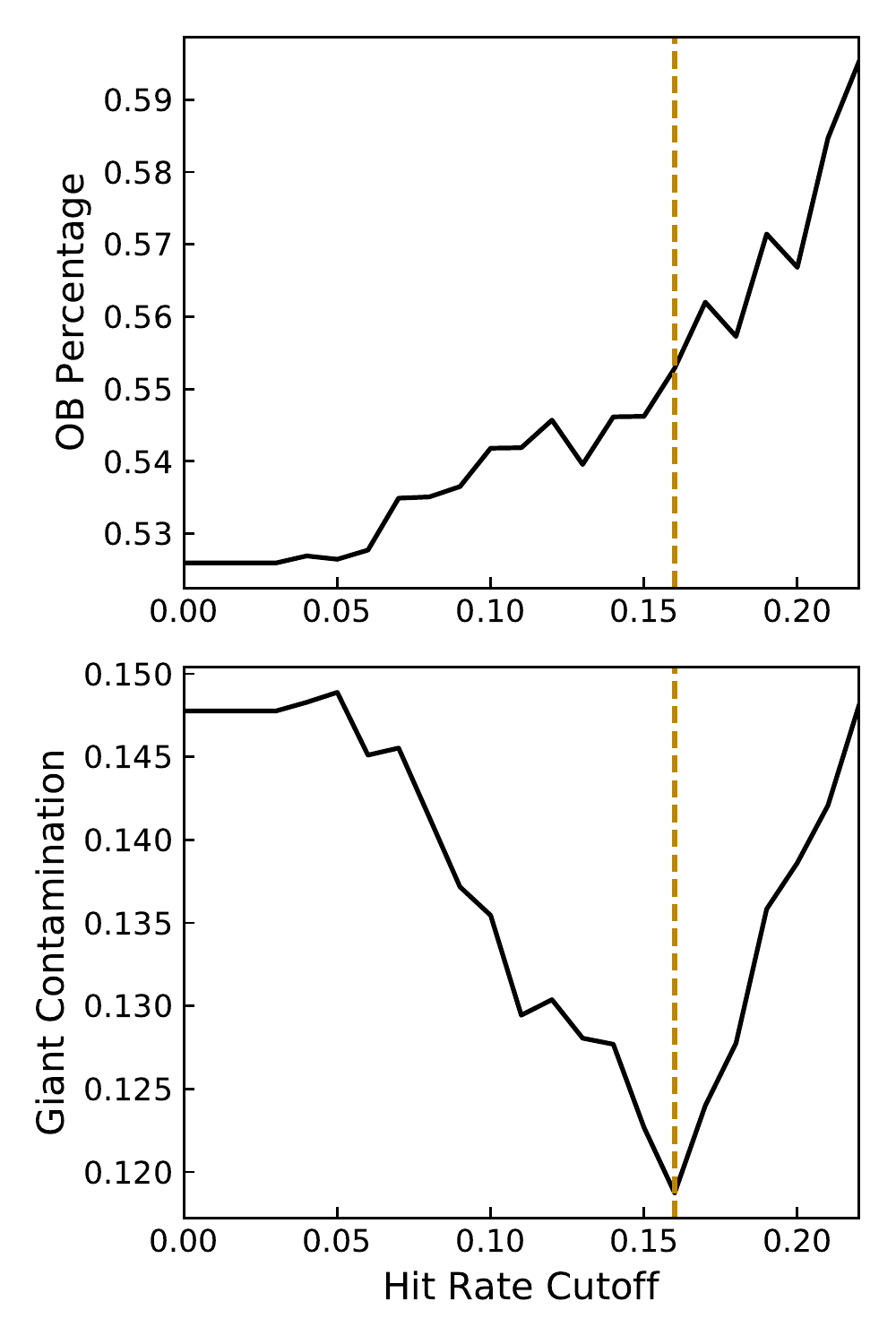}
\caption{Curves displaying the percentage of driving star clusters consistent with the main sequence (top) and giant branch (bottom) reddening vectors. The dashed golden-brown line is placed at the minimum (HR=0.16) for giant percentage and sets the hit rate floor for the bow shock catalog. 
\label{fig:OBvsgiants}}
\end{figure}

To determine the minimum acceptable HR3 for inclusion of newly-discovered, unmatched BDSCs in the DR2 catalogue we analysed the 2MASS colours of all BDSCs. Figure~\ref{fig:CCDs} shows the locations of all 2MASS stars matched to BDSCs of the $J-H$ vs. $H-K_s$ colour-colour diagram. In almost all cases the bow shocks that are bright enough at 24~\um~ to be observed in MWP should be driven by OB stars (\citealp{KCP18,KCP19}),
hence we expect the large majority of BDSCs to be located near the locus of reddened OB stars. However, since the GLIMPSE images of the inner Galactic plane are dominated by IR-bright KM giants, we might worry about giant star contamination in our BDSC catalog. Using reddening vectors \citep{1985ApJ...288..618R}
for the main sequence and giants in $J-H$ vs. $H-K_s$ colour space, we measured the population of stars consistent with each vector for an array of hit rate cutoffs. 
We determine consistency by checking if the 2MASS colour error ellipses of our stars intersect the OB and/or Giant loci. To minimize the giant contamination of our catalogue we iterated hit rate cutoff values by 0.01 from our minimum recorded hit rate value ($\sim 0.04$) to our high reliability hit rate cutoff ($0.22$). We then selected the hit rate where the percentage of stars consistent with the giant locus was minimal. As the hit rate cutoff was increased, the OB percentage also increased. However, there is a sample of high HR3 BDSCs consistent with both the OB and giant loci. Beyond a certain hit rate threshold, the overall size of the catalogue was reduced faster than the number of stars aligned with the Giant locus. We found that a threshold  ${\rm HR3}\ge \gamma=0.16$ minimized contamination from stars consistent with the reddened giant locus and used this as the hit rate floor for the bow shock catalogue (Figure~\ref{fig:OBvsgiants}). 
 
 As we did for bubbles, we break down the BDSCs into two subsamples based on reliability flags, but the quantitative definitions for these flags and hit rate cutoff values are different:
 \begin{enumerate}
     \item \textit{More reliable subset}: New MWP BDSCs that were not found by K16 were assigned an `R' flag only if ${\rm HR3}\ge \beta$. All BDSCs matched to the K16 catalogue were assigned an `R' flag.
     \item \textit{More complete sample}: New, unmatched MWP BDSCs were assigned a `C' reliability flag if $\gamma\le{\rm HR3}<\beta$.
     \item \textit{Reject}: Unmatched BDSCs with  ${\rm HR3}<\gamma$ were rejected from the catalog.
 \end{enumerate}
Table~\ref{reliabilityclass} lists the numbers of BDSCs in the more reliable and more complete subsets of the DR2 catalog, as well as the number of rejected BDSC clusters.

\subsection{Assigning bubble hierarchy flags}

Bubble catalogs have been used for statistical studies of star formation triggered by massive stellar feedback \citep{T+12,2012ApJ...755...71K}.
CP06 and CWP07 identified that a small fraction of the bubbles in these catalogs were part of hierarchies, where small `daughter' bubbles were located on or within the rims of larger, `parent' bubbles.
SPK12 reported that $29\%$ of the DR1 bubbles were part of hierarchies.

Following the cuts and visual review process in $\S3.6.1$, hierarchical bubbles in the DR2 catalogue are assigned a hierarchy flag of either parent (`P') or daughter (`D'). The DR2 bubble catalogue was sorted by size and compared with itself to identify pairs of bubbles with the separation between the central coordinates smaller than the size of the largest bubble. 
If $R_{\rm eff}$ of the smaller bubble was smaller than 50$\%$ of the larger bubble, the pair of bubbles was declared a hierarchy, and the smaller bubble is given a `D' flag while the larger bubble was given a `P' flag. It is possible for multiple daughter bubbles to be assigned to the same parent bubble.

\subsection{DR2 workflow summary}
\begin{figure*} 
	\includegraphics[width=0.95 \textwidth]{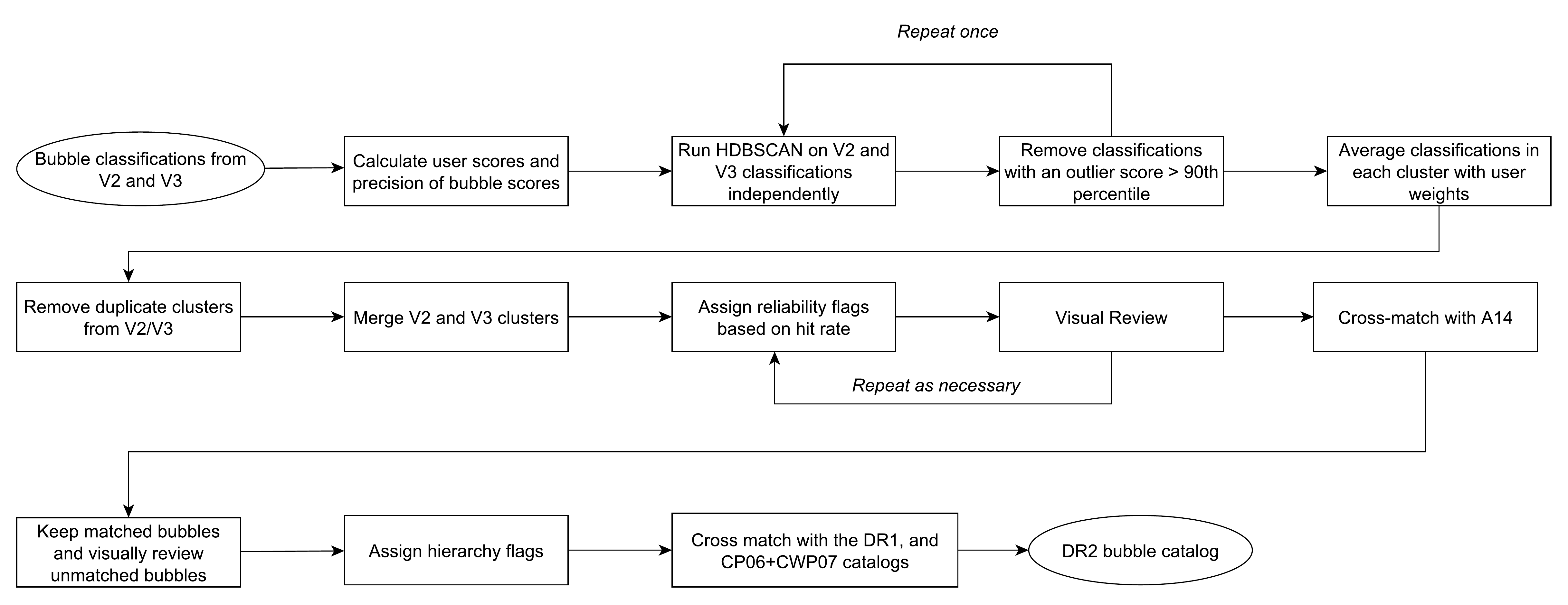}
	\caption{Flowchart illustration of the bubble catalogue creation process.}
	\label{bubflow}
\end{figure*}

\begin{figure*} 
	\includegraphics[width=0.9 \textwidth]{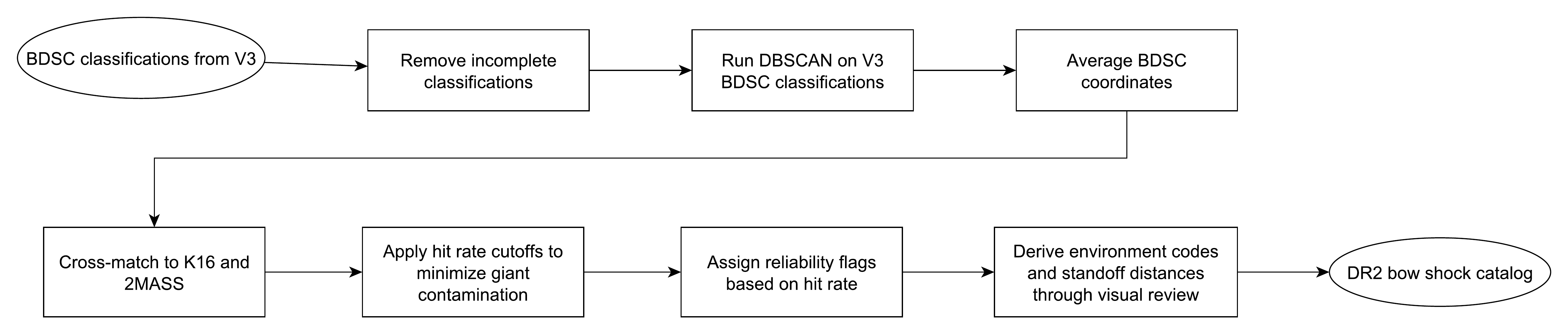}
	\caption{Flowchart illustration of the bow shock catalogue creation process.}
	\label{bowshockflow}
\end{figure*}

The overall processes of creating the bubbles and BDSC catalogs are summarized in Figures \ref{bubflow} and \ref{bowshockflow} respectively.

We have improved upon the bubble catalogue construction process described in SPK12 by utilizing a more robust hierarchical clustering algorithm ({\scshape hdbscan}) over the clustering algorithm used to create the DR1 catalog. Due to the flexibility in the clustering radius afforded by  {\scshape hdbscan}, we ran the clustering algorithm on the entire sample, instead of binning the classifications by location, and homogeneously clustered the set of classifications in V2 and V3. We implement various checks to remove spurious clusters and introduced bubble reliability flags that use classification data from both V2 and V3. The implementation of reliability flags in the bubble catalogue allows for the identification of a primary, higher-reliability sample of 1394 bubbles identified in both V2 and V3, plus an additional 1206 bubbles in the more complete sample.

We have also produced the first citizen-science enabled catalogue of bow shocks.  Volunteer classifications of bow shocks were automatically checked for completeness before being considered for clustering. Bow shock classifications that did not contain a reticle to mark a star and/or a closed polygon to trace the impacted infrared arc emission were discarded before running {\scshape dbscan}. The resulting average cluster positions were cross-matched to the K16 and 2MASS catalogues. In order to determine reliability flags, we derived cuts on the hit rate based on the overlap with the K16 catalogue and potential giant contamination based on 2MASS colours. Through visual review, we also determined environment codes and standoff distances for the new MWP BDSCs.

\section{The MWP DR2 Bubble catalogue Properties}
The final bubble catalogue contains 2600 bubbles visually identified by MWP volunteers.  Each bubble has been independently identified and measured by at least 5 MWP users, and matched bubbles (all of which have `R' flags) were identified by at least 10 users. Table \ref{bubcat} lists the columns in the MWP DR2 bubbles catalog. The DR2 bubble catalogue is available online through \textit{Vizier} and as supporting information with the electronic version of the paper. 

\begin{table*}
	\caption{Description of the columns in the MWP DR2 bubbles catalog}
	\label{bubcat}
	\begin{tabular}{ll}
		\hline
		Column & Description \\
		\hline
		MWP ID & Unique MWP identifier (MWP2GLLLllll+BBBbbbb)\\
		$l$ & Galactic longitude (degrees) \\
		$b$  & Galactic latitude (degrees)\\
		$\sigma_{lb}$ & Dispersion of the central coordinates (in arcmin)\\
		$r_{\rm maj}$& Semi-major axis (arcmin) \\
		$r_{\rm min}$& Semi-minor axis (arcmin)\\
		$R_{\rm eff}$ & Effective radius of the bubble in (arcmin)\\
		$\sigma_{r}$ & Error in $R_{\rm eff}$ (arcmin)\\
		$\theta$ & Orientation angle, defined as degrees from Galactic north toward increasing $l$\\
		$\sigma_{\theta}$& Uncertainty in orientation angle (degrees)\\
		$e$ & Eccentricity  \\        
		HR2  & MWP V2 hit rate  \\
		HR3  & MWP V3 hit rate \\
		Reliability & Reliability flag: `R' = more reliable subset, `C' = more complete sample \\		
		Hierarchy flag & Hierarchy flag: `P' = parent and `D'=daughter in bubble hierarchy\\
		DR1 Match & Identifier of matched MWP DR1 bubble\\
		A14 Association & Identifier of the matched A14 \hii region(s)\\
		A14 Distance & Kinematic distance (in kpc) associated with the matched A14 \hii region (if available)\\
		CP06/CWP07 Association & Identifier(s) of matched CP06 and CWP07 bubble(s) \\		
		\hline
	\end{tabular}
\end{table*}

\subsection{Bubble size distribution}
\begin{figure} 
	\includegraphics[width=\columnwidth]{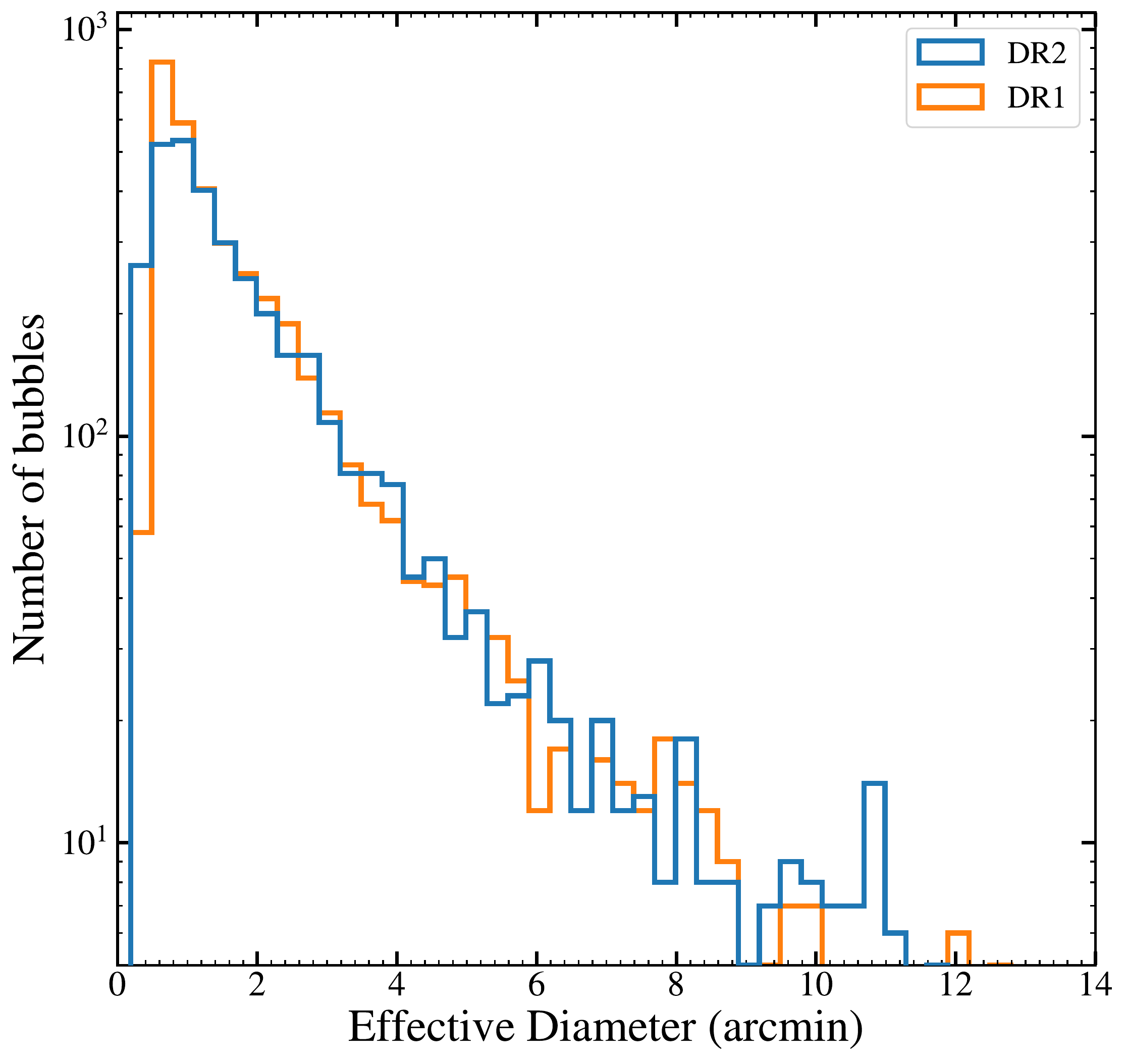}
	\caption{Distribution of bubble diameters in the GLIMPSE survey. Left: The number of bubbles decay as a power law with increasing angular
 diameter for both the DR2 and DR1 (large bubble) catalogs.
 }
 
\label{deff}
\end{figure}

Much like the CP06+CWP07, MWP DR1 and A14 catalogs, the size distribution of DR2 bubbles follows a decreasing power law with increasing angular diameter (Figure \ref{deff}). When comparing the DR2 catalogue with the DR1 large bubbles catalog, we recalculated $R_{\rm eff}$ for the DR1 bubbles using the measurements of the inner bubble rim.  At its peak, the DR1 size distribution piles up bubbles towards the limit of measurement in DR1 ($D_{\rm eff}=0.45'$). This indicates a slight overestimation of the sizes of the smallest bubbles in the DR1 large bubbles catalogue due to limitations in the drawing tool. The DR2 distribution probes bubbles smaller than the DR1 limit of measurement, down to $D_{\rm eff}=0.22'$, which is close to the resolution limit for these extended structures in the MIPS 24~\um~ images. The DR2 catalogue lists 136 bubbles with angular diameters smaller than 0.45'. The need for a separate small bubbles catalogue is hence eliminated in DR2. 

\begin{figure} 
	\includegraphics[width=\columnwidth]{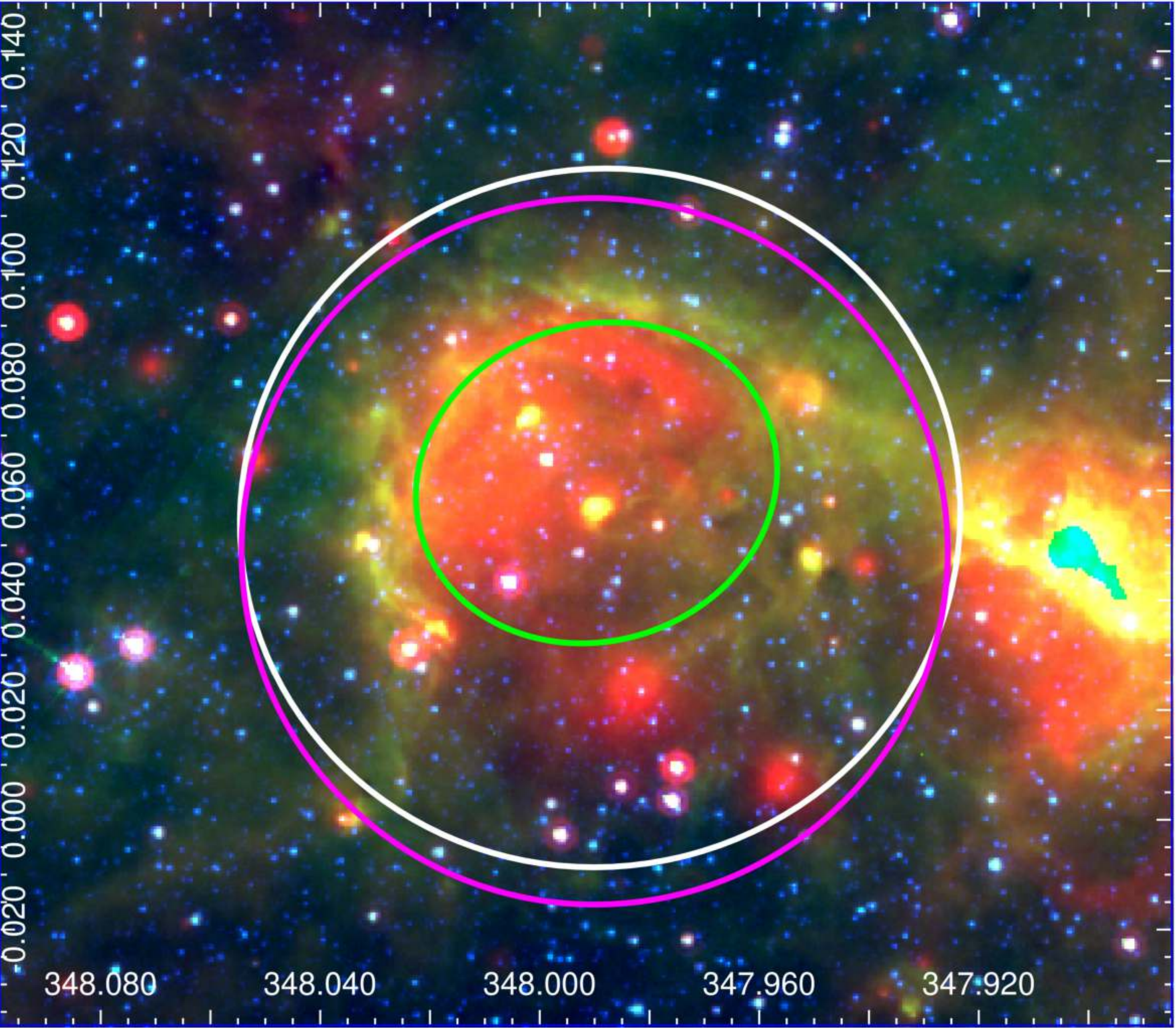}
	\caption{Comparing the DR2 (green), DR1 (white) and A14 (magenta) size measurements for the bubble MWP2G3479896+0006139. Other bubbles in the FOV of this image are not shown to reduce clutter.}
	\label{dr12compare}
\end{figure}

Visual review of the DR2 catalogue by multiple authors suggests that DR2 better constrains the size and shape measurements of bubbles previously listed in DR1 and A14 (Figure \ref{dr12compare}).

\begin{figure} 
	\includegraphics[width=\columnwidth]{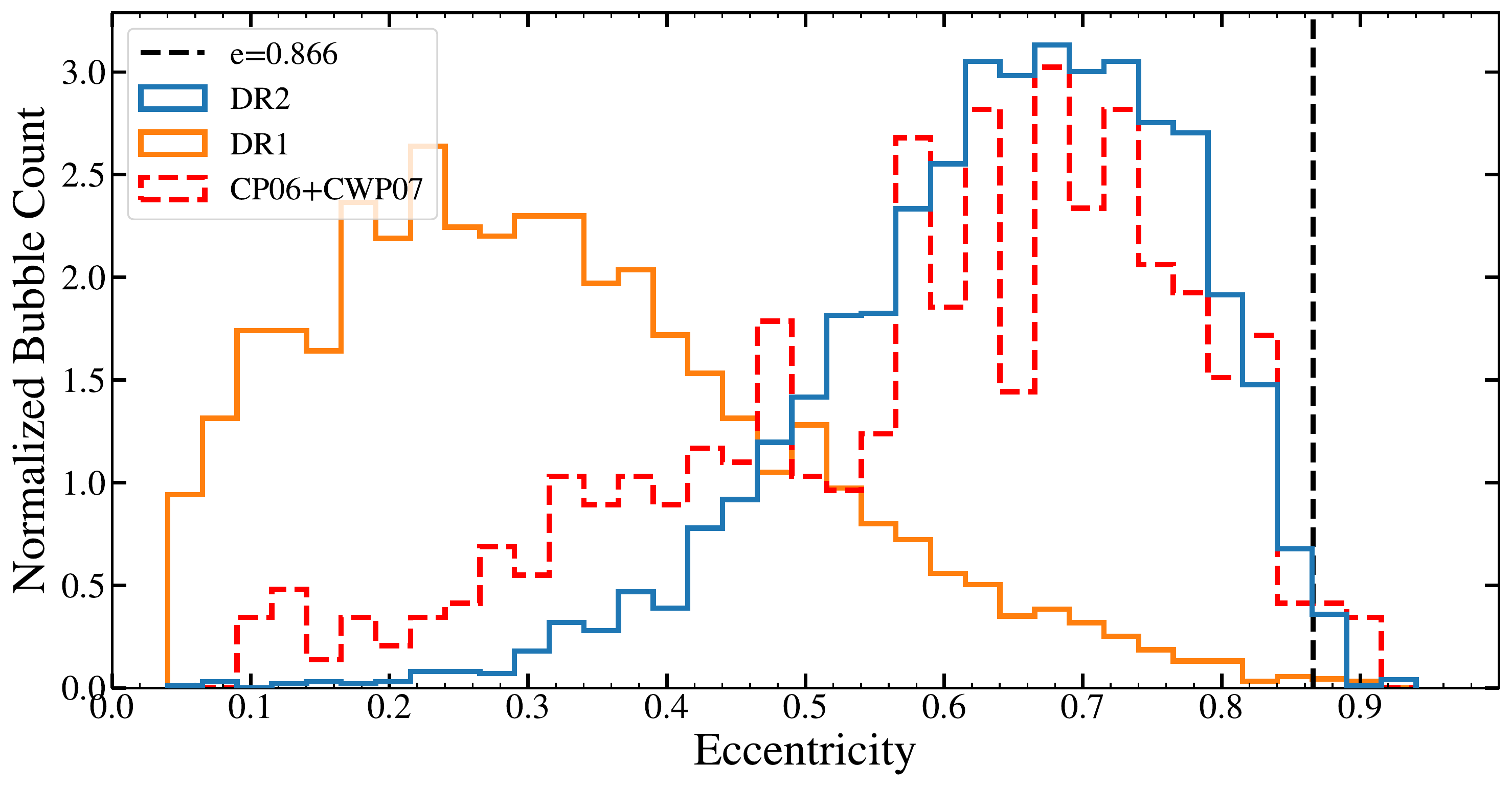}
	\caption{Distribution of eccentricities. A default bubble has an eccentricity e=0.866 corresponding to a 2:1 axis ratio. The notable scarcity of bubbles at this eccentricity highlights the elimination of biases with regard to imprecise bubble drawings.}
	\label{fig9}
\end{figure}

Among all bubble parameters measured, the eccentricity distribution of DR2 bubbles is the most drastic departure from DR1 (Figure \ref{fig9}).
The DR2 eccentricity distribution closely resembles that of CP06+CWP07 bubbles, which was measured by a small group of `expert' classifiers. By contrast, DR1 bubbles are biased toward more circular shapes.
	
\subsection{The spatial distribution of MWP bubbles}
\begin{figure} 
	\includegraphics[width=\columnwidth]{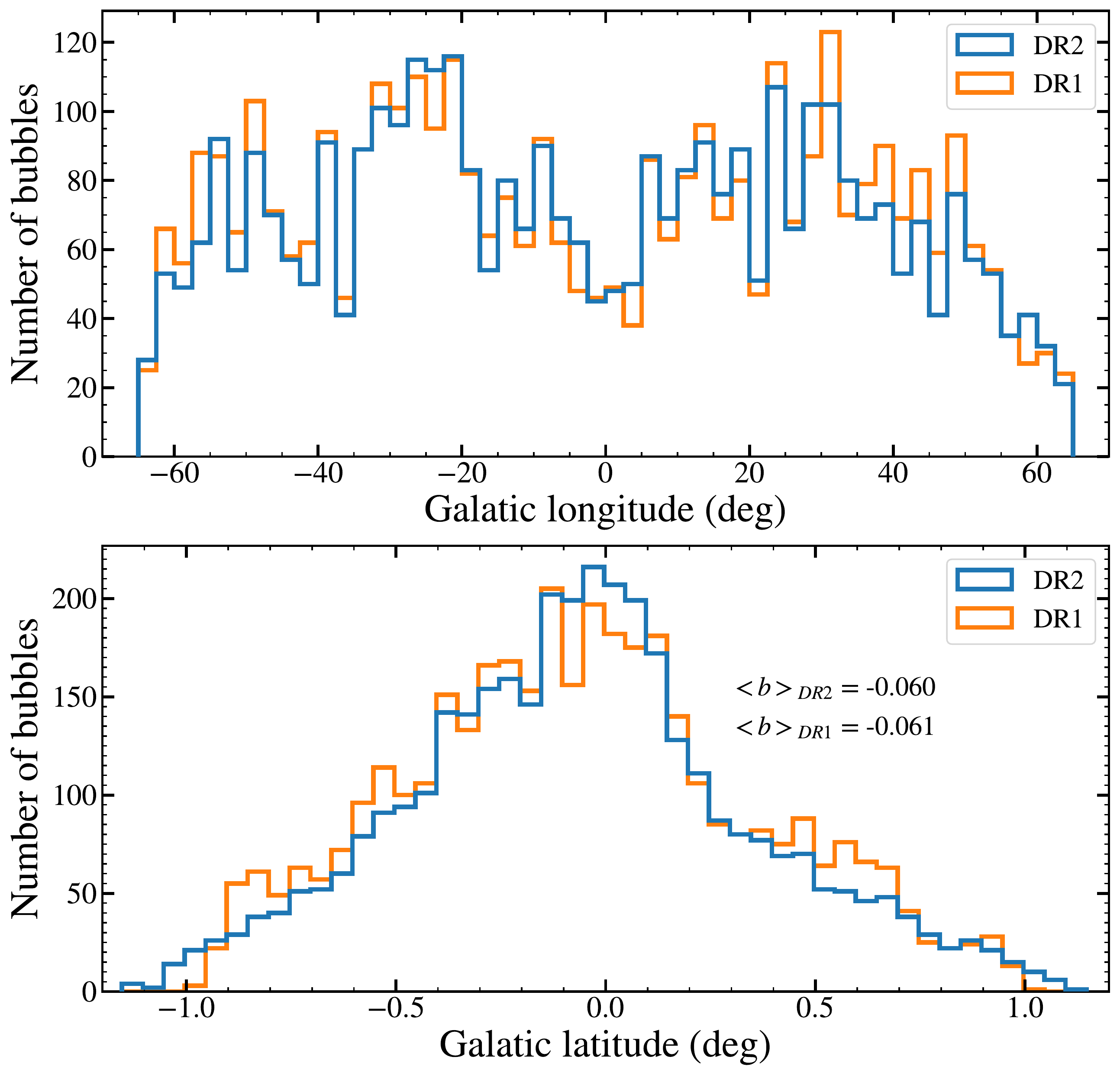}
	\caption{\textit{Top}: Distribution of MWP bubbles with Galactic longitude. This histogram is binned with 2.5\degree~bins. \textit{Bottom}: Distribution of MWP bubbles with Galactic latitude. This histogram is binned with 0.05 \degree~bins.}
	\label{GLIMPSELon}
\end{figure}

The  distribution of MWP bubbles with Galactic longitude (restricted to the bounds of the GLIMPSE I+II survey area) is shown in the top panel of Figure \ref{GLIMPSELon}. The characteristic dearth of bubbles towards the Galactic centre is clearly visible in both the DR1 and DR2 bubbles. 

The distribution of MWP bubbles with latitude (restricted to $|b|\le 1\degree$ within the GLIMPSE I+II survey area) is shown in the bottom panel of Figure \ref{GLIMPSELon}. To find the scale height, we calculated the second moment of the distribution. We also computed the mean and second moments for the latitude distributions of the DR1 catalog, the CP06+CWP07 catalogue and the A14 catalogue (summarized in Table \ref{latitude}). Our results are generally consistent with those from the DR1, CP06+CWP07 and A14 catalogs, although the DR2 bubbles and the A14 \hii regions display a marginally lower Galactic scaleheight compared to the DR1 and CP06+CWP07 samples.

\begin{figure}
	\includegraphics[width=\columnwidth]{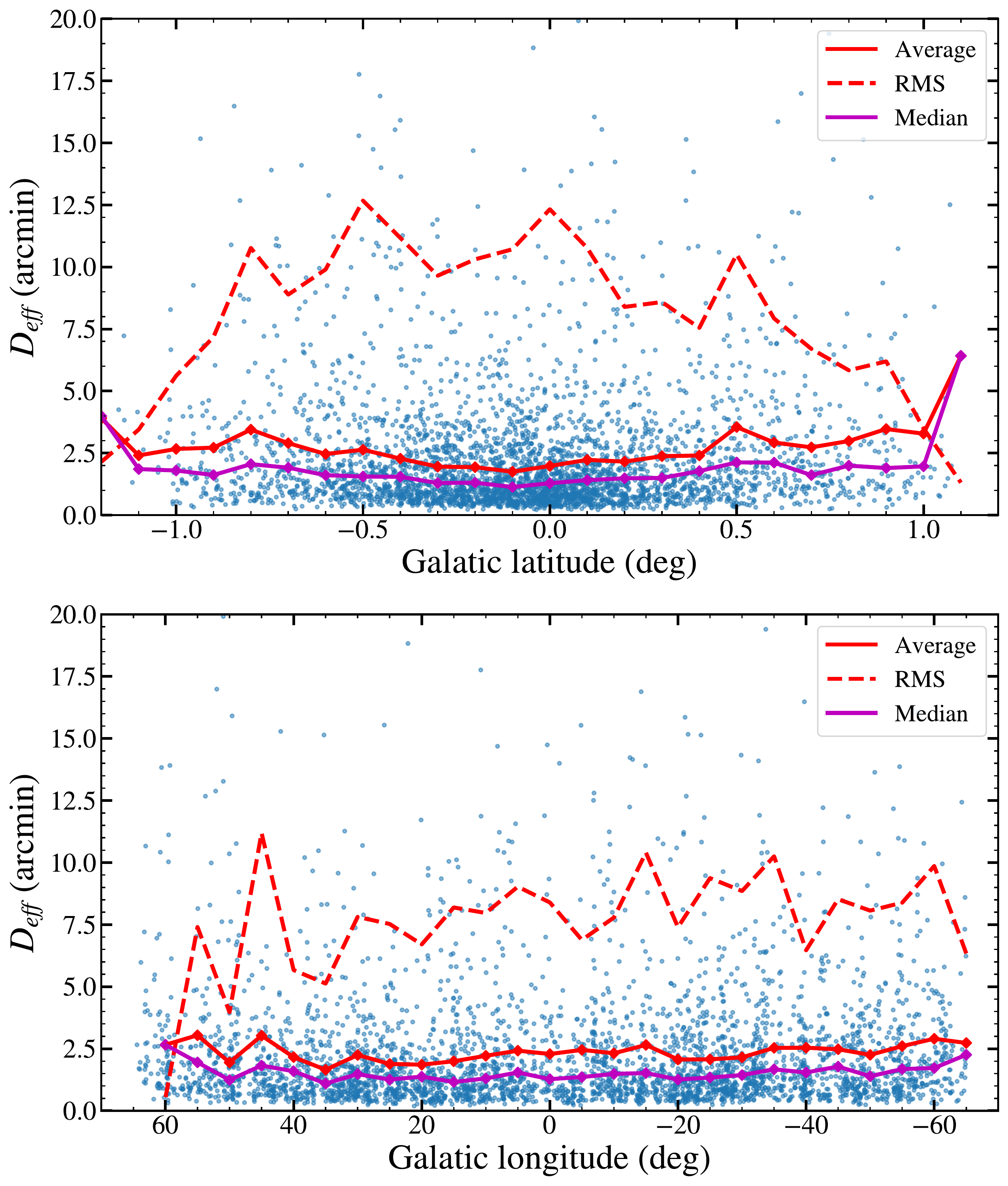}	
	\caption{The Spatial distribution of bubbles with diameter. \textit{Top}: $D_{\rm eff}$ against latitude, \textit{Bottom}: $D_{\rm eff}$ against longitude }
	\label{scatterspatial}
\end{figure}

We examine the spatial distribution of DR2 bubbles with bubble size in Figure \ref{scatterspatial}. The running mean, RMS and median distributions were plotted with 0.1$\degree$ and 5$\degree$ bins for the latitude and longitude respectively. The running mean and median distributions are generally constant across different bins. In contrast, the RMS distribution highlights the differences in large bubbles across Galactic latitude and longitude. We would expect that physically larger bubbles tend to be fainter for the same luminosity and more distant bubbles to be concentrated toward the midplane. This is not noticeable in the latitude distribution for DR2. Even though we are unable to completely rule this out, we do not see any obvious bias towards our ability to detect bubbles of different sizes with location in the survey. We have begun a new data-collection phase of the MWP using both synthetic and transplanted bubbles designed to test our survey sensitivity and completeness with Galactic location. We also identify a slope in the distribution of RMS size with longitude and find that we get systemically larger bubbles on average from left-to-right in the longitude distribution in Figure \ref{scatterspatial}, which could possibly be a signature of spiral arm structure. Spiral arms swing closer to the Sun at negative Galactic longitudes and this could potentially create a shift in the distribution of apparent bubble sizes with longitude.

We identified a total of 121 hierarchical bubbles (${\sim}5\%$ of the DR2 catalog), of which 52 were flagged as `parent' bubbles with 69 `daughter' bubbles (hence a few parents had multiple daughters). Of these 121 bubbles, 61 are listed in the more reliable sample. These results reveal that a significant fraction of the DR1 bubble hierarchies involved false bubble candidates, particularly towards \textit{busy} star forming regions. \citet{2012ApJ...755...71K} used the DR1 catalogue to study massive star formation associated with infrared bubbles and identified a positive correlation between massive young stellar objects and \hii regions.  \citet{2016ApJ...825..142K} subsequently found a strong correlation between cold, dense material in and around the DR1 bubbles, which suggests that the overdensity of the cold, turbulent dust clumps along bubble rims is associated with star formation.

\begin{table}
	\caption{Mean latitude and scale height of DR2 bubbles and bow shocks}
	\label{latitude}
	\begin{tabular}{ccc}
		\hline
		Distribution & $\langle b\rangle$  [\degree] & Second Moment [\degree]\\
		\hline
		DR1 & $-0.061$ & 0.175\\
		DR2 all & $-0.060$ & 0.143\\
		DR2 more reliable & $-0.055$ & 0.153\\	
		DR2 more complete & $-0.066$ & 0.132\\
		A14 & $-0.071$  & 0.145\\
		CP06+CWP07 & $-0.074$ & 0.184\\	
		\hline
        DR2 bow shocks & $-0.029$ & 0.226\\
        K16 GLIMPSE bow shocks & $-0.022$ & 0.206\\
		\hline
	\end{tabular}
	
\end{table}

\subsection{Cross-matching with existing catalogues}

\begin{figure} 
	\includegraphics[width=\columnwidth]{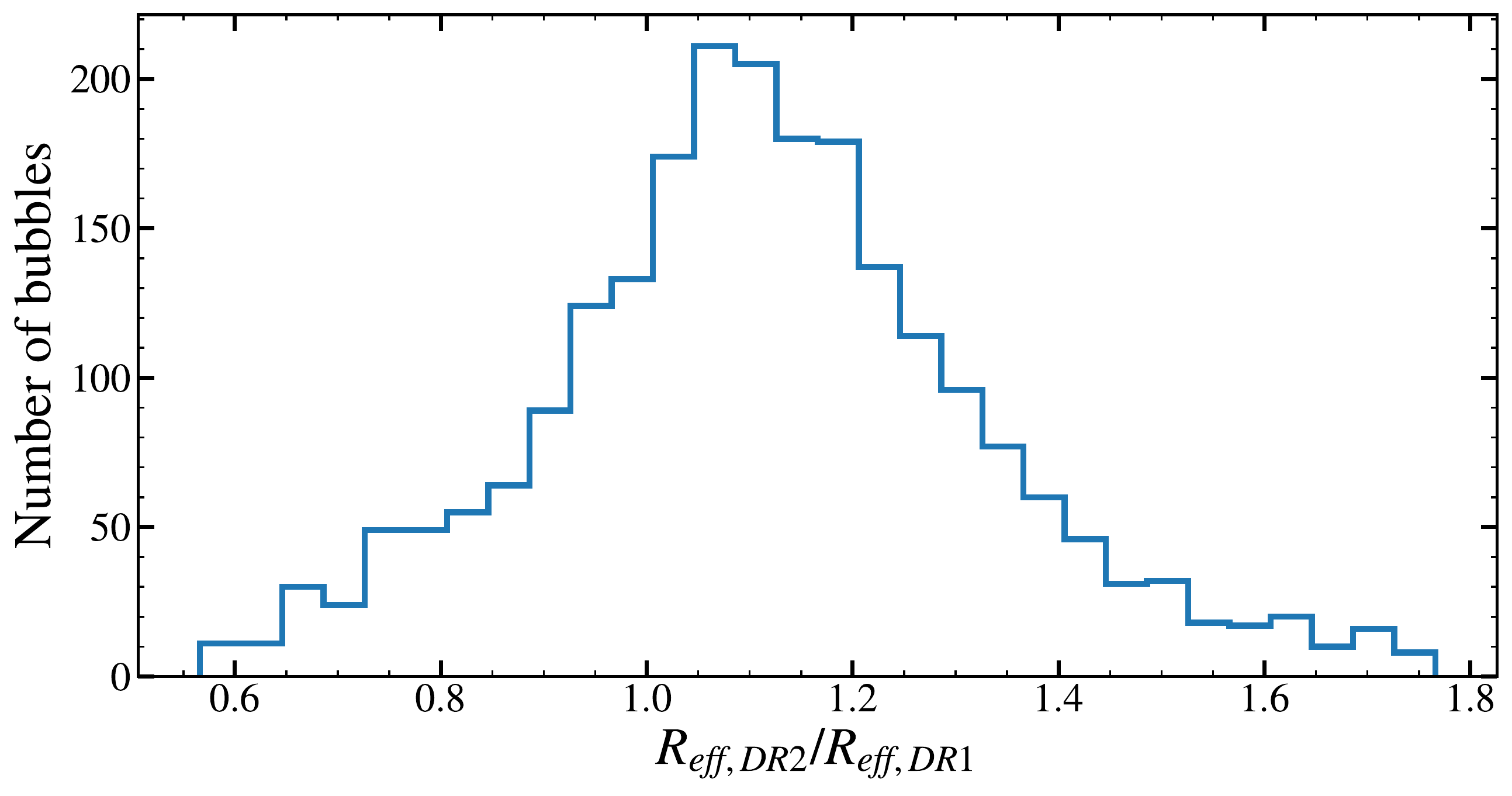}
	\caption{Distribution of the ratio of the effective radii for DR2 bubbles matched to DR1 bubbles.}
	\label{sizeratios}
\end{figure}

 A bubble listed in the MWP DR2 catalogue was identified as matching a DR1 (large) bubble when the central coordinate of the DR1 bubble lay within the radius of the DR2 bubble and when the effective radii of the two bubbles agreed to within twice the (highest) global fractional uncertainty in $R_{\rm eff}$ across the V2 and V3 bubble clusters. 71.1$\%$ of the bubbles in the DR2 catalogue had a match in the DR1 catalog. Figure \ref{sizeratios} shows the ratio of the effective radii $R_{eff,DR2}/R_{eff,DR1}$ between matching DR2/DR1 bubbles. DR2 bubbles are on average 12$\%$ larger than their DR1 counterparts. Because the DR2 bubbles also measured smaller sizes than  DR1, the overall dynamic range in the size distribution is greater in DR2 than in DR1 (Figure~\ref{deff}).

A DR2 bubble was identified as coincident with the CP06+CWP07 catalogs when the central coordinate of a catalogued bubble lies within the radius of the DR2 bubble. 86.7$\%$ of the bubbles in the CP06+CWP07 catalogs were matched to the DR2 catalog. All the possible CP06+CWP07 matches are included in the final DR2 bubble catalog. 

\begin{figure} 
	\includegraphics[width=\columnwidth]{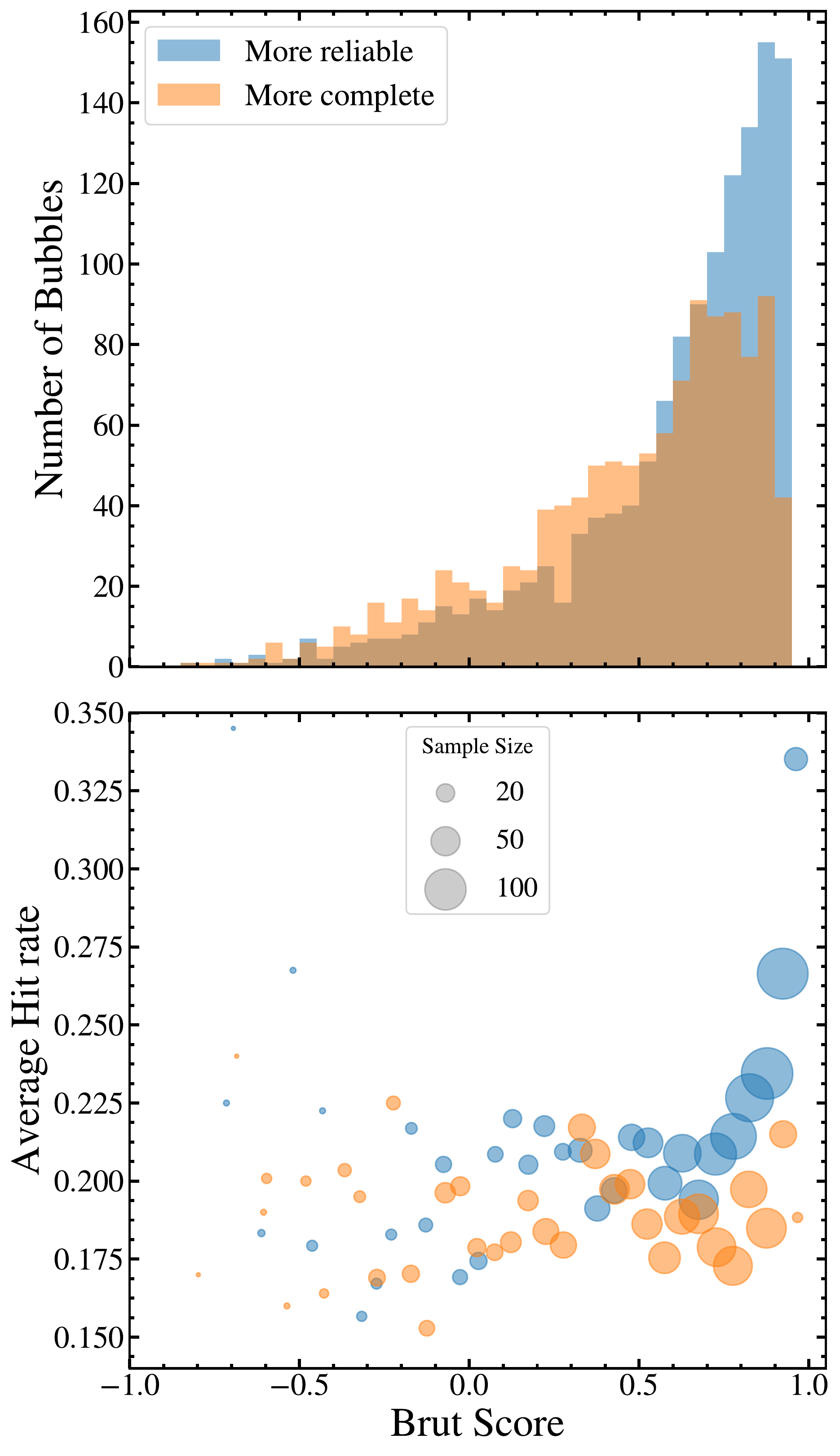}
	\caption{Results of applying the retrained {\scshape Brut} machine-learning algorithm \citep{Xu+Offner17} to the DR2 bubbles catalog. {\it Top:} Distributions in {\scshape Brut} score for the more-complete and the more-reliable subsets of bubbles. {\it Bottom:} Average hit rate versus average binned {\scshape Brut} score.}
	\label{brutscores}
\end{figure}

 Using a machine learning algorithm ({\scshape Brut}) designed to locate and identify \hii region bubble morphologies in 3-colour GLIMPSE+MIPSGAL images, \citet{2014ApJS..214....3B} assigned a {\scshape Brut} score to each DR1 bubble. The {\scshape Brut} score is defined as $2P-1$, where $P$ is the probability that the structure is a real bubble, based on a training set of 486 high-confidence DR1 bubbles. A {\scshape Brut} score of 0.2 was considered as the minimum acceptable score for bubbles, hence \citet{2014ApJS..214....3B} reported that roughly 30$\%$ of the DR1 catalogue consisted of random ISM structures that were incorrectly identified as bubbles. 
 \citet{Xu+Offner17} supplemented the original {\scshape Brut} training set with synthetic images of bubbles created from simulations of intermediate-mass stars launching winds into turbulent molecular clouds. We applied this `retrained' {\scshape Brut} algorithm to the DR2 catalog, and the results are presented in Figure \ref{brutscores} (the analogous plots for DR1 large bubbles are presented in Fig.~14 of \citealt{Xu+Offner17}). The distributions of {\scshape Brut} scores for DR2 more-reliable bubbles are more strongly skewed toward higher values compared to bubbles in the more-complete sample (top panel of Figure~\ref{brutscores}). 
 
 The bottom panel of Figure \ref{brutscores} shows that the hit rate correlates strongly with positive values of {\scshape Brut} score for DR2 bubbles in the more-reliable subset, while this correlation is marginal for bubbles in the more-complete sample. For the relatively small numbers of DR2 bubbles with negative {\scshape Brut} scores we observe large scatter in the hit rate with respect to the machine-learning results. This may simply be statistical noise, or it could reflect some small set of patterns in the imaging data for which the human and the machine-learning classifications systematically diverge. Overall, 15$\%$ of the bubbles in the DR2 catalogue had a {\scshape Brut} score smaller than 0.2, a factor of 2 reduction when compared to the DR1 catalog.

\subsection{Uncertainties}

To assess the performance of this pipeline, we analysed four bubbles from the more-reliable subset of the DR2 catalogue, all of which were previously listed in the DR1 catalogue. Two small bubbles ($R_{\rm eff}<1'$) and two large bubbles ($R_{\rm eff}>2.5'$) were chosen in order to assess the differences in the weighted averaging process at two different size ranges. Relevant parameters of these four bubbles are listed in Table~\ref{err}.

\begin{figure*}
	\includegraphics[width=\textwidth]{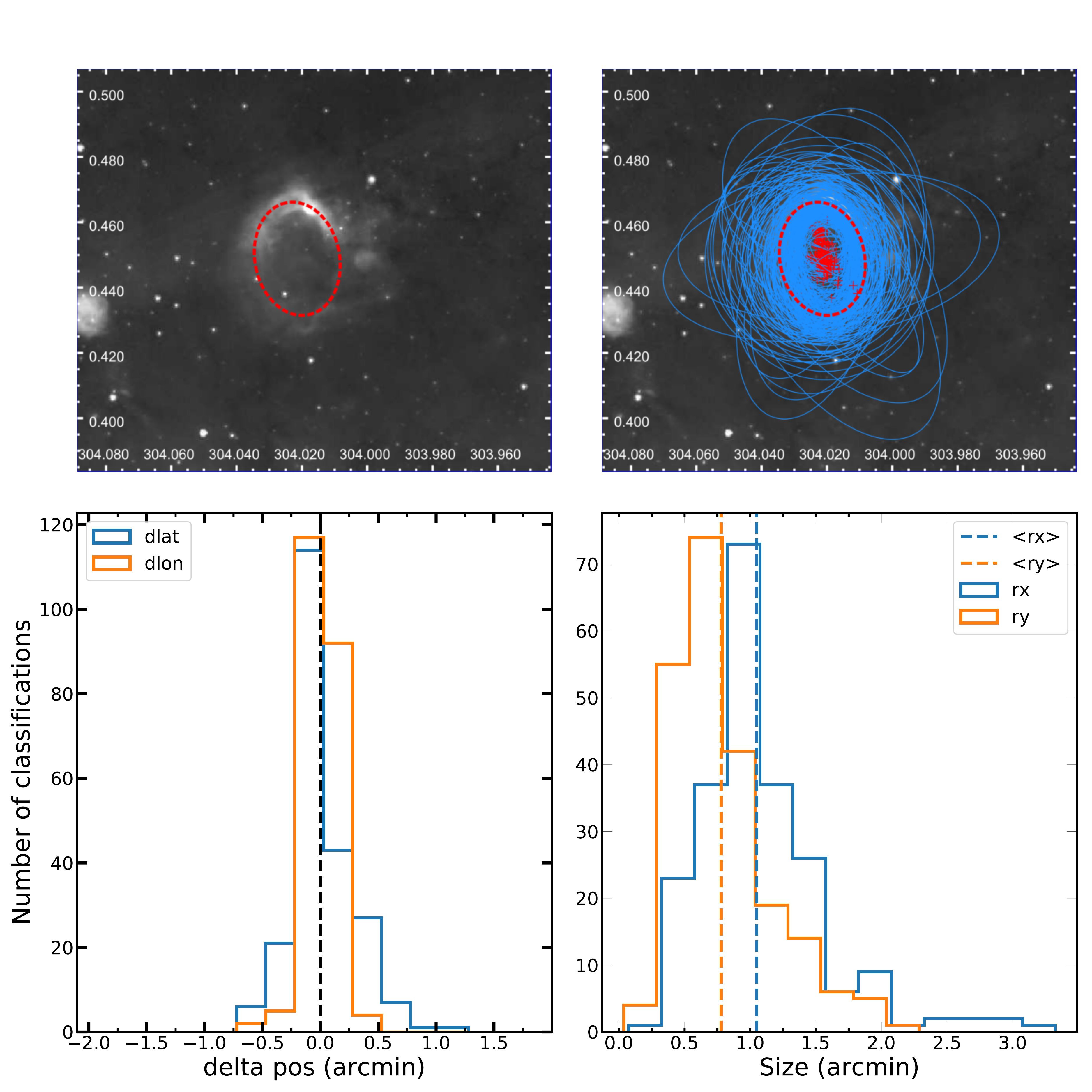}	
 	\caption{High reliability bubble MWP2G3040214+0044881. This bubble has a V2 hit rate of 0.400 and a V3 hit rate of 0.346, along with a dispersion of 0.38 arcmin. The top figures illustrate the final, reduced bubble alongside the raw bubble drawings. The bottom figures show dispersions in the measurements of position and size.}
	\label{buba}
\end{figure*}

\begin{figure*}
	\includegraphics[width=\textwidth]{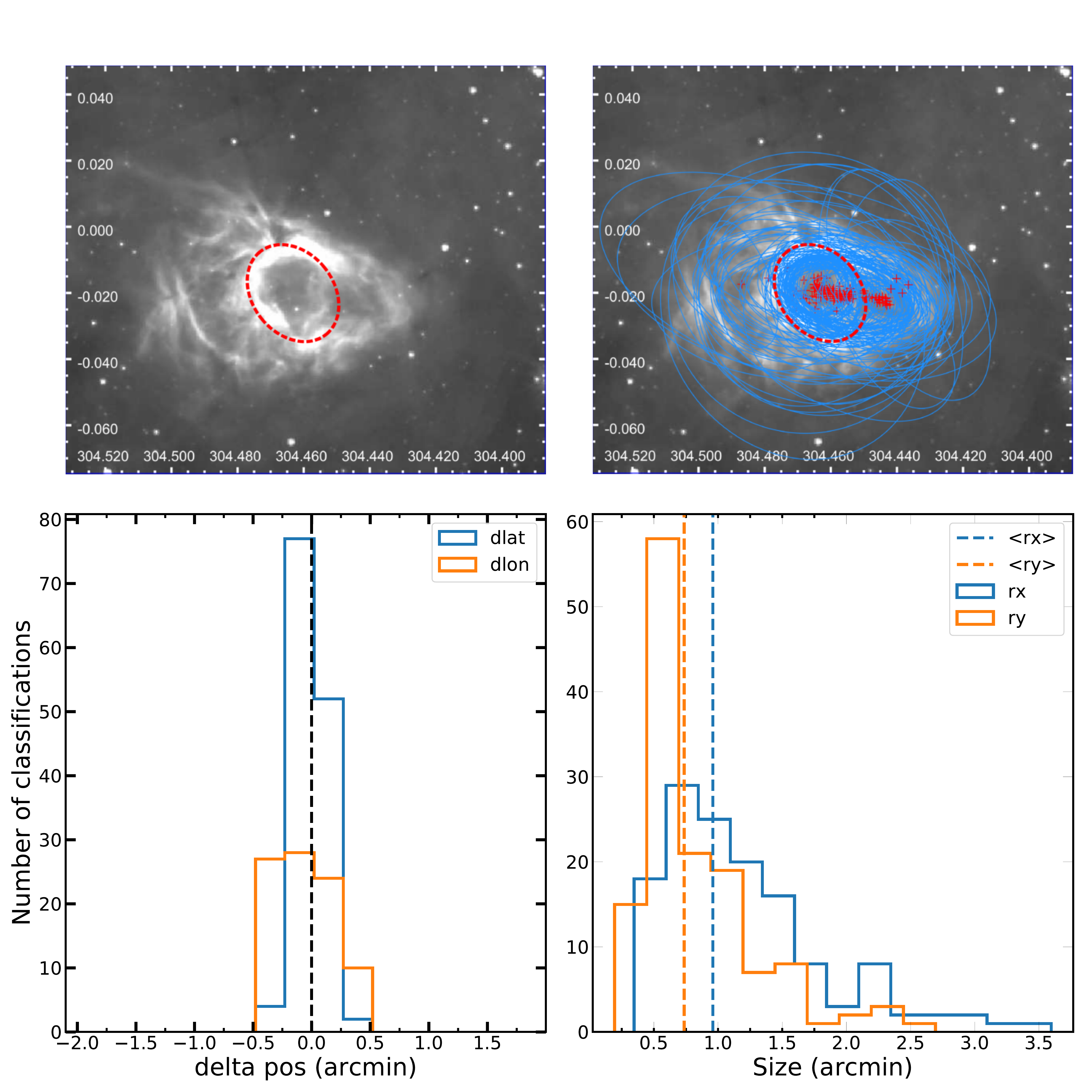}	
 	\caption{High reliability bubble MWP2G3044632-0002005. This bubble has a V2 hit rate of 0.333 and a V3 hit rate of 0.467, along with a dispersion of 0.54 arcmin. The top figures illustrate the final, reduced bubble alongside the raw bubble drawings. The bottom figures show dispersions in the measurements of position and size.}
	\label{bubb}
\end{figure*}

\begin{figure*}
	\includegraphics[width=\textwidth]{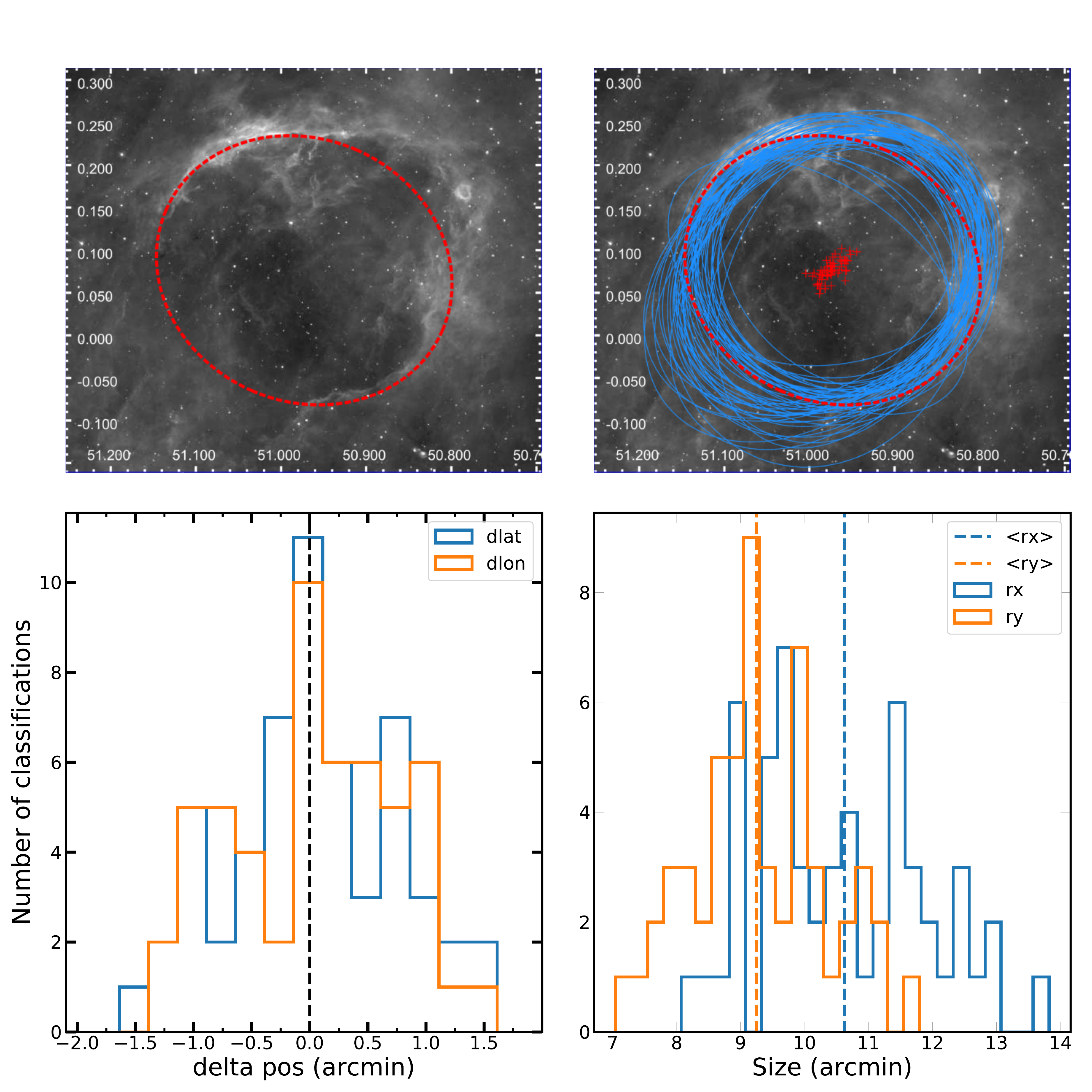}	
 	\caption{High reliability bubble MWP2G0509727+0007641. This bubble has a V2 hit rate of 0.145 and a V3 hit rate of 0.376, along with a dispersion of 1.45 arcmin. The top figures illustrate the final, reduced bubble alongside the raw bubble drawings. The bottom figures show dispersions in the measurements of position and size.}
	\label{bubc}
\end{figure*}

\begin{figure*}
	\includegraphics[width=\textwidth]{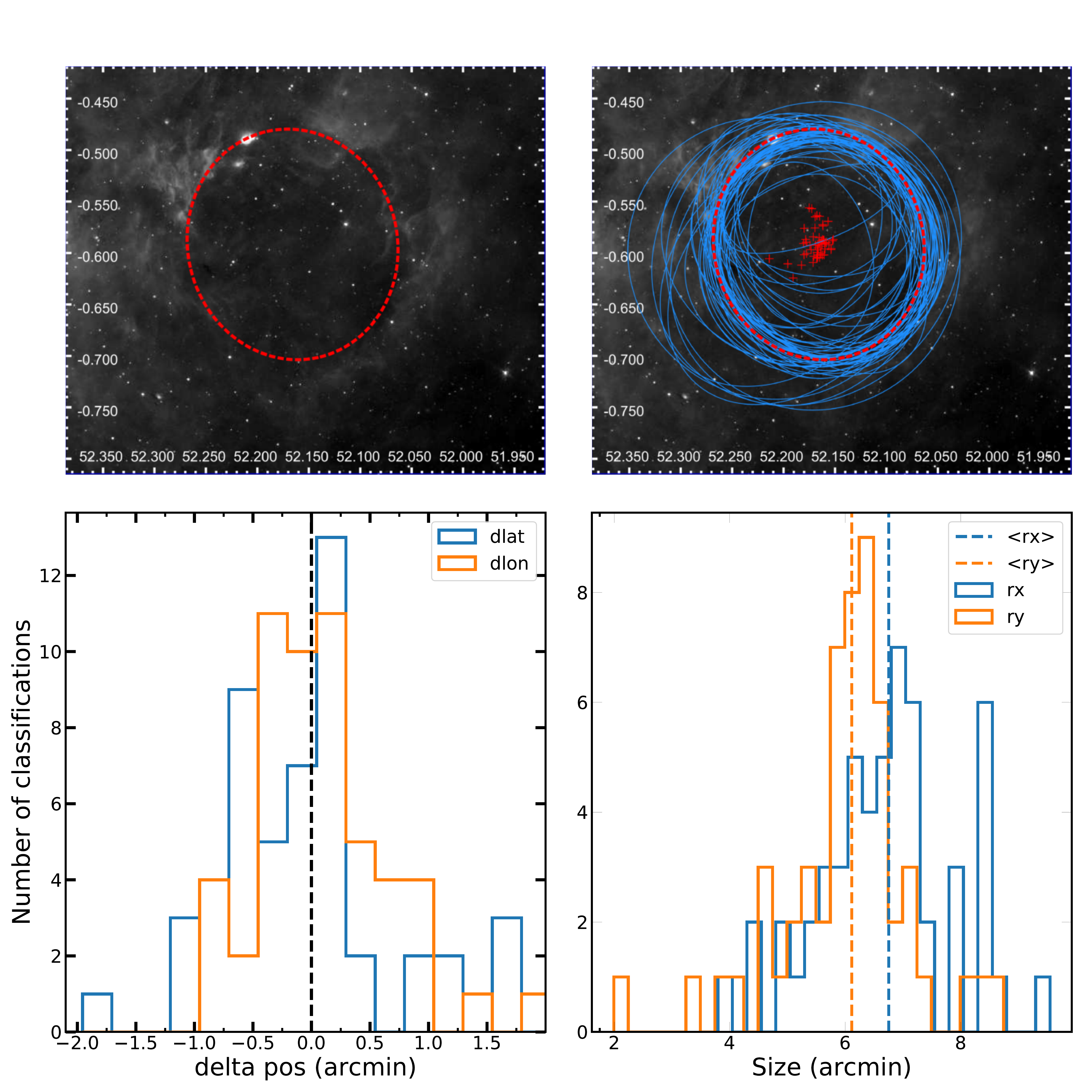}	
 	\caption{High reliability bubble MWP2G0521657-0059170. This bubble has a V2 hit rate of 0.205 and a V3 hit rate of 0.316, along with a dispersion of 1.36 arcmin. The top figures illustrate the final, reduced bubble alongside the raw bubble drawings. The bottom figures show dispersions in the measurements of position and size.}
	\label{bubd}
\end{figure*}

The central coordinates of smaller bubbles (Figures \ref{buba} and \ref{bubb}) are very well constrained and can be approximated by a Gaussian distribution. In contrast, larger bubbles have a greater spread in the central coordinates (Figures \ref{bubc} and \ref{bubd}), making the localization of the central coordinates much more difficult. User weighting helps in this regard, with the bubble drawings by MWP volunteers with a greater precision bubble fraction given more weight than drawings made by less careful users. 

The size parameters for bubble drawings cannot be well described by a Gaussian distribution. Frequently, multiple peaks in the size distribution are observed, regardless of the size of the bubble. This is inherently due to the way MWP volunteers perceive bubbles. Often, volunteers pick out different patterns in the 8 $\mu$m emission to demarcate the bubble drawing. This results in varying sizes in the user drawings for a given bubble. 

We note that while we report uncertainties in the various bubble parameters as standard deviations, the underlying distributions of these parameters may not be well described by a Gaussian distribution. As we noted here, non-Gaussianity is more common in the distribution of the size parameters. Nevertheless, the reported uncertainties are useful estimates in the errors of the various bubble parameters when used with caution.

\begin{table*}
	\caption{Parameters of the small and large bubbles used to assess the performance of the DR2 data reduction pipeline}
	\label{err}
	\begin{tabular}{cccccccc}
		\hline
		DR2 ID & DR1 ID & Classifications & HR2  & HR3 & $\sigma_{lb}$ [arcmin]  & $R_{\rm eff}$  [arcmin] & $\sigma_{r}$ [arcmin]\\
		\hline
		MWP2G3040214+0044881 & MWP1G304021+004503  & 242 & 0.400 & 0.346  & 0.38  & 0.92 & 0.38 \\
		MWP2G3044632-0002005 & MWP1G304463-000217  & 190& 0.333 & 0.467   & 0.54 & 0.86  & 0.30 \\
		MWP2G0509727+0007641 & MWP1G050955+001074  & 67 & 0.145  & 0.376  & 1.45 & 9.96  & 1.13 \\		
		MWP2G0521657-0059170 & MWP1G052171-005832  & 73 & 0.205  & 0.316  & 1.36 & 6.44  & 1.08 \\
		\hline
	\end{tabular}
	
\end{table*}

\section{The MWP Bow Shock catalogue Properties}

The final bow shock catalogue contains 599 BDSCs visually identified by MWP volunteers. Table \ref{table:bs_table} lists the columns in the MWP DR2 bow shock catalog. The DR2 bow shock catalogue is available online through \textit{Vizier} and as supporting information with the electronic version of the paper.

We cross matched the BDSCs with the K16 and 2MASS catalogs, the results of which were edited as necessary during visual review (see Section 3.6.2). From visual review, we also include standoff distances, position angles, `yes/no' flags for detectable 8~$\mu m$ emission and categorical codes describing the local environment (as defined by K16). By percentage, the distribution of local environment codes in the BDSC catalogue is 9\% FB, 12\% FH, 7\% H and 72\% I. Much like the K16 catalogue, most of the MWP bow shocks are located far from known \hii regions.
We also assigned reliability flags (`R' \& `C') for the BDSCs. BDSCs with K16 matches were identified and classified through two different bow shock searches with substantial methodological differences, thus we assign these BDSCs to the more reliable sample (`R'), while unmatched BDSCs were only assigned to the more reliable sample when $HR3 > (\beta=0.22$). Of the 599 BDSCs in the catalog, 453 are assigned a more reliable (`R') flag and 288 of these are matched to the K16 catalog.

\begin{table*}
	\caption{MWP DR2 BDSC catalogue columns}
	\label{table:bs_table}
	\begin{tabular}{ll} 
		\hline
		Column & Description\\
		\hline
		MWP ID & Unique MWP identifier (MWP2GLLLllll+BBBbbbb)\\
		$l$& Galactic longitude (degrees) \\
		$b$& Galactic latitude (degrees)\\
        $\sigma_{lb}$ & Dispersion of the central coordinates  (arcsec)\\
        HR3 & Hit rate 	\\
        Reliability & Reliability flag: `R' = more-reliable subset, `C' = more-complete sample \\
		K16 DSC & ID of matched K16 driving star candidate \\
		K16 Arc & ID of matched K16 bow shock arc\\
		MWP Bubble & ID of MWP bubble containing BDSC \\
		R.A. & Right ascension of matched 2MASS star (deg)\\		
		Dec. & Declination of matched 2MASS star (deg)\\
		Sep & Offset of matched 2MASS star (arcsec)\\
		$J$ & $J$-band magnitude of matched 2MASS star\\		
		$H$ & $H$-band magnitude of matched 2MASS star\\
		$K_s$  & $K_s$-band magnitude of matched 2MASS star\\
		$R_0$ & Distance from BDSC to apsis of arc (arcsec) \\
        P.A. & Position angle of vector joining apsis of identified arc (deg)  \\
		$[8.0]$ & Flag indicating arc detected in 8~$\mu$m emission\\
		Env. & Code describing local environment and orientation, as in K16 \\
		\hline
	\end{tabular}
\end{table*}

\begin{figure}
 \includegraphics[width=\columnwidth]{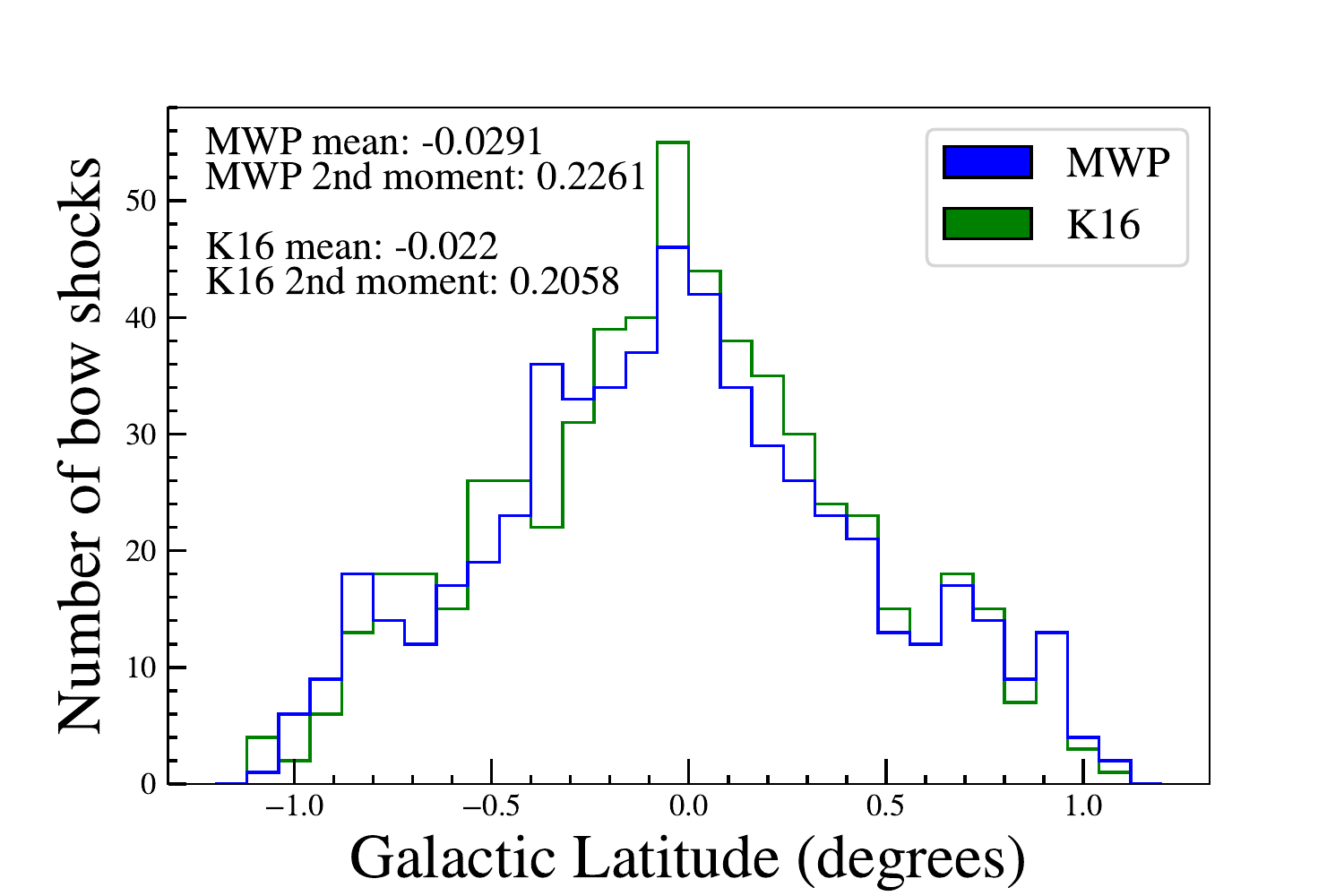}
 \centering
 \caption{Latitude distribution of bow shocks for both the MWP and K16 catalogue within the GLIMPSE+MIPSGAL footprint ($|l|\le 65\degree$, $|b|\le 1.2\degree$). \label{fig:BS_lat}}
 \end{figure}

\begin{figure}
 \includegraphics[width=\columnwidth]{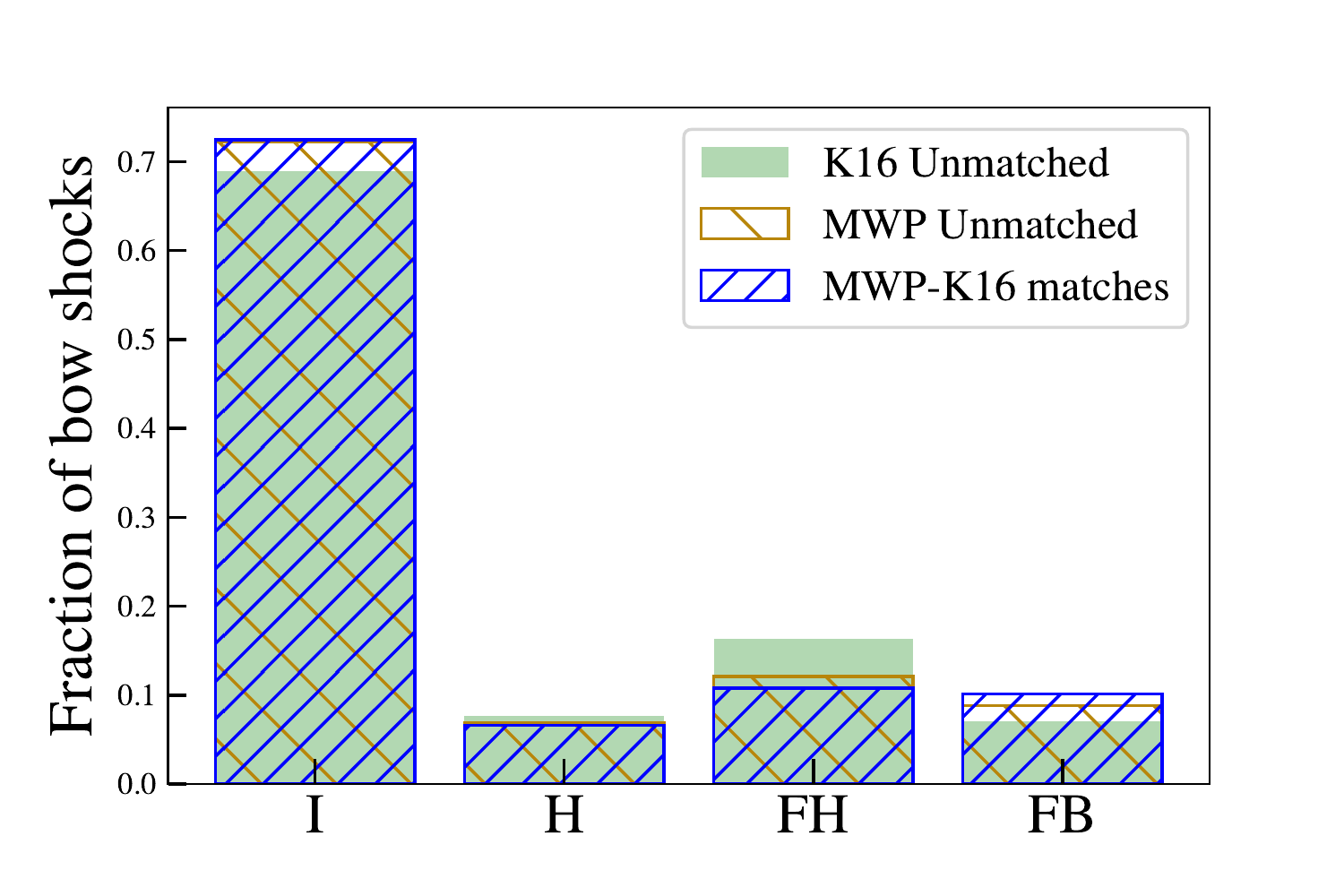}
 \centering
 \caption{Distribution of environment labels between MWP and K16, separated into samples of unmatched BDSCs (blue), K16 matched BDSCs (golden-brown) and unmatched K16 stars (green).\label{fig:BS_env}
 }
 \end{figure}
 
 \begin{figure}
 \includegraphics[width=\columnwidth]{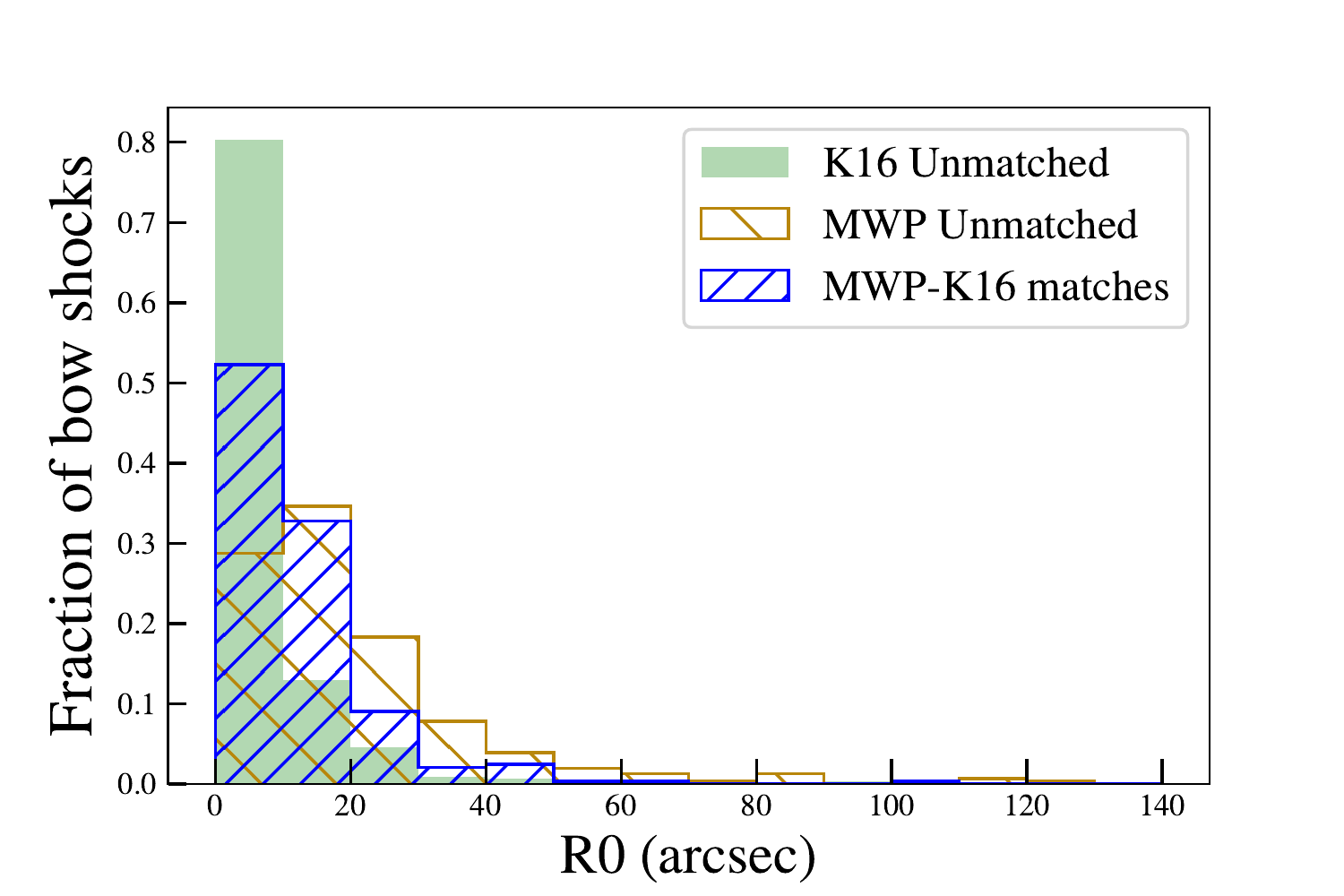}
 \centering
 \caption{Distribution of standoff distances beteween MWP and K16, separated into samples of unmatched BDSCs (blue), K16 matched BDSCs (golden-brown) and unmatched K16 stars (green).\label{fig:BS_R0}}
 \end{figure}

\section{Discussion}
\subsection{Performance of citizen scientists versus visual searches by `expert' astronomers }

The MWP DR2 and DR1 bubble catalogs each identified almost an order of magnitude more bubbles than CP06 and CWP07, which were constructed by visual identifications made by a small group of trained `expert' astronomers. However, we note that the CP06+CWP07 catalogs were constructed without the use of MIPS 24~$\mu$m data. We also find that MWP V2 bubble classifications were largely unreliable on their own, revealing that the experts outperformed the MWP citizen scientists when finding bubbles that were hard to spot in the image assets that lacked MIPS 24~$\mu$m data (i.e., these bubbles were identified by the 8~$\mu$m rings only). With the MIPS 24~$\mu m$ data, bubbles are easily identified, and when combined with our improved tutorials in V3, MWP citizen scientists performed much better than the experts. This is evident in the distribution of bubble eccentricities in Figure \ref{fig9}; MWP citizen scientists identified bubbles in a similar fashion to the experts in DR2. 

The size and shape measurements of the bubbles in DR2 tend to be more reliable than those identified by the experts. The MWP DR2 bubbles are drawn from a distribution of bubble drawings made by at least 5 individuals, compared to the 1 or 2 experts who identified bubbles in the CP06, CWP07 and A14 catalogs. We are also able to report uncertainties in these measurements unlike these catalogues (see Section $4.3$), which should be useful when using the DR2 bubbles as photometric apertures or comparing the locations of bubble rims with other signposts of star formation \citep[e.g.][]{T+12,2012ApJ...755...71K,2016ApJ...825..142K}. Since A14 never attempted to make precise measurements of the shapes and sizes of bubbles, the cross-matches between the DR2 catalogue with the A14 catalogue provides useful new measurements that supplement the A14 \textit{WISE} \hii region catalogue.

The bow shocks catalogued by K16 were identified by a small group of expert astronomers and research students, who in a similar fashion to MWP primarily searched in a set of MIPSGAL and GLIMPSE mosaics. Of the 709 bow shocks in K16, 609 lie within the MWP survey area. The intersection between both catalogues consists of just over half of the BDSC catalogue, with 288 (48\%) driving star matches plus 27 (4.5\%) bow shock arc matches. MWP has hence discovered 284 new bow shock candidates.

We compare spatial, environmental and size distributions between the MWP DR2 BDSC and K16 catalogues to check for biases in the different methods of classification. We select the driving stars in the GLIMPSE/MIPS survey for both catalogues and compare their Galactic latitude distributions (Figure~\ref{fig:BS_lat}). We calculate the mean and second moment for both distributions and find they are similar, but there is an  excess of K16 bow shocks compared to MWP BDSCs closer to the midplane, leading to a smaller second moment of the K16 latitude distribution. The distribution of environment codes between the DR2 BDSC catalogue and K16 are divided into unmatched and matched samples and plotted in Figure~\ref{fig:BS_env}. The distribution of environment codes is broadly similar across these subsamples, but there is a surplus of a few percent of K16 bow shocks facing \hii regions that were not rediscovered by MWP balanced by a few percent deficit of K16 unmatched, isolated bow shocks. This indicates a small but significant level of selection bias. We compare the sizes of bow shocks in the same samples used in our environment distributions using recorded standoff distances and find a strong bias in MWP against small bow shocks with $R_0<10\arcsec$ (Figure~\ref{fig:BS_R0}). 

Upon visual inspection of the unmatched K16 bow shocks, we find some likely causes for MWP selection bias. A number of bow shocks near bright infrared sources were scaled too dimly to be reasonably identified by volunteers. Many of the sources responsible were \hii regions, which helps to explain both the bias in the FH environment code and the relative deficit of MWP BDSCs at low Galactic latitudes, where the MIR nebular background emission is the brightest. The smallest bow shocks are barely resolved from their BDSCs, making it necessary for a MWP user trying to make a complete bow shock classification to place a BDSC reticle overlapping the bezier tracing the arc, which could lead volunteers to skip classifying these objects. 
 
 One co-author (HAK) performed an `expert' visual review of MWP bow shocks to investigate why such a large fraction were not previously discovered by K16. We determined that most of the new bow shocks were simply missed by the small number of expert classifiers. Additionally, we concluded that the K16 catalogue was biased against large bow shocks with extended arc emission, particularly in cases where the suspected driving star could be the principal ionising source of an \hii region. Three brand-new bow shocks were discovered and examined by multiple ``experts'' during this review process. These bow shocks (MWP2G0077756+0002274, MWP2G0249613+0022734 \& MWP2G0250137+0014294) were given the high reliability flag (`R') and added to the final DR2 catalogue.

Overall, the MWP DR2 and K16 bow shock catalogues are mostly consistent with and complement one another. K16 contains bow shocks that MWP missed near bright IR sources and/or small standoff distances, while  MWP discovered new bow shocks that were missed due to the lack of expert manpower and bias against extended arc emission.

\subsection{Bubbles versus bow shocks (and bow shocks within bubbles)}

 The Galactic latitude distribution of MWP DR2 bow shock candidates is virtually identical to that of the (larger) sample of K16 bow shocks within the GLIMPSE+MIPSGAL footprint (Figure~\ref{fig:BS_lat}). As K16 noted, both bubbles and bow shocks exhibit similarly tight distributions about the Galactic midplane, with the bow shock distribution showing a marginally larger second moment (Table~\ref{latitude}). This supports our assertion that the majority of both BDSCs and bubbles are produced by massive, OB stars, which are known to have a low Galactic scaleheight (CP06,K16).\footnote{We attempted to model the latitude distributions of bubbles and bow shocks with both Gaussian and reflected-exponential distribution functions but were unable to achieve satisfactory fits, hence we choose to report second moments in lieu of exponential scaleheights.} Follow-up spectroscopy of the K16 BDSCs has confirmed that the large majority (>85\%) are indeed OB stars (\citealp{KCP18,KCP19}, W. Chick et al. 2019, in prep.). 
 
 MWP DR2 provides, for the first time, automated matches between individual BDSCs and bubbles. Nevertheless, similarly to K16 we still find that ${\sim}70\%$ of BDSCs are not associated with any bubble or other obvious massive star-forming region in the {\it Spitzer} images (Figure~\ref{fig:BS_env}). If this supermajority of BDSCs were actual runaway OB stars, we would expect to observe a much larger Galactic scaleheight for IR bow shocks. \citet{KCP19} found an average proper motion of 4~mas~yr$^{-1}$ for a sample of 70 BDSCs with reliable kinematic and parallax data from the Gaia DR2 catalogue \citep{GaiaDR2}. This proper motion would move an OB star several degrees across the sky during its main-sequence lifetime, yet the median peculiar velocity of 11~km~s$^{-1}$ for these stars falls short of the 30~km~s$^{-1}$ threshold expected for runaways. 
 
 It is therefore clear that the observed spatial distribution of BDSCs does not trace the actual Galactic distribution of runaway OB stars, raising questions about the origins of these apparently isolated massive stars. The essential ingredients to produce both IR bubbles and bow shocks are hot, luminous stars surrounded by a {\it dusty} ISM. (Feedback from the stars has sufficiently cleared the local environment that they are no longer embedded, hence hyper/ultracompact \hii regions are excluded.)
 The Galactic latitude distributions of both bubbles and bow shocks are comparable to the molecular cloud scaleheight (Figure~\ref{fig:BS_lat} and CP06), which is a good proxy for the dense, dusty ISM in the thin disk midplane. High-velocity runaways would quickly move clear of the dense, dusty ISM of their natal environments, and without nearby dust they could not produce IR-bright bow shocks.

 \begin{figure}
\includegraphics[width=\columnwidth]{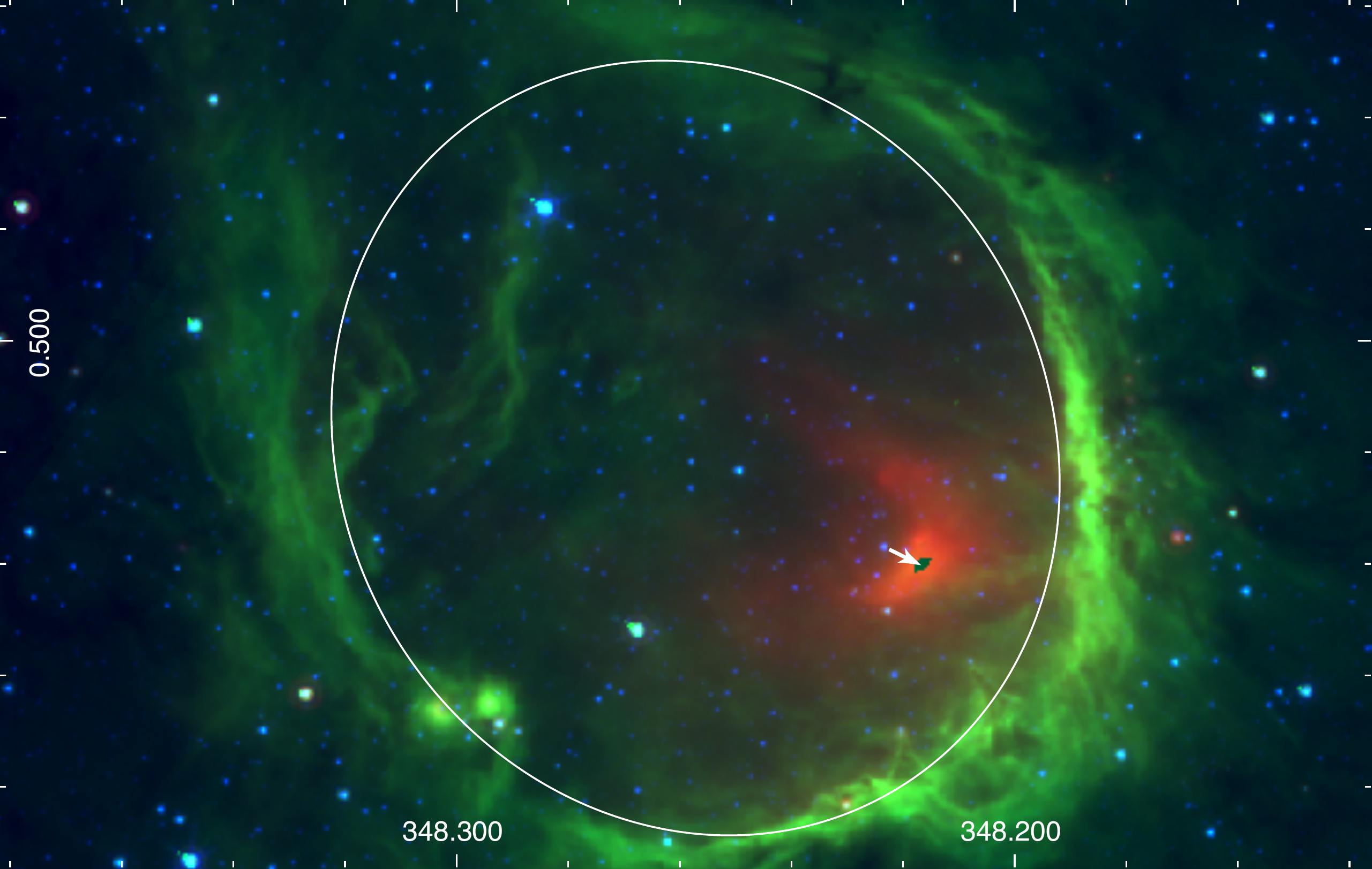}
\caption{GLIMPSE+MIPSGAL image of RCW 120 (MWP2G3482572+0048076; white ellipse), the prototypical example of a BDSC as the principal ionising star (MWP2G3482232+0046283) of a bubble \hii region lies at the base of the white vector representing standoff distance and position angle. In this example two pillars can be seen extending from the bubble rim at lower right, both pointing toward the projected position of the ionising star. 
}
\label{fig:perfect}
\end{figure} 
 Some BDSCs inside bubbles appear to be the principal ionising star of the \hii region they reside within. In these cases the arcs tend to be larger than average, with more extended 24~\um\ emission in the region between the convex side of the arc and the bright 8~\um\ bubble rim. Such objects are of particular interest as they appear to exemplify the transition between wind-blown bubble and bow shock morphologies \citep{Mackey+15,Mackey+16} and may provide laboratories for studying the physics of photoevaporative flows originating from the interface between hot ionised and cold molecular gas (`PEF' bow shocks; K16).
 We consider the prototypical bow shock candidate within a bubble to be RCW 120 (Figure~\ref{fig:perfect}),  a `perfect' bubble due to its near-circular morphology \citep{Deharveng+09}. RCW 120, or MWP2G3482572+0048076 in the DR2 bubbles catalogue (also known as G348.261+00.485 in A14, MWP1G348228+004692 in DR1 and S7 in CP06), has been extensively studied as a possible site of star formation triggered by the interaction of the expanding bubble with a surrounding molecular cloud \citep{Z07_RCW120}. The distance to RCW 120 was recently revised using Gaia DR2 parallaxes of 29 stellar constituents to $d=1.68^{+0.13}_{-0.11}$~kpc \citep{K19_Gaia}, which is significantly greater than the 1.3~kpc distance typically assumed in prior studies.
 
The ionising star of RCW 120 is CD-38 11636, with a reported spectral type of O8 V/III \citep{GG70,M+10}.  This star is MWP2G3482233+0046284 in the BDSC catalogue, and it has a high HR3 = 0.25. It was not selected for inclusion in the K16 catalog, probably because it is associated with highly extended, bright 24~\um\ emission filling the bubble. The star and its associated 24~\um\ arc are significantly displaced from the geometric centre of the circular bubble. \citet{MW19} suggested that CD-38~11636 has a high velocity relative to the bubble based on its Gaia DR2 proper motion, hence the star moved from the centre of the bubble to its current location. 

However, the Gaia DR2 proper-motion measurement appears to be unreliable for this star. Following the guidance of Lindegren (2018)\footnote{Technical note GAIA-C3-TN-LU-LL-124-01 available from https://www.cosmos.esa.int/web/gaia/public-dpac-documents.}, we calculate a renormalized unit-weight error (RUWE) of 6.39. ${\rm RUWE}<1.4$ is required to be confident in a measured Gaia proper motion, as higher values indicate that the single-star assumption used in the astrometric solution did not provide a good fit. It is possible that CD-38~11636 is an unresolved binary/multiple system \citep{MOXC2}.

A high peculiar velocity for the BDSC is not necessary to explain the morphology of the 24~\um\ arc within RCW 120. Simulations by  \citet{Mackey+16} reproduce its 24~\um\ morphology with a subsonic relative velocity between the star and the surrounding ISM.
We suggest an {\it in situ} PEF bow shock is fully consistent with available multiwavelength imaging data. 
The 8~\um\ bubble rim is clearly broken in the upper-left quadrant as viewed in Galactic coordinates (Figure~\ref{fig:perfect}), revealing a pathway for photons to leak out of the \hii region and depressurize the bubble.
\citet{MOXC2} presented a map of X-ray diffuse emission tracing hot, ionised plasma within the bubble. This reveals a second leak, this one in the lower-left quadrant where the diffuse plasma apparently passes in front of two embedded young clusters on the bubble rim (appearing as yellow-green knots in Figure~\ref{fig:perfect}). Using H$\alpha$ emission, \citet{2015ApJ...800..101A} also showed that ${\sim}5\%$ of the ionising photons leaked into the ISM through holes in the RCW 120 photo-dissociation region. These leaks indicate a large-scale density gradient running across RCW 120 in (approximately) the direction of decreasing Galactic longitude. The ionising star(s) most strongly irradiate the dense molecular cloud on the nearest (rightmost) part of the bubble rim. Gas and dust ablated by the process then become entrained in a photoevaporative flow directed down this density gradient toward the leaks on the opposite side of the bubble. As this flow passes by the ionising star(s), they interact with the stellar wind(s) to form the observed 24~\um\ bow shock structure. Numerous other instances of this morphology exist within the area surveyed by MWP, another beautiful example being the O9.5V BD+57 2513 (BDSC MWP2G1045667+0128085) ionising the S135 \hii region (bubble MWP2G1045663+0122810) in the outer Galaxy (imaged as part of the SMOG survey).

\subsection{Serendipitous bow shock discoveries from MWP V2}
\begin{figure}
\includegraphics[width=\columnwidth]{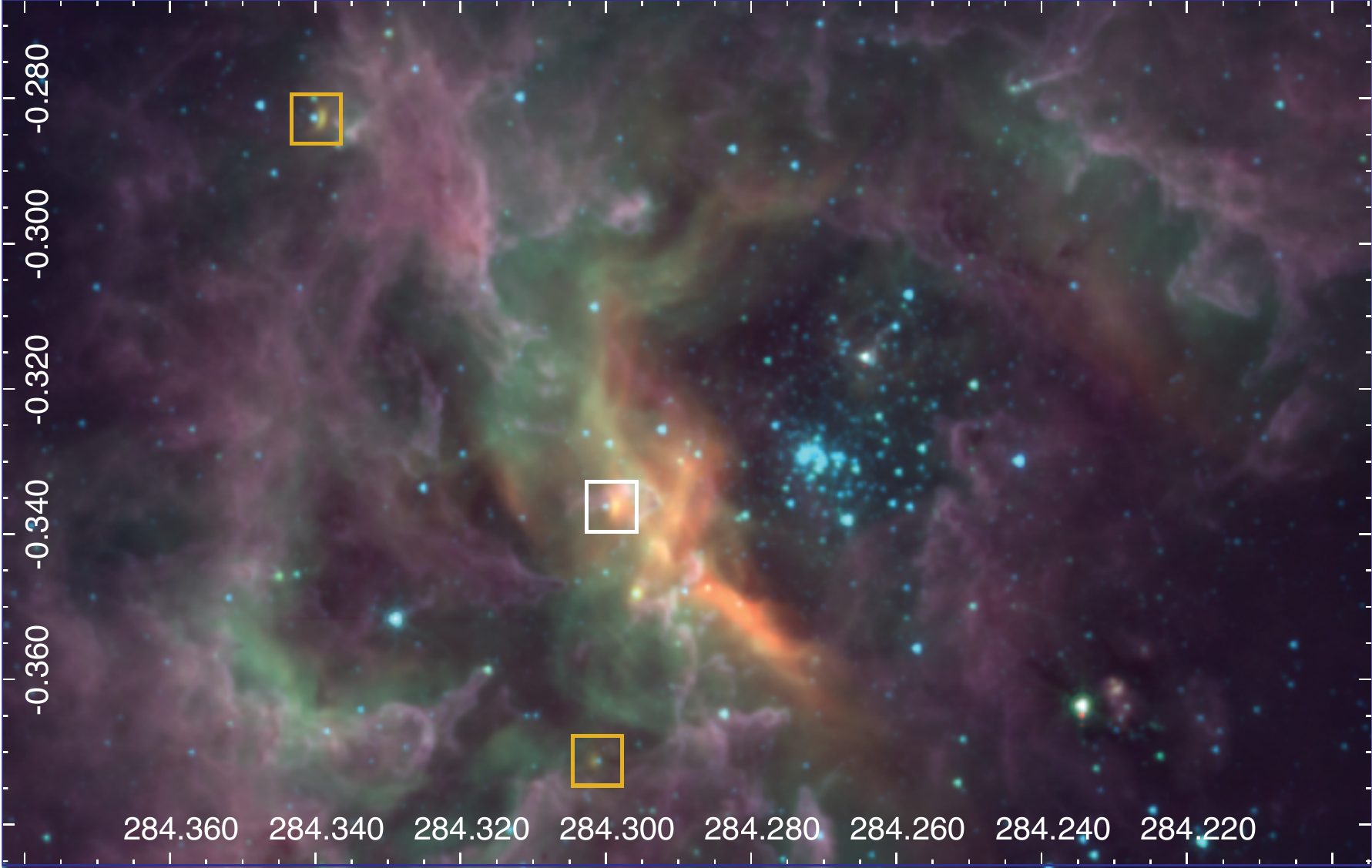}
\includegraphics[width=\columnwidth]{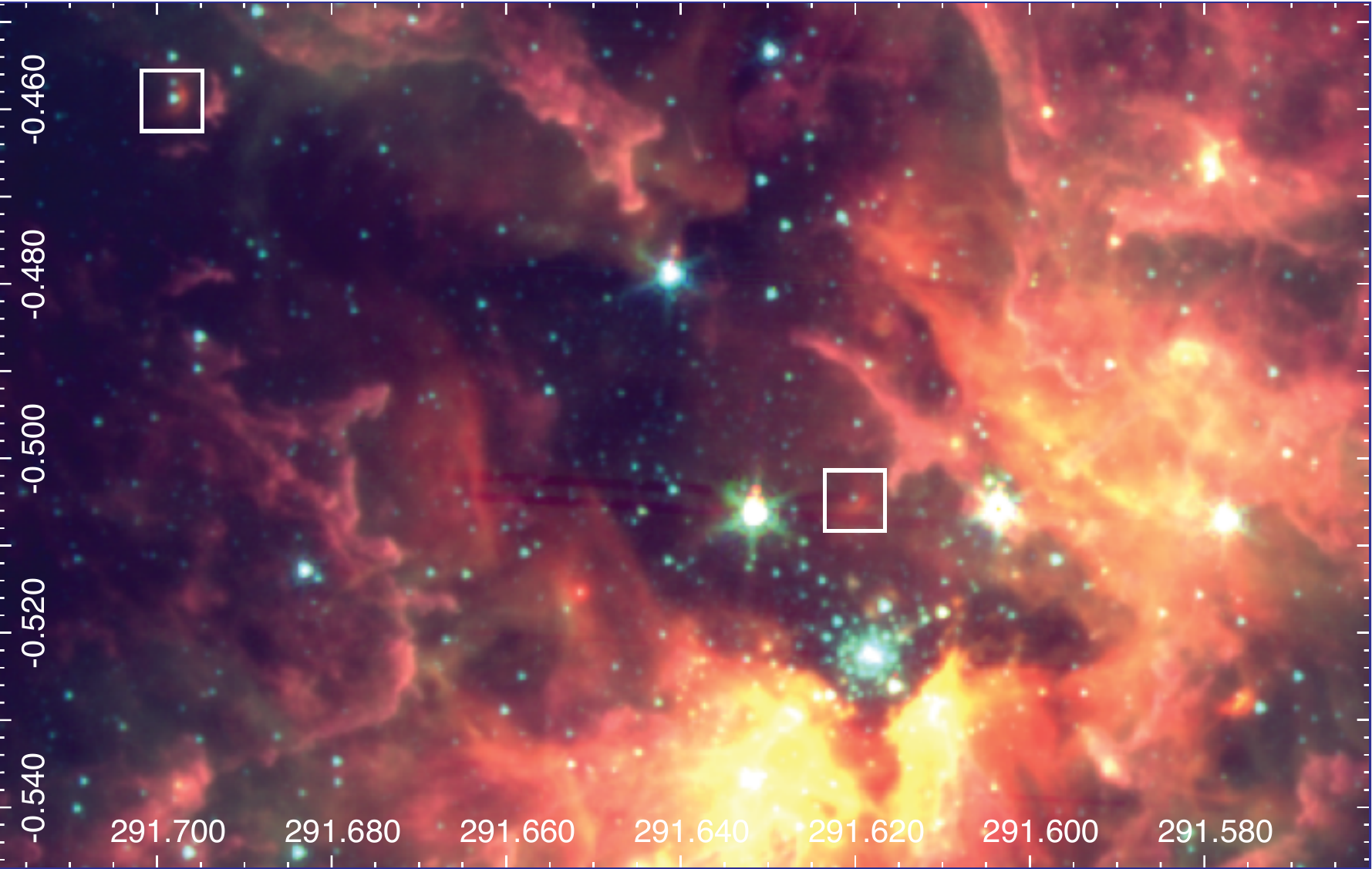}
\caption{Serendipitous discoveries of bow shocks seen only at 8.0~$\mu$m in MWP. {\it Top:} The RCW 49 \hii region containing the Westerlund 2 star cluster with one new bow shock candidate, MWP2G2842999-0033601 (white box), and two bow shocks previously discovered by \citet[][orange boxes]{Povich+08}. {\it Bottom:} The NGC 3603 \hii region with two new bow shocks, MWP2G2916980-0045894 and MWP2G2916201-0050459 (boxed).  In both images red, green and blue are assigned to 8~$\mu$m, 4.5~$\mu$m and 3.6~$\mu$m, respectively.\label{fig:legacy_bowshocks}}
\end{figure}

Three bow shocks  associated with areas of intense star formation were  discovered serendipitously during MWP V2 and manually added to the DR2 catalog. These bow shocks were identified as compact 8.0~\um\ arcs in Galactic starburst regions, where the 24~\um\ nebular background emission ranges from extremely bright to saturated. 

The first, MWP2G2842999-0033601, was identified by MWP users in an image of the RCW 49 giant \hii region and brought to researchers' attention via in the MWP talk discussion board. The bow shock lies in projection near the area of the brightest mid-IR emission in the entire nebula, where feedback from the very massive Westerlund 2 cluster impacts a dense molecular cloud (Figure ~\ref{fig:legacy_bowshocks}, top panel). Its arc, observed in all three of the [4.5], [5.8], and [8.0] bandpasses, appears to be oriented in the direction of Westerlund 2, making it a prime example of an {\it in situ} bow shock with an FH (facing \hii region) environment flag. Two other nearby bow shocks in RCW 49 were discovered by \citet{Povich+08}, but this one was missed. This highlights an important pitfall of using small groups of experts to search astronomical images, because relatively small/faint objects in close proximity to large/bright objects are easy to miss when manually panning, zooming, and scaling astronomical images with high dynamic range. This bow shock was listed in K16 as G284.2999-00.3359. 

The two other serendipitous bow shocks, MWP2G2916980-0045894 and MWP2G2916201-0050459, were found by MWP researchers while examining IRAC images of the NGC 3603 giant \hii region from the Vela-Carina survey (Figure ~\ref{fig:legacy_bowshocks}, bottom panel). \citet{G+13_NGC3603} previously reported the discovery of a 24~\um\ bow shock possibly associated with a runaway O6~V star from NGC 3603. In contrast, these new MWP bow shocks are closer (in projection) to the ionising cluster, more compact, observed at 8.0~\um\ but not at 24~\um, and both arcs are oriented toward, rather than away from, the ionising cluster (FH environment flags).
The nebular morphology of NGC 3603 exhibits a central cavity surrounding the dense, young massive cluster. The cavity has a narrow opening, providing a channel for high-pressure gas to leak out of the \hii region (RCW 49 displays a similar morphology; Figure~\ref{fig:legacy_bowshocks}). Indeed, \citet{MOXC1} observe very bright diffuse X-ray emission filling the cavity around the cluster and extending outward through this channel, which strongly suggests a `champagne flow' of hot, ionised plasma. The interaction of the winds from the two BDSCs with this gas flow could produce these {\it in situ} 8.0~\um\ bow shocks.  Exterior to the compact, red 8.0~\um\ arc, MWP2G2916980-0045894 exhibits two additional, pink flocculent arc fragments (just outside the white box in the upper-left corner of the bottom image in Figure~\ref{fig:legacy_bowshocks}). The pink colour of these fragments suggests a greater relative contribution of 3.6~\um\ (blue) than in the bow shock itself, indicative of the PAH emission that traces PDRs elsewhere in this image of NGC 3603. MWP2G2916980-0045894 therefore appears to be the remains of a bubble around the BDSC that has been almost completely destroyed by the powerful, feedback-driven flow of plasma driven by the NGC 3603 ionising cluster.\footnote{One caveat must be given about the association of MWP2G2916980-0045894 with NGC 3603. This star is listed in the Gaia DR2 catalogue with a parallax of $0.30\pm
0.05$~mas. Including systematic uncertainties, this is inconsistent with the 7-kpc distance of NGC~3603 \citep{HEM08} at the $2\sigma$ level.}

\subsection{The coffee ring nebula: a dark bubble?}

\begin{figure}
\includegraphics[width=\columnwidth]{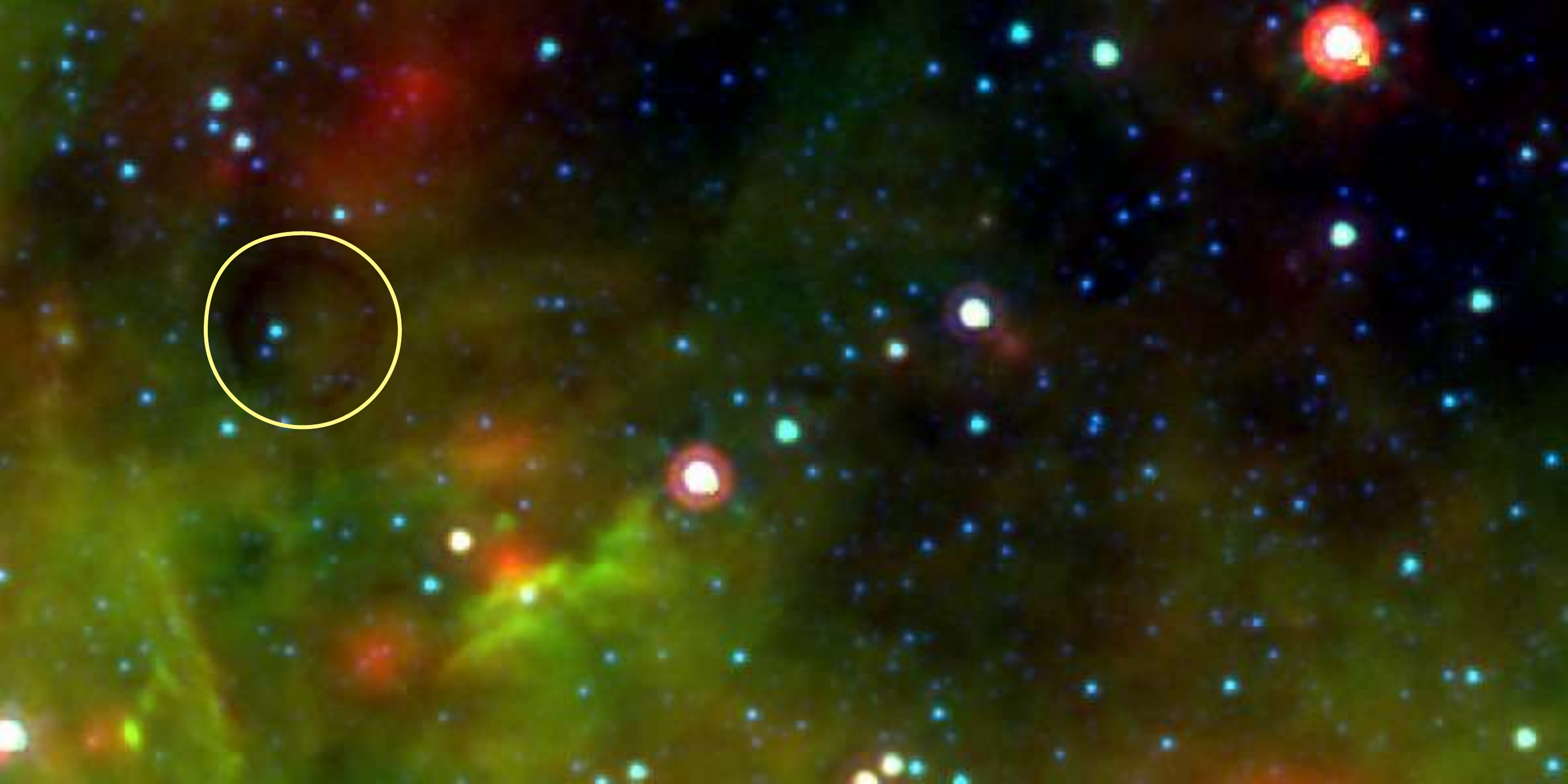}
\includegraphics[width=\columnwidth]{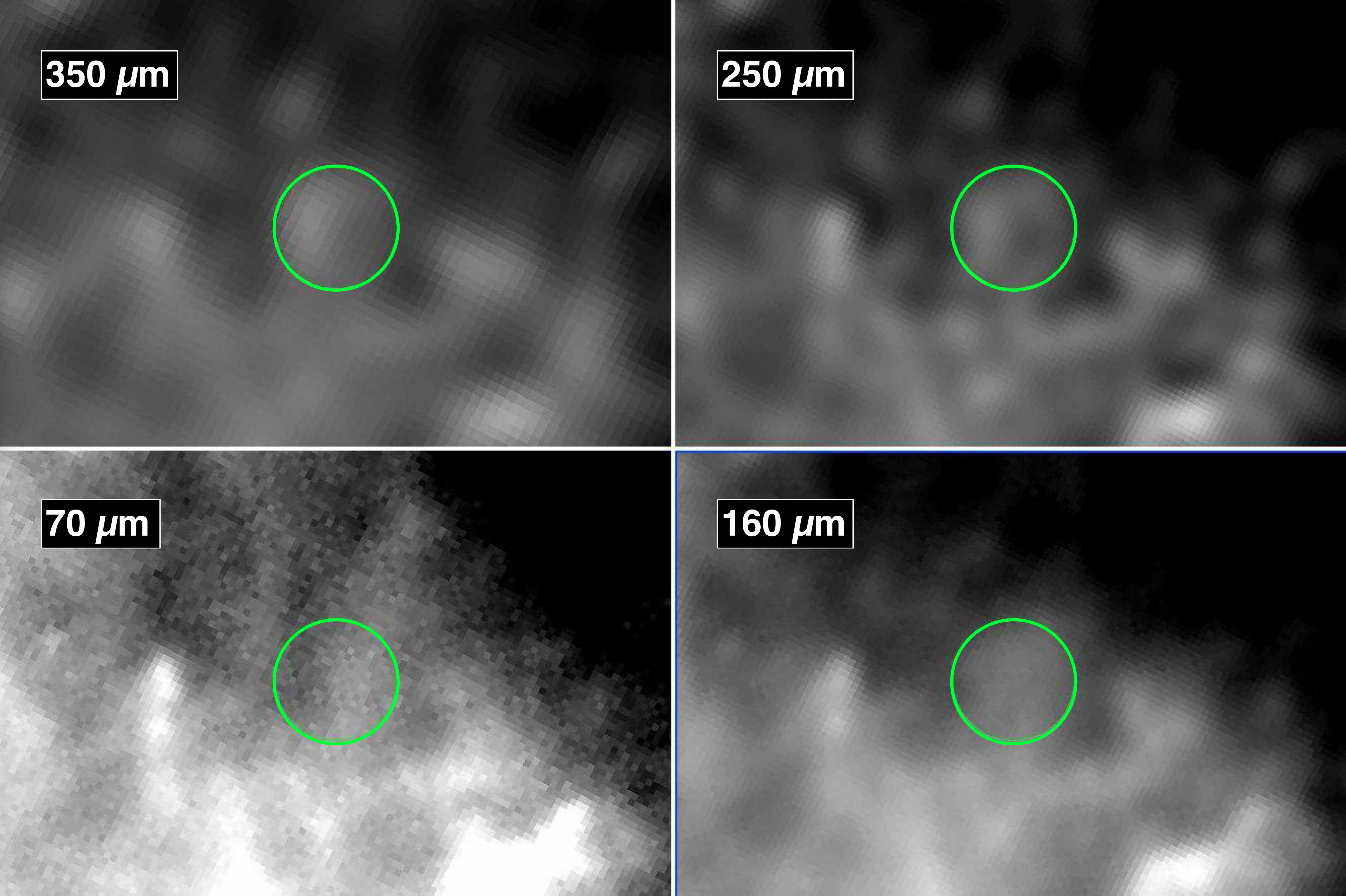}
\caption{The `coffee ring nebula.' {\it Top:} MWP V3 image asset centred at $(l,b)=(30.0968\degree, 0.2218\degree)$ with dimensions $0.15\degree\times0.075\degree$. {\it Bottom: Herschel} images centred on the ring (green circle) at (clockwise from upper-left panel) 350~$\mu$m, 250~$\mu$m, 160~$\mu$m and 70~$\mu$m.\label{fig:coffee}}
\end{figure}

MWP volunteers have proven themselves very capable of identifying unusual patterns and objects.
An excellent example of this is the `coffee ring nebula,' which looks like an almost perfectly circular IR dark cloud, ${\sim}1'$ in diameter (Figure~\ref{fig:coffee}). \citet{at11} first noted this object and remarked that its almost perfectly circular morphology is suggestive of an evolved circumstellar shell seen in absorption. 
However, the two brightest stars within the coffee ring are off-centre, and there is no star located close enough to the geometric centre of the ring in the GLIMPSE, 2MASS or Gaia DR2 catalogs to explain its  symmetric shape.

MWP volunteers classified the coffee ring as a bubble in DR1 (MWP1G030143+002282), however it was (correctly) omitted from our improved DR2 catalog. A MWP image containing the coffee ring was flagged on the MWP V3 talk forum by user @ZUCCO66 in September 2016 and has since been discussed by more than 19 individual users, some of whom have searched in vain for other examples of this type of object in the MWP image assets.

The coffee ring is not apparent in the \textit{Herschel}/PACS image at 70~\um, but it is clearly visible in emission at 250 and 350~$\mu$m in \textit{Herschel} SPIRE images \citep{2010A&A...518L...3G}, suggesting that it consists of cold ($T<20$~K) dust. The cold dust continuum highlights an asymmetry in the ring, which is brighter and thicker on the left-hand side (toward increasing Galactic longitude) than on the right; this asymmetry is also apparent in the 8~\um\ absorption (Figure~\ref{fig:coffee}). One of us (LDA) performed follow-up $^{13}$CO observations of the coffee ring, but these data do not provide a confident detection of any molecular line emission. The true nature of this object remains mysterious. 

\section{Summary and conclusions}
We have presented the second data release (DR2) from the Milky Way Project after aggregating nearly 3 million classification drawings made by citizen scientists during the years 2012 -- 2018. The DR2 catalogue contains 2600 IR bubbles and 599 bow shocks. 
The reliability of bubble identifications is assessed by comparison to the DR1 catalogue and the results of scoring by a machine-learning algorithm, while the reliability of IR bow shocks is assessed by comparison to visual identifications by trained `experts' and the locations of candidate bow shock driving stars found in the {\it 2MASS} point-source catalogue on the $J-H$ versus $H-K_s$ colour-colour diagram. We hence identify `highly-reliable' subsets of 1394 DR2 bubbles and 453 bow shocks.

An updated bubble drawing tool allowed MWP users to fit ellipses instead of elliptical annuli to bubble morphologies. This resulted in improved bubble shape and size measurements in DR2 compared to DR1. MWP users had access to a much larger set of image cutouts with a maximum zoom level that was twice that employed in DR1, enabling the identification of small bubbles to a greater degree of precision. The DR2 catalogue also eliminates the known issue of duplicated bubbles between the DR1 large and small bubble catalogues. We cross-matched the DR2 catalogue with the A14 catalogue of \textit{WISE} \hii regions to minimize the number of spurious bubbles. We retained only those DR2 bubbles with matches to the A14 catalogue and the unmatched bubbles that passed expert visual verification. The eccentricity distribution of the DR2 bubbles closely resembles the eccentricity distribution of the bubbles catalogued by CP06+CWP07 (unlike the DR1 bubbles that have lower eccentricities), suggesting that the DR2 catalogue more accurately captures bubble shapes. While the A14 catalogue is more complete and covers a larger patch of the sky, the MWP DR2 catalogue provides better size and shape measurements. 
Uncertainties on object coordinates and bubble size/shape parameters are also calculated and included in the DR2 catalogue.

With DR2, the results of applying the {\scshape Brut} machine-learning algorithm exhibit greater convergence with the work of MWP citizen scientists than was achieved with DR1. Machine-learning classifications have also validated our hit-rate criteria for identifying a more-reliable subset of DR2 bubbles. \citet{2014ApJS..214....3B} noted that up to 30\% of the 3744 large bubbles identified in DR1 could be spurious. The DR2 catalogue includes 2600 bubbles, which is ${\sim}30$\% fewer than the DR1 large bubble catalog, hence we conclude that DR1 bubbles eliminated from DR2 were most likely spurious. 

Our 2MASS colour analysis for BDSCs checks for consistency with the reddened OB locus and giant branch evolutionary states. Iterative removal of BDSCs through cuts made on hit rate shows an increase in the percentage of OB stars with a corresponding decrease in contamination from red giants. This trend confirms that the hit rate of a MWP BDSC is an appropriate measure of reliability. 

 MWP has produced a self-consistent and reproducible set of BDSCs, including 311 new candidate BDSCs associated with 284 newly-identified IR bow shock arcs that augment the K16 bow shock catalogue. Additionally, the MWP bow shock catalogue records associations to DR2 bubbles to help identify cases where a bow shock driving star is also the principal ionising source of a bubble.
Combining both the MWP DR2 and K16 IR bow shock catalogues, we have expanded the total number of these objects known in the Galaxy to nearly one thousand. 

The MWP DR2 catalogue expands upon and supplements existing records of \hii regions and bow shocks. MWP bubbles that have AT14 matches should be considered the highest-quality \hii region candidates, with the MWP providing more accurate positions, shapes and sizes for the interface between the ionised gas and surrounding PDRs. The MWP BDSC catalogue is of comparable size to K16, with over half the composing driving stars being unmatched to K16, displaying the ability of citizen scientists to contribute to bow shock discovery and validation. Because BDSCs found in both MWP and K16 represent independent rediscoveries, they are prime targets for spectroscopic follow-up to confirm or reject their OB types. Followup studies of these MWP objects will continue to improve our understanding of massive star formation throughout the Milky Way.

\section*{Acknowledgments}
We thank the referee, Alberto Noriega-Crespo, for providing a timely, positive review of this work. We thank the MWP moderators, Melina Th\` evenot, Barbara T\'egl\'as, Dennis Stanescu, Julia Wilkinson and Elisabeth Baeten, for their work facilitating discussion in the MWP Talk forums and bringing serendipitous discoveries (including the Coffee Ring Nebula and new RCW 49 bow shock) to the attention of the research team. We thank J. E. Andrews for his work during the preliminary phases of the MWP bow shock search and for his help spotting the new bow shock candidates associated with NGC 3603. 

This work was supported by the U.S. National Science Foundation through grants CAREER-1454333, AST-1412845, and AST-1411851. DX and SSRO are supported by NSF grant AST-1812747. This work is based on observations made with the {\it Spitzer Space Telescope}, which is operated by the Jet Propulsion Laboratory, California Institute of Technology under a contract with NASA. This publication uses data generated via the Zooniverse.org platform, development of which is funded by generous support, including a Global Impact Award from Google, and by a grant from the Alfred P. Sloan Foundation.




\bibliographystyle{mnras}




\bsp	
\label{lastpage}


\end{document}